\documentclass[11pt]{article}
\usepackage[margin=24mm]{geometry}
\usepackage{amsmath,amssymb,bm,booktabs,longtable,array,graphicx,microtype,etoolbox}
\usepackage[T1]{fontenc}
\usepackage[colorlinks=true,linkcolor=blue,citecolor=blue,urlcolor=blue]{hyperref}
\DeclareMathOperator{\sn}{sn}\DeclareMathOperator{\cn}{cn}\DeclareMathOperator{\dn}{dn}
\newcommand{\anti}[2]{\{#1,#2\}}
\DeclareMathOperator{\tr}{tr}
\allowdisplaybreaks
\title{Integrable Orbital Exchange between\\Ellipsoidally Deformed $p$, $d$, and $f$ Electron States\\[5pt]\large Two-Angle Sklyanin Representations and Fused Elliptic $R$-Matrices}
\author{Vladimir Mikhalev}
\date{}
\makeatletter
\patchcmd{\@maketitle}{\vskip 2em}{\vskip 0pt}{}{}
\patchcmd{\@maketitle}{\vskip 1.5em}{\vskip .75em}{}{}
\patchcmd{\@maketitle}{\vskip 1em}{\vskip 0pt}{}{}
\patchcmd{\@maketitle}{\vskip 1.5em}{\vskip .5em}{}{}
\newcommand{\compactcontents}{%
  \begingroup\footnotesize\setlength{\parskip}{0pt}\setcounter{tocdepth}{1}%
  \fontsize{9}{10.5}\selectfont\setlength{\parskip}{0pt}%
  \renewcommand*\l@section{\@dottedtocline{1}{0em}{1.8em}}%
  \tableofcontents\endgroup}
\makeatother
\hypersetup{pdftitle={Integrable Orbital Exchange between Ellipsoidally Deformed p, d, and f Electron States},
  pdfauthor={Vladimir Mikhalev}}
\begin{document}\maketitle
\begin{abstract}
We construct integrable orbital exchange models from ellipsoidally deformed $p$, $d$, and $f$ multiplets. A two-angle realization of the Sklyanin algebra gives explicit Jacobi difference operators $D,L_i$, complete distortion spectra and eigenfunctions, and a positive inner product transported from Sklyanin's integral. In the $p$ sector the coordinate functions carry an exact spin-one action, and applying $D$ componentwise maps the confocal ellipsoid to a sphere. Elliptic fusion gives explicit $R$-matrices in the angular generators on both factors. Their Yang--Baxter relations yield commuting charges and exchange Hamiltonians for homogeneous and alternating chains, as well as mixed-pair interactions that can act between angular degrees of freedom on the same center. Exchange changes angular projections and correlations while preserving the two local Casimirs of each constituent, which fix its angular-momentum-squared and distortion-plus-rotation kinetic-energy invariants. The $p$--$p$ model recovers Fateev's elliptic spin-one interaction and reduces to the Takhtajan--Babujian Hamiltonian in the rational isotropic limit.
\end{abstract}
\compactcontents\clearpage

\section{Introduction}\label{sec:introduction}
In highly anisotropic molecules and crystals, local distortion can change both the angular states of an electron and the interactions that couple its orbital degrees of freedom to others. Spherical symmetry imposes selection rules that make some orbital-resolved matrix elements vanish. Lowering the local symmetry can activate additional contributions and modify existing hopping and exchange pathways. Jahn--Teller coupling provides one microscopic route to distortion by coupling an orbitally degenerate electronic state to nuclear displacements. Pseudo-Jahn--Teller coupling mixes distinct electronic states and can destabilize a high-symmetry configuration. The resulting changes in orbital shape, orientation, and hybridisation motivate two related questions. How can we construct angular bases adapted to a finite distortion, and how can we use them to describe integrable exchange that mixes the angular states while preserving the local angular-momentum and energy invariants?

\begin{figure}[!htbp]
\centering
\includegraphics[width=0.87\textwidth]{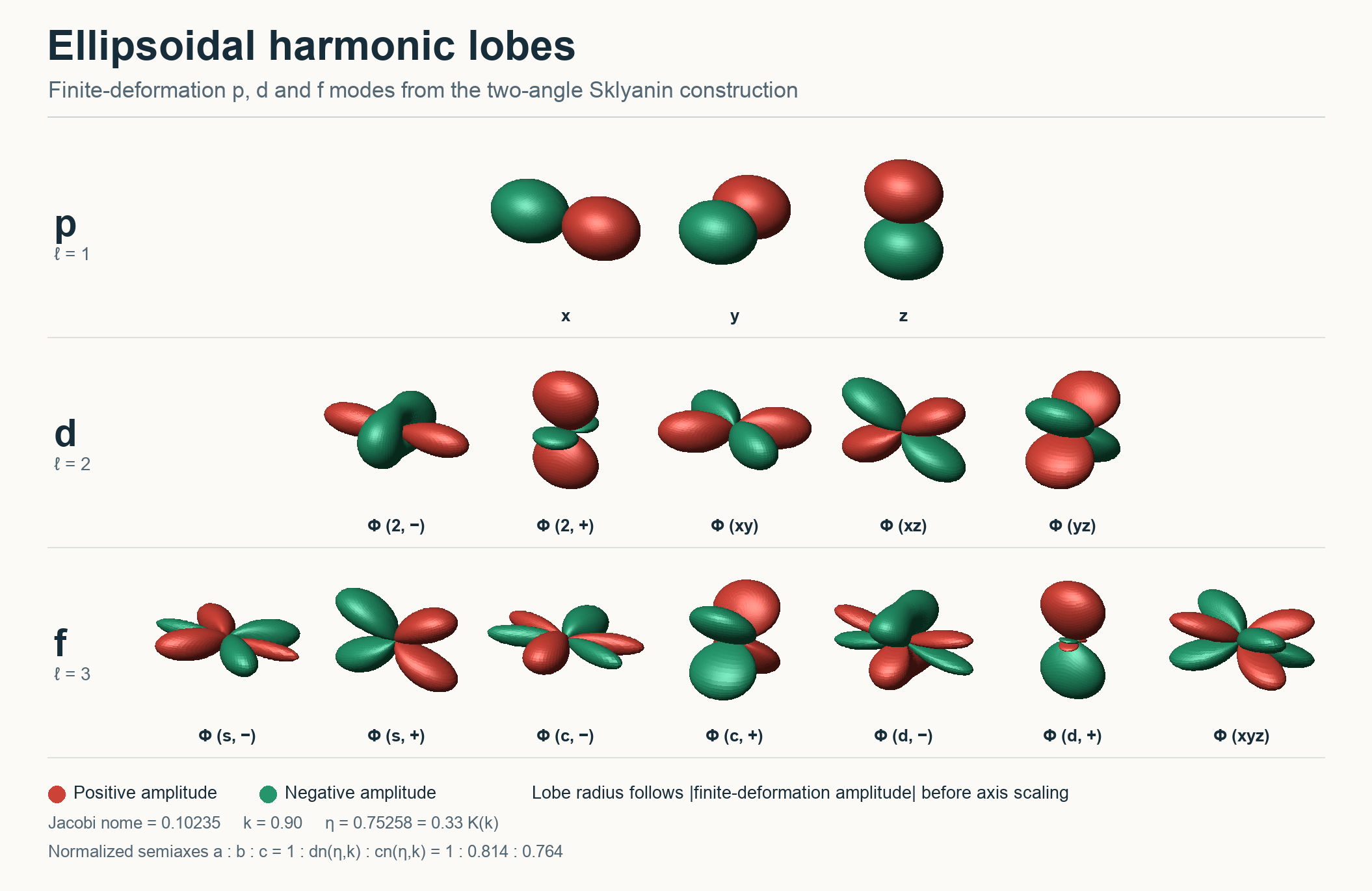}
\caption{Lobe plots of the two-angle Sklyanin bases for the $p$, $d$, and $f$ sectors. Each mode is scaled to unit peak magnitude. The distance from the origin before axis scaling is proportional to the magnitude of its angular amplitude, and red and green indicate opposite signs. The plotted coordinates use the confocal semiaxis ratios $1:\dn(\eta,k):\cn(\eta,k)$ from Eq.~\eqref{eq:surface}. The angular functions and the geometry use $k=0.90$ and $\eta=0.33K(k)$, with Jacobi nome $q_{\mathrm J}=\exp(-\pi K'/K)\simeq0.10235$. These values illustrate the geometry and lie outside the sufficient common positivity domain in Eq.~\eqref{eq:positive-branch}.}
\label{fig:harmonic-lobes}
\end{figure}

The Sklyanin algebra was introduced in the study of the Yang--Baxter equation and elliptic quantum integrable systems~\cite{sklyanin1982,sklyanin1983}. Its difference-operator realizations admit finite-dimensional invariant spaces. We start from the two-variable realization of Krichever and Zabrodin, whose variables obey a difference constraint~\cite{krichever1998vacuum}, and convert it to Jacobi form in the ellipsoidal angles $\theta,\varphi$ in Appendix~\ref{app:construction}. The result is an explicit set of difference operators $D,L_1,L_2,L_3$ with deformation parameters $k,\eta$. The orbital degree $\ell=0,1,2,3$ indexes the $s,p,d,f$ sectors, respectively. The $s$ sector is a scalar representation, while the $p$, $d$, and $f$ sectors have three, five, and seven angular states, respectively. Figure~\ref{fig:harmonic-lobes} shows the three nontrivial sectors at finite deformation.

Sections~\ref{sec:p-spectrum}--\ref{sec:f-spectrum} give the four operators in each nontrivial sector, together with the complete distortion eigenvalues and angular eigenfunctions. The $p$ eigenstates have simple coordinate forms and exact three-by-three angular-momentum matrices. The $d$ and $f$ eigenstates exhibit deformation-dependent mixing within reflection sectors and give five and seven distortion levels, respectively. The angular bases connect continuously to the corresponding spherical sectors. The higher-sector generators generally have deformed commutation relations at finite step and recover the ordinary differential $su(2)$ generators as the step vanishes. The triplet also has an exact spin-one matrix identification at finite step. Because the elliptic generators act by finite differences, the usual differential construction $\bm L=\bm r\times\bm p$ does not carry over directly. The existence of an appropriate translational momentum operator and its relation to the deformed generators remain open.

These bases allow the angular contributions to low-symmetry ligand- and crystal-field splittings and field-induced orbital mixing to be evaluated with finite matrix Hamiltonians. Quantitative electronic spectra require radial functions, a microscopic potential, and the relevant electron-spin and electron-interaction terms. Strong anisotropy may also require coupling between different orbital sectors.

The first quadratic Casimir, $C_1=\sum_i L_i^2$, fixes the local angular-momentum-squared invariant. The second, $C_2=D^2+\sum_i J_i^2L_i^2$, combines distortion and weighted rotation and gives the local Casimir kinetic energy when multiplied by a fixed energy scale. Both are scalars in each irreducible sector. Anisotropic or external-field perturbations can drive dynamics within a sector, as illustrated by the coherent reversal of a triplet's angular-momentum projection in Section~\ref{sec:triplet-dynamics}. In the coupled models, exchange can change individual angular projections and redistribute the distortion and rotational contributions at fixed local Casimir values. Equation~\eqref{eq:casimirs} and Section~\ref{sec:casimir-preservation} state the invariants and their conservation under exchange.

We define the inner product by transporting Sklyanin's positive integral through the finite-dimensional intertwiner and express the density in Jacobi form. The conversion of the integral is exact. The identification of the displayed $d$ and $f$ generators with the corresponding transported operators has been checked numerically at the specified parameters. The positive coefficient metric determines which angular-basis observables are Hermitian and gives the norm for unitary evolution. Section~\ref{sec:metric} and Appendix~\ref{app:jacobimetric} give the metric and the scope of the finite-sector checks.

Baxter's elliptic $R$-matrix formalism turns these angular representations into local quantum spaces for integrable models of interacting atoms or localized orbital states. With the same angular translation step on both factors and in the fusion string, we construct explicit elliptic $R$-matrices in the angular generators $D,L_i$ on each factor~\cite{kulish1981fusion,takebe1996,konno2005}. The mixed $p$--$d$, $p$--$f$, and $d$--$f$ matrices describe interactions between angular degrees of freedom on different sites or on the same atomic center. Each tensor factor represents one angular degree of freedom, so several factors can share a center. At finite distortion, cross-sector overlaps in a common one-particle embedding are governed by the surviving symmetry and can be nonzero for different orbital sectors. Section~\ref{sec:onsite-pairs} describes the onsite interpretation and the role of the cross-sector overlap matrix.

A concrete motivation for the mixed $R^{(d,f)}$ family is CeCoSi. Matsumura et al.~\cite{matsumura2026cecosi} associate acoustic softening with a $(yz+zx)$-type distortion and propose that mixing of the $5d$ crystal-field states activates the matrix element $\langle 4f\text{-}\Gamma_{7(1)}|\mathcal H_{fd}|5d\text{-}\Gamma_6\rangle$, which vanishes in the tetragonal reference configuration. Here $\mathcal H_{fd}$ is the local hybridisation operator and $\Gamma_{7(1)},\Gamma_6$ label crystal-field doublets. The Ce sites already lack inversion symmetry, so the distortion opens an additional channel within the existing $f$--$d$ hybridisation. The $35\times35$ matrix $R^{(d,f)}$ supplies a candidate integrable description of the associated effective orbital exchange. The Ce doublets require a spin--orbit and crystal-field embedding.

{\clubpenalty=10000
A complementary motivation for $R^{(p,d)}$ comes from BaTiO$_3$. In the TiO$_6$ cluster, Sah et al.~\cite{sah2025batio3} describe a polar displacement of $t_{1u}$ symmetry that couples the \({}^{1}A_{1g}\) ground state to \({}^{1}T_{1u}\) excited states through the pseudo-Jahn--Teller effect. These many-electron cluster states have opposite parity, and their mixing is forbidden at the centrosymmetric reference geometry. The polar displacement supplies the inversion-odd component needed for the additional mixing channel within the existing Ti--O covalency. An inversion-preserving ellipsoidal deformation retains the parity selection rule. In this oxygen-$2p$/titanium-$3d$ setting, the $15\times15$ matrix $R^{(p,d)}$ gives a candidate model for the angular part of distortion-dependent ligand--metal interactions. A quantitative application would relate the exchange to virtual charge-transfer processes and identify the physical distortion with the model parameters.\par}

The auxiliary $R^{(\mathrm{aux},p)}$ matrix could also provide an effective angular model for $sp$ hybridisation in distorted molecules and crystals, once the auxiliary channel is embedded in the physical orbital configurations. Both $R^{(\mathrm{aux},p)}$ and $R^{(p,p)}$ could describe angular interactions in valence-bond distortions. Section~\ref{sec:hybridisation} explains the physical interpretation of the auxiliary channel.

Quantum integrability follows from the quantum inverse-scattering construction~\cite{faddeev1996}. The spectral-parameter-dependent fused matrices satisfy the Yang--Baxter equation and generate commuting transfer matrices. Their logarithmic derivatives give mutually commuting charges, including the exchange Hamiltonians. This construction is the quantum counterpart of the classical Lax construction. The resulting models include finite pair spectra, homogeneous chains with nearest-neighbor densities, and alternating chains with two- and three-site densities. The $p$--$p$ member is related by an explicit angular-basis transformation to Fateev's elliptic spin-one model in the $21$-vertex family~\cite{fateev1981,konno2005}. In the rational isotropic limit it gives the Takhtajan--Babujian bilinear--biquadratic Hamiltonian. Section~\ref{sec:fateev-hamiltonian} specifies the coupling convention.

These Hamiltonians give controlled reference models for orbital exchange and a starting point for material-specific perturbations. Whether a given distortion is energetically stable depends on the electronic and elastic energies. Section~\ref{sec:physical-model} defines the observables and exchange scales, and Section~\ref{sec:integrable-exchange} gives the local densities and their isotropic reference. Sections~\ref{sec:fundamental}--\ref{sec:fused-higher} construct the seed and the explicit matrices, including the two-space generator polynomial and the higher-row entry formulas. The appendices provide the angular derivations, basis maps, and scalar normalization. The higher-spin exchange construction uses a regular irreducible branch. The numerical checks of the angular realization and fused matrices establish the accuracy of the finite implementations at the stated parameters, with general-deformation closure retaining the qualification in Section~\ref{sec:angular-verification}.

\section{Deformation parameters and the distortion spectrum}\label{sec:parameters}
The elliptic modulus $0<k<1$ and the shift $\eta$ control the angular deformation. Equivalently, the modulus can be specified by the Jacobi nome
\begin{equation}
q_{\mathrm J}=\exp[-\pi K(k')/K(k)],
\qquad k'=\sqrt{1-k^2}.
\end{equation}
Here $K=K(k)$ and $K'=K(k')$, where $K(k)$ is the complete elliptic integral of the first kind. We write the modulus as the second argument of a Jacobi function. The coordinate formulas use the regular real branch $0<3\eta<K$, with coordinate and shift poles excluded. For exchange with the positive integral metric we use the stronger common sufficient restriction
\begin{equation}
0<k<1,\qquad 0<7\eta/2<K.
\label{eq:positive-branch}
\end{equation}
The orbital degree $\ell=0,1,2,3$ corresponds to $s,p,d,f$, respectively. The angular space $\mathcal V_\ell$ is spanned by the displayed eigenfunctions, and the numerical indices on $D_\ell$ and $L_i^{(\ell)}$ refer to that degree.

We write
\begin{equation*}
\begin{aligned}
s_n&=\sn(n\eta,k),\quad c_n=\cn(n\eta,k),\\
d_n&=\dn(n\eta,k),\\
(s_\varphi,c_\varphi,d_\varphi)&=(\sn,\cn,\dn)(\varphi,k),\\
(s_\theta,c_\theta,d_\theta)&=(\sn,\cn,\dn)(\theta,k').
\end{aligned}
\end{equation*}
The common angular coefficients are
\begin{equation}
J_1=c_1d_1,\qquad J_2=c_1,\qquad J_3=d_1.
\label{eq:J}
\end{equation}
We use the commutator $[A,B]=AB-BA$ and anticommutator $\anti{A}{B}=AB+BA$. In our normalization the quadratic relations are
\begin{align}
[D,L_i]&=i(J_j-J_h)\anti{L_j}{L_h},\label{eq:alg1}\\
\anti{D}{L_i}&=-i(J_j+J_h)[L_j,L_h],\label{eq:alg2}
\end{align}
where $(i,j,h)$ is a cyclic permutation of $(1,2,3)$. Appendix~\ref{app:operators} gives the conversion from the standard generators.

We construct two-angle eigenfunctions with matching distortion eigenvalues for the two one-angle operators,
\begin{equation}
D_\ell^\varphi\Phi_{\ell,a}=D_\ell^\theta\Phi_{\ell,a}=\lambda_{\ell,a}\Phi_{\ell,a}.
\label{eq:spectralproblem}
\end{equation}
On $\mathcal V_\ell$, $D_\ell$ denotes the common restriction of $D_\ell^\varphi$ and $D_\ell^\theta$. The eigenvalues $\lambda_{\ell,a}$ characterize distortion. The angular-momentum and energy invariants are
\begin{equation}
\begin{aligned}
C_1&=L_1^2+L_2^2+L_3^2,\\
C_2&=D^2+J_1^2L_1^2+J_2^2L_2^2+J_3^2L_3^2,
\end{aligned}
\label{eq:casimirs}
\end{equation}
Both commute with all four generators. The conserved total angular-momentum squared is $\hbar^2C_1$. For an energy scale $\varepsilon$, the unperturbed Hamiltonian is
\begin{equation}
H_0=\varepsilon C_2
=\varepsilon D^2+\varepsilon\sum_iJ_i^2L_i^2.
\label{eq:casimir-energy}
\end{equation}
The first term is the distortion contribution and the second is the rotational contribution. Their sum is conserved under dynamics generated by any polynomial in the represented generators. At fixed distortion, the rotational part has the weighted quadratic form of a rigid-body Hamiltonian.

In each irreducible sector, $C_1$ and $C_2$ are scalar matrices. The $p$ sector has $C_1=2I$ and $C_2=(J_1^2+J_2^2+J_3^2)I$. The $d$ and $f$ values depend on the deformation, with $C_1\to6I$ and $12I$, respectively, in the spherical limit. Thus all distortion eigenstates in one sector share the same unperturbed energy. A separately specified perturbation $gD_\ell$ contributes shifts $g\lambda_{\ell,a}$, where $g$ is its energy coefficient. The following eigenstate catalogues provide a basis for this perturbation and for additional field, spin, and interaction terms.

The common angular denominator is $\Delta=1-k^2s_\varphi^2-k'^2s_\theta^2$. The translations in the formulas below act as
\begin{equation}
\begin{aligned}
\cosh(\eta\partial_\varphi)F&=\tfrac12[F(\theta,\varphi+\eta)+F(\theta,\varphi-\eta)],\\
\sinh(\eta\partial_\varphi)F&=\tfrac12[F(\theta,\varphi+\eta)-F(\theta,\varphi-\eta)],\\
\cos(\eta\partial_\theta)F&=\tfrac12[F(\theta+i\eta,\varphi)+F(\theta-i\eta,\varphi)],\\
\sin(\eta\partial_\theta)F&=\tfrac1{2i}[F(\theta+i\eta,\varphi)-F(\theta-i\eta,\varphi)].
\end{aligned}
\label{eq:main-translations}
\end{equation}
All coefficient functions multiply the translations from the left. In each sector the angular states obey the difference constraint $(D_\ell^\varphi-D_\ell^\theta)F=0$. The following sections give the explicit two-angle operators and their distortion eigenstates. Appendix~\ref{app:operators} collects the algebraic normalization and spherical limits.

\section[The p distortion spectrum and angular momentum]{The $p$ distortion spectrum and angular momentum}\label{sec:p-spectrum}
\subsection[Two-angle D and angular-momentum operators]{Two-angle $D$ and angular-momentum operators}
The explicit $p$ representation is
\begin{equation}
\begin{aligned}
\mathcal A_1(\varphi)&=1-k^2s_1^2s_\varphi^2,&
\mathcal B_1(\theta)&=1-s_1^2d_\theta^2,
\end{aligned}
\label{eq:p-weights}
\end{equation}
\begin{equation}
D_1^\varphi=\mathcal A_1\cosh(\eta\partial_\varphi),
\qquad D_1^\theta=\mathcal B_1\cos(\eta\partial_\theta),
\label{eq:p-D}
\end{equation}
\begin{equation}
\boxed{\begin{aligned}
L_1^{(1)}&=-\frac{i}{s_1\sqrt{c_1d_1}\,\Delta}
\left[\mathcal A_1 k'^2s_\theta c_\theta s_\varphi\,\sinh(\eta\partial_\varphi)
+\mathcal B_1 d_\theta c_\varphi d_\varphi\,\sin(\eta\partial_\theta)\right],\\
L_2^{(1)}&=-\frac{i}{s_1\sqrt{c_1}\,\Delta}
\left[\mathcal A_1 s_\theta d_\theta c_\varphi\,\sinh(\eta\partial_\varphi)
-\mathcal B_1 c_\theta s_\varphi d_\varphi\,\sin(\eta\partial_\theta)\right],\\
L_3^{(1)}&=\frac{i}{s_1\sqrt{d_1}\,\Delta}
\left[\mathcal A_1 c_\theta d_\theta d_\varphi\,\sinh(\eta\partial_\varphi)
+\mathcal B_1 k^2s_\theta s_\varphi c_\varphi\,\sin(\eta\partial_\theta)\right].
\end{aligned}}
\label{eq:p-operators}
\end{equation}

\subsection[The three p eigenstates]{The three $p$ eigenstates}
The separated factors belong to two distinct one-angle spaces, $\mathcal W_\ell^\varphi$ for functions of $\varphi$ and $\mathcal W_\ell^\theta$ for functions of $\theta$. Each has dimension $2\ell+1$. The $\varphi$ bases are listed with the spectra. The corresponding $\theta$ spaces are obtained by the complementary substitution in Appendix~\ref{app:separation}. The restrictions of the operators in Eq.~\eqref{eq:constraint} act on $f\in\mathcal W_\ell^\varphi$ and $g\in\mathcal W_\ell^\theta$ as
\begin{equation}
\begin{aligned}
(D_\ell^\varphi f)(\varphi)
 &=\frac{\mathcal A_\ell(\varphi)}{2}
   \bigl[f(\varphi+\eta)+f(\varphi-\eta)\bigr],\\
(D_\ell^\theta g)(\theta)
 &=\frac{\mathcal B_\ell(\theta)}{2}
   \bigl[g(\theta+i\eta)+g(\theta-i\eta)\bigr].
\end{aligned}
\label{eq:D-on-W}
\end{equation}
Each operator preserves its corresponding space. For the $p$ sector these spaces are
\begin{equation}
\mathcal W_1^\varphi=\operatorname{span}\{s_\varphi,c_\varphi,d_\varphi\},
\qquad
\mathcal W_1^\theta=\operatorname{span}\{d_\theta,c_\theta,s_\theta\}.
\label{eq:W1}
\end{equation}
The Jacobi addition identities give the paired one-angle actions
\begin{equation}
\begin{aligned}
D_1^\varphi s_\varphi&=J_1s_\varphi, & D_1^\theta d_\theta&=J_1d_\theta,\\
D_1^\varphi c_\varphi&=J_2c_\varphi, & D_1^\theta c_\theta&=J_2c_\theta,\\
D_1^\varphi d_\varphi&=J_3d_\varphi, & D_1^\theta s_\theta&=J_3s_\theta.
\end{aligned}
\label{eq:D1-on-W1}
\end{equation}
Multiplying the paired factors with the same eigenvalue gives the two-angle eigenstates. Each operator acts on its own factor with the other angle held fixed.

The functions
\begin{equation}
x=d_\theta s_\varphi,\qquad y=c_\theta c_\varphi,\qquad z=s_\theta d_\varphi
\label{eq:unitcoords}
\end{equation}
satisfy $x^2+y^2+z^2=1$. They are Cartesian coordinate functions on the unit sphere, parametrized by spheroconical angles. Each is a product of functions of $\theta$ and $\varphi$.

The three distortion eigenstates and their eigenvalues are
\begin{equation}
\begin{array}{c|c|c}
\text{state} & \Phi & \lambda\\\hline
p_x & x=d_\theta s_\varphi & J_1=c_1d_1\\
p_y & y=c_\theta c_\varphi & J_2=c_1\\
p_z & z=s_\theta d_\varphi & J_3=d_1
\end{array}.
\label{eq:tripleteigen}
\end{equation}
On each displayed product $\Phi$, $D_1^\varphi\Phi=D_1^\theta\Phi=\lambda\Phi$.
Thus $\mathcal V_1=\operatorname{span}\{x,y,z\}$ lies in $\ker\mathcal N_1$. The imaginary shift and the factor $1-\sn^2(\eta,k)\dn^2(\theta,k')$ are essential to this equality.

\subsection{Action on the triplet}
Use the rescaled coordinate basis
\begin{equation}
e_1=x,\qquad e_2=y/\sqrt{d_1},\qquad e_3=z/\sqrt{c_1}.
\label{eq:tripletbasis}
\end{equation}
The $p$ operators in Eq.~\eqref{eq:p-operators} act on these states as
\begin{equation}
\begin{array}{c|ccc}
&e_1&e_2&e_3\\\hline
L_1&0&i e_3&-i e_2\\
L_2&-i e_3&0&i e_1\\
L_3&i e_2&-i e_1&0
\end{array}.
\label{eq:actiontable}
\end{equation}
In this basis $D_1=\operatorname{diag}(J_1,J_2,J_3)$ and
\begin{equation}
\begin{aligned}
L_1&=\begin{pmatrix}0&0&0\\0&0&-i\\0&i&0\end{pmatrix},\quad
L_2=\begin{pmatrix}0&0&i\\0&0&0\\-i&0&0\end{pmatrix},\\
L_3&=\begin{pmatrix}0&-i&0\\i&0&0\\0&0&0\end{pmatrix}.
\end{aligned}
\label{eq:spinonematrices}
\end{equation}
Columns represent images of basis vectors. These matrices satisfy both Eqs.~\eqref{eq:alg1}--\eqref{eq:alg2} and the ordinary spin-one relations
\begin{equation}
[L_i,L_j]=i\epsilon_{ijh}L_h,\qquad \sum_iL_i^2=2I.
\label{eq:spinonecasimir}
\end{equation}
The extra $su(2)$ identities concern this representation; they are not identities on the entire function space or the abstract Sklyanin algebra.

For the triplet dynamics below we use the common normalization of Eq.~\eqref{eq:normalizedtriplet}, which leaves these matrices unchanged. Section~\ref{sec:metric} identifies the positive inner product through transport of Sklyanin's integral metric. This also supplies the basis for the $d$- and $f$-sector Hermiticity checks. It is distinct from an unweighted angle integral or the unchanged spherical area measure. In the spherical limit $\eta\to0$, however, $\cn(\eta,k),\dn(\eta,k)\to1$, and the triplet metric tends to $(4\pi/3)I$, exactly the Gram matrix of $(x,y,z)$ under the standard round-sphere measure $d\Omega$. The normalized states therefore tend to $\sqrt{3/(4\pi)}(x,y,z)$, the usual real degree-one spherical harmonics. The geometric sphere metric itself is independent of $\eta$; it is the deformed inner product on the functions that approaches the spherical one.

\section[The d two-angle representation and distortion spectrum]{The $d$ two-angle representation and distortion spectrum}
\subsection[Two-angle D and angular-momentum operators]{Two-angle $D$ and angular-momentum operators}
The explicit $d$ representation is
\begin{equation}
\begin{aligned}
\mathcal A_2(\varphi)&=\frac{1-k^2s_2^2s_\varphi^2}{1-k^2s_1^2s_\varphi^2},&
\mathcal B_2(\theta)&=\frac{1-s_2^2d_\theta^2}{1-s_1^2d_\theta^2},
\end{aligned}
\label{eq:d-weights}
\end{equation}
\begin{equation}
D_2^\varphi=\mathcal A_2\cosh(\eta\partial_\varphi),
\qquad D_2^\theta=\mathcal B_2\cos(\eta\partial_\theta),
\label{eq:d-D}
\end{equation}
\begin{equation}
\boxed{\begin{aligned}
L_1^{(2)}&=-\frac{i}{s_1\sqrt{c_1d_1}\,\Delta}
\left[\mathcal A_2 k'^2s_\theta c_\theta s_\varphi\,\sinh(\eta\partial_\varphi)
+\mathcal B_2 d_\theta c_\varphi d_\varphi\,\sin(\eta\partial_\theta)\right],\\
L_2^{(2)}&=-\frac{i}{s_1\sqrt{c_1}\,\Delta}
\left[\mathcal A_2 s_\theta d_\theta c_\varphi\,\sinh(\eta\partial_\varphi)
-\mathcal B_2 c_\theta s_\varphi d_\varphi\,\sin(\eta\partial_\theta)\right],\\
L_3^{(2)}&=\frac{i}{s_1\sqrt{d_1}\,\Delta}
\left[\mathcal A_2 c_\theta d_\theta d_\varphi\,\sinh(\eta\partial_\varphi)
+\mathcal B_2 k^2s_\theta s_\varphi c_\varphi\,\sin(\eta\partial_\theta)\right].
\end{aligned}}
\label{eq:d-operators}
\end{equation}

\subsection[The five d eigenstates]{The five $d$ eigenstates}\label{sec:d-spectrum}
The $d$ sector contains five angular eigenstates of the distortion operator. Their eigenvalues and eigenfunctions are paired explicitly by
\begin{equation}
\begin{array}{c|c}
\text{eigenfunction} & D_2\text{ eigenvalue}\\\hline
\Phi_{2,-},\ \Phi_{2,+} & \lambda_{2,-},\ \lambda_{2,+}\\
\Phi_{xy} & \lambda_{sc}\\
\Phi_{xz} & \lambda_{sd}\\
\Phi_{yz} & \lambda_{cd}
\end{array}
\label{eq:d-spectrum}
\end{equation}
Each two-angle state satisfies $D_2^\varphi\Phi=D_2^\theta\Phi=\lambda\Phi$. The labels $xy,xz,yz$ identify the reflection parities of the three off-diagonal modes. The two remaining states mix the diagonal $d$ patterns. Their eigenvalues and mixing coefficients are
\begin{equation}
\begin{aligned}
\lambda_{2,\pm}
&=\frac{(M_2)_{11}+(M_2)_{22}
\pm\sqrt{[(M_2)_{11}-(M_2)_{22}]^2+4(M_2)_{12}(M_2)_{21}}}{2},\\
r_\pm&=\frac{\lambda_{2,\pm}-(M_2)_{11}}{(M_2)_{12}}.
\end{aligned}
\label{eq:d-mixing}
\end{equation}
The explicit two-state matrix $M_2$ is given in Appendix~\ref{app:blocks}. Its entries depend only on $k$ and $\eta$.
The three off-diagonal eigenvalues are
\begin{equation*}
\begin{aligned}
\lambda_{sc}&=\frac{d_1(1-2s_1^2+k^2s_1^4)}{(1-k^2s_1^4)^2},\\
\lambda_{sd}&=\frac{c_1(1-2k^2s_1^2+k^2s_1^4)}{(1-k^2s_1^4)^2},\\
\lambda_{cd}&=\frac{c_1d_1}{1-k^2s_1^4}.
\end{aligned}
\end{equation*}
The corresponding five unnormalized angular eigenfunctions are
\begin{equation}
\begin{aligned}
\Phi_{2,\pm}&=\frac{(1+r_\pm s_\varphi^2)(1+r_\pm d_\theta^2/k^2)}{[1-k^2\sn^2(\eta,k)\sn^2(\varphi,k)][1-\sn^2(\eta,k)\dn^2(\theta,k')]},\\
\Phi_{xy}&=\frac{s_\varphi c_\varphi c_\theta d_\theta}{[1-k^2\sn^2(\eta,k)\sn^2(\varphi,k)][1-\sn^2(\eta,k)\dn^2(\theta,k')]},\\
\Phi_{xz}&=\frac{s_\varphi d_\varphi s_\theta d_\theta}{[1-k^2\sn^2(\eta,k)\sn^2(\varphi,k)][1-\sn^2(\eta,k)\dn^2(\theta,k')]},\\
\Phi_{yz}&=\frac{c_\varphi d_\varphi s_\theta c_\theta}{[1-k^2\sn^2(\eta,k)\sn^2(\varphi,k)][1-\sn^2(\eta,k)\dn^2(\theta,k')]}.
\end{aligned}
\label{eq:fivebasis}
\end{equation}
For reference, the one-variable $d$ basis spans
\begin{equation*}
\mathcal W_2^\varphi=\frac{1}{[1-k^2\sn^2(\eta,k)\sn^2(\varphi,k)]}\operatorname{span}
\{1,s_\varphi^2,s_\varphi c_\varphi,s_\varphi d_\varphi,c_\varphi d_\varphi\}.
\end{equation*}

These states approach the five ordinary $d$ harmonics as $\eta\to0$. Their normalization is specified in Sec.~\ref{sec:metric}.

\section[The f two-angle representation and distortion spectrum]{The $f$ two-angle representation and distortion spectrum}\label{sec:f-spectrum}
\subsection[Two-angle D and angular-momentum operators]{Two-angle $D$ and angular-momentum operators}
The explicit $f$ representation is
\begin{equation}
\begin{aligned}
\mathcal A_3(\varphi)&=\frac{(1-k^2s_3^2s_\varphi^2)(1-k^2s_1^2s_\varphi^2)}{1-k^2s_2^2s_\varphi^2},&
\mathcal B_3(\theta)&=\frac{(1-s_3^2d_\theta^2)(1-s_1^2d_\theta^2)}{1-s_2^2d_\theta^2}.
\end{aligned}
\label{eq:f-weights}
\end{equation}
\begin{equation}
D_3^\varphi=\mathcal A_3\cosh(\eta\partial_\varphi),
\qquad D_3^\theta=\mathcal B_3\cos(\eta\partial_\theta),
\label{eq:f-D}
\end{equation}
\begin{equation}
\boxed{\begin{aligned}
L_1^{(3)}&=-\frac{i}{s_1\sqrt{c_1d_1}\,\Delta}
\left[\mathcal A_3 k'^2s_\theta c_\theta s_\varphi\,\sinh(\eta\partial_\varphi)
+\mathcal B_3 d_\theta c_\varphi d_\varphi\,\sin(\eta\partial_\theta)\right],\\
L_2^{(3)}&=-\frac{i}{s_1\sqrt{c_1}\,\Delta}
\left[\mathcal A_3 s_\theta d_\theta c_\varphi\,\sinh(\eta\partial_\varphi)
-\mathcal B_3 c_\theta s_\varphi d_\varphi\,\sin(\eta\partial_\theta)\right],\\
L_3^{(3)}&=\frac{i}{s_1\sqrt{d_1}\,\Delta}
\left[\mathcal A_3 c_\theta d_\theta d_\varphi\,\sinh(\eta\partial_\varphi)
+\mathcal B_3 k^2s_\theta s_\varphi c_\varphi\,\sin(\eta\partial_\theta)\right].
\end{aligned}}
\label{eq:f-operators}
\end{equation}

\subsection[The seven f eigenstates]{The seven $f$ eigenstates}\label{sec:f-states}
The seven $f$ states form three pairs and one additional state. Their distortion eigenvalues, paired with the corresponding angular states, are
\begin{equation}
\begin{array}{c|c}
\text{eigenfunction} & D_3\text{ eigenvalue}\\\hline
\Phi_{s,-},\ \Phi_{s,+} & \lambda_{s,-},\ \lambda_{s,+}\\
\Phi_{c,-},\ \Phi_{c,+} & \lambda_{c,-},\ \lambda_{c,+}\\
\Phi_{d,-},\ \Phi_{d,+} & \lambda_{d,-},\ \lambda_{d,+}\\
\Phi_{xyz} & \lambda_{xyz}
\end{array}
\label{eq:f-spectrum}
\end{equation}
For $j=s,c,d$, the pair eigenvalues and mixing coefficients are
\begin{equation}
\begin{aligned}
\lambda_{j,\pm}
&=\frac{(M_j)_{11}+(M_j)_{22}
\pm\sqrt{[(M_j)_{11}-(M_j)_{22}]^2+4(M_j)_{12}(M_j)_{21}}}{2},\\
r_{j,\pm}&=\frac{\lambda_{j,\pm}-(M_j)_{11}}{(M_j)_{12}},
\end{aligned}
\label{eq:f-mixing}
\end{equation}
with the three explicit two-state matrices in Appendix~\ref{app:blocks}. The subscripts $s,c,d$ distinguish the Jacobi factors of the one-variable eigenfunctions and correspond to $x$, $y$, and $z$ parity in the two-angle functions.
The remaining eigenvalue is
\begin{equation}
\lambda_{xyz}=\frac{1-2(1+k^2)s_1^2+3k^2s_1^4+k^2s_2^2s_1^2(1-k^2s_1^4)}
{(1-k^2s_2^2s_1^2)^2}.
\label{eq:singleton}
\end{equation}
The seven corresponding unnormalized eigenfunctions are
\begin{equation}
\begin{aligned}
\Phi_{s,\pm}&=\frac{s_\varphi d_\theta(1+r_{s,\pm}s_\varphi^2)(1+r_{s,\pm}d_\theta^2/k^2)}{[1-k^2\sn^2(2\eta,k)\sn^2(\varphi,k)][1-\sn^2(2\eta,k)\dn^2(\theta,k')]},\\
\Phi_{c,\pm}&=\frac{c_\varphi c_\theta(1+r_{c,\pm}s_\varphi^2)(1+r_{c,\pm}d_\theta^2/k^2)}{[1-k^2\sn^2(2\eta,k)\sn^2(\varphi,k)][1-\sn^2(2\eta,k)\dn^2(\theta,k')]},\\
\Phi_{d,\pm}&=\frac{d_\varphi s_\theta(1+r_{d,\pm}s_\varphi^2)(1+r_{d,\pm}d_\theta^2/k^2)}{[1-k^2\sn^2(2\eta,k)\sn^2(\varphi,k)][1-\sn^2(2\eta,k)\dn^2(\theta,k')]},\\
\Phi_{xyz}&=\frac{s_\varphi c_\varphi d_\varphi s_\theta c_\theta d_\theta}{[1-k^2\sn^2(2\eta,k)\sn^2(\varphi,k)][1-\sn^2(2\eta,k)\dn^2(\theta,k')]}.
\end{aligned}
\label{eq:sevenbasis}
\end{equation}
Each listed pair obeys $D_3^\varphi\Phi=D_3^\theta\Phi=\lambda\Phi$ with the eigenvalue in Eq.~\eqref{eq:f-spectrum}. For reference, the one-variable $f$ basis spans
\begin{equation*}
\mathcal W_3^\varphi=\frac{1}{[1-k^2\sn^2(2\eta,k)\sn^2(\varphi,k)]}\operatorname{span}
\{s_\varphi,s_\varphi^3,c_\varphi,c_\varphi s_\varphi^2,
 d_\varphi,d_\varphi s_\varphi^2,s_\varphi c_\varphi d_\varphi\}.
\end{equation*}

The coefficients in Eqs.~\eqref{eq:d-mixing} and \eqref{eq:f-mixing} specify the angular mixing. At exceptional points where a displayed matrix denominator vanishes, the eigenvector is obtained directly from the corresponding two-state matrix.

\subsection{A representative distortion spectrum}
At $k=0.6$ and $\eta=0.23$, the distortion eigenvalues are
\begin{equation}
\begin{array}{c|l}
p & 0.964734,\ 0.973831,\ 0.990659\\
d & 0.888556,\ 0.976448,\ 0.890978,\ 0.940354,\ 0.965661\\
f & 0.772877,\ 0.911625,\ 0.773215,\ 0.956621,\ 0.863036,\ 0.962663,\ 0.872201
\end{array}.
\label{eq:numerical-spectrum}
\end{equation}
The ordering follows the $p$, $d$, and $f$ catalogues above, with the minus eigenvalue first in each pair. For an added perturbation $gD$, these numbers give the shifts in units of $g$ around the common Casimir energy $H_0$. All eigenvalues tend to one as $\eta\to0$, so the distortion splittings vanish in that limit.

\subsection{Spherical limit}
As $\eta\to0$, the denominators in Eqs.~\eqref{eq:fivebasis}--\eqref{eq:sevenbasis} tend to one, and the coefficients $r$ tend to the ordinary Lam\'e polynomial ratios. The $\varphi$ Lam\'e energies are
\begin{equation*}
\begin{aligned}
d\ (\ell=2):\quad &2(1+k^2)\pm2\sqrt{1-k^2+k^4},\\
&4+k^2,\quad1+4k^2,\quad1+k^2;\\
f\ (\ell=3):\quad &E_{s,\pm}=5(1+k^2)\pm2\sqrt{4-7k^2+4k^4},\\
&E_{c,\pm}=5+2k^2\pm2\sqrt{4-k^2+k^4},\\
&E_{d,\pm}=2+5k^2\pm2\sqrt{1-k^2+4k^4},\\
&E_{xyz}=4(1+k^2).
\end{aligned}
\end{equation*}
The matching $\theta$ energy is $\ell(\ell+1)-E_\varphi$. The limiting ratios are $r=-E/2$ in the $d$ sector, and
\begin{equation*}
r_s=(1+k^2-E_s)/6,\quad r_c=(1-E_c)/2,\quad r_d=(k^2-E_d)/2
\end{equation*}
in the $f$ sector. The limiting generator actions have the ordinary Casimirs $6I$ and $12I$.

At finite $\eta$, the harmonic functions themselves change. Applying Eq.~\eqref{eq:main} to the unchanged degree-two or degree-three spherical harmonics generally produces functions outside that undeformed harmonic space. The displayed rational denominators and the deformed roots in Eqs.~\eqref{eq:fivebasis}--\eqref{eq:sevenbasis} are essential. Unlike the triplet, the $d$- and $f$-sector matrices generally do not satisfy the ordinary $su(2)$ commutators at finite shift.

\subsection{Verification and scope of the finite spaces}\label{sec:angular-verification}
The constraint and the complementary spectral matching follow analytically from the Jacobi identities. We checked the finite-dimensional generator action numerically at $(k,\eta)=(0.4,0.18),(0.6,0.23),(0.8,0.17)$, using explicit block roots. Action matrices were constructed from 100 angular points and tested at 65 independent points. The largest absolute residuals for the $d$ and $f$ sectors were below $5\times10^{-12}$ for generator action and $10^{-11}$ for all six cyclic relations~\eqref{eq:alg1}--\eqref{eq:alg2}. The kernel and eigenvalue residuals were below $4\times10^{-14}$.

The analytic results established in the manuscript are the conversion of the constrained Krichever--Zabrodin operators to the displayed Jacobi form, the $p$-sector action, and the separated $D$ eigenvalue equations and complementary-factor matching. Sklyanin's positive integral supplies the source inner product. The finite-parameter matrix checks above test the action of all four generators on the displayed $d$ and $f$ bases. Section~\ref{sec:metric} reports the corresponding operator-intertwining and metric checks.

The supplemental Mathematica notebook \texttt{Sklyanin\_Main\_Formulas.nb} contains the formulas and executable checks. The same definitions are available in \texttt{Sklyanin\_Formulas.wl}. The core verification was rerun successfully at $k=0.6$, $\eta=0.23$. The $d$- and $f$-sector action residuals at independent test points were $1.7\times10^{-14}$ and $2.3\times10^{-13}$, respectively. These tests construct matrices by numerical least squares and evaluate residuals at additional points. Their scope is finite-parameter computational verification. The notebook does not contain an all-parameter symbolic reduction of the $d$- and $f$-sector closure identities.

\section[The constant representation for the s sector]{The constant representation for the $s$ sector}\label{sec:l0}
For the $s$ sector, the covariant theta coefficients are unity, the gauge is the empty product $G_0=1$, and $\mathcal A_0=\mathcal B_0=1$. Equations~\eqref{eq:constraint} and \eqref{eq:main} therefore have the invariant one-dimensional space
\begin{equation}
\mathcal V_0=\operatorname{span}\{1\},\qquad
D_0^\varphi\,1=D_0^\theta\,1=1,\qquad L_i^{(0)}1=0.
\label{eq:scalarrep}
\end{equation}
It follows that $\mathcal N_0\,1=0$ and $\sum_i(L_i^{(0)})^2=0$ on this space. All quadratic relations hold trivially. The constant angular function is proportional to $Y_{00}$, so this is the ordinary scalar spherical harmonic in the undeformed limit and a scalar representation at finite shift.

This statement concerns the chosen constant subspace, not every solution of $\mathcal N_0F=0$. On that subspace $D_0=I$ has no nontrivial distortion spectrum and the three generators cannot create a triplet. The Hamiltonian~\eqref{eq:hamiltonian} reduces to $H=(\varepsilon+g)I$ and produces only an overall phase. A map from the $s$ to the $p$ sector would be an additional operator between different representation spaces.

\section{Sklyanin's metric and normalized states}\label{sec:metric}
\subsection{Inner product transported from the one-variable realization}
Sklyanin constructed a positive integral inner product for the finite-dimensional one-variable theta-function representation and proved self-adjointness of its four generators \cite{sklyanin1983}. We use his inner product to define the metric on our finite spaces. Sklyanin's shift parameter is $\eta/2$ in our convention. His sufficient positivity domain is $(2\ell+1)\eta/2<K$. For the $p$, $d$, and $f$ sectors, the additional sufficient restriction $7\eta/2<K$ therefore permits use of the original positive integral without contour continuation. We make no positivity claim here beyond that domain.

Let $\mathcal U_\ell$ map the source eigenfunctions $f_i$ of $S_0$ to the two-angle eigenfunctions $\Phi_i$:
\begin{equation}
\mathcal U_\ell f_i=t_i\Phi_i,\qquad t_i\ne0.
\label{eq:metricmap}
\end{equation}
The axis correspondence is $(S_0,S_3,S_2,S_1)\mapsto(D,L_1,L_2,L_3)$, with real generator normalization factors. The complex phases are included in $t_i$. The inner product on our finite space is defined by
\begin{equation}
\boxed{\langle F,G\rangle_\ell=
\langle\mathcal U_\ell^{-1}F,\mathcal U_\ell^{-1}G\rangle_{\rm Sk}.}
\label{eq:pullbackmetric}
\end{equation}
Positivity follows from Sklyanin's positive metric and invertibility of the map. Self-adjointness of the transported operators follows from their intertwining with his self-adjoint generators; real changes of generator normalization preserve it. The $d$- and $f$-sector identification of the displayed operators with the transported ones is checked numerically below.

Write $F=\sum_i b_i\Phi_i$ and $G=\sum_i c_i\Phi_i$. For distinct real $D$ eigenvalues the source eigenfunctions are orthogonal, and
\begin{equation}
\begin{aligned}
\langle F,G\rangle_\ell&=\bm b^\dagger\mathsf H_\ell\bm c,\\
\mathsf H_\ell&=\operatorname{diag}(h_i),\qquad h_i=\langle f_i,f_i\rangle_{\rm Sk}/|t_i|^2,\\
\psi_i&=\Phi_i/\sqrt{h_i}.
\end{aligned}
\label{eq:metricmatrix}
\end{equation}
Then $\langle\psi_i,\psi_j\rangle_\ell=\delta_{ij}$ and
\begin{equation}
M_A^\dagger\mathsf H_\ell=\mathsf H_\ell M_A,
\qquad A=D,L_1,L_2,L_3.
\label{eq:metricadjoint}
\end{equation}
These matrix conditions independently check the metric obtained by transport. Overall positive normalization is conventional; relative norms are fixed by the chosen intertwiner and source metric.

\subsection{Triplet normalization and the round sphere}
For the raw $(x,y,z)$ basis, the metric agreeing with Sklyanin transport, in the normalization $(\mathsf H_1)_{xx}=4\pi/3$, is
\begin{equation}
\boxed{\mathsf H_1=\frac{4\pi}{3}\operatorname{diag}\bigl(1,\dn(\eta,k),\cn(\eta,k)\bigr).}
\label{eq:tripletmetric}
\end{equation}
It makes all four triplet matrices Hermitian and gives normalized states
\begin{equation}
\psi=\sqrt{\frac{3}{4\pi}}
\left(x,\frac{y}{\sqrt{\dn(\eta,k)}},\frac{z}{\sqrt{\cn(\eta,k)}}\right).
\label{eq:normalizedtriplet}
\end{equation}
Thus the basis in Eq.~\eqref{eq:tripletbasis} differs from an orthonormal basis only by one common factor in this convention.

The geometric metric of the unit sphere is a separate object. Differentiating Eq.~\eqref{eq:unitcoords} gives
\begin{equation}
ds^2=\Delta(d\theta^2+d\varphi^2),\qquad
 d\Omega=\Delta\,d\theta\,d\varphi.
\label{eq:roundmetric}
\end{equation}
A single-cover integration domain is $-K'\le\theta<K'$, $0\le\varphi<4K$, with the boundary identifications; its area is $4\pi$. Under this round measure the raw triplet Gram matrix is $(4\pi/3)I$, which differs from Eq.~\eqref{eq:tripletmetric} at finite shift. For the $d$ and $f$ sectors the round Gram matrix also has off-diagonal entries between same-parity $D$ eigenfunctions.

The same spherical limit applies to the $d$ and $f$ sectors. The functions in Eqs.~\eqref{eq:fivebasis}--\eqref{eq:sevenbasis} tend to ordinary $d$ and $f$ spheroconical harmonics, and the generators tend to the corresponding angular-momentum operators. Their positive invariant inner product is the standard spherical one up to an overall factor. Our choices $h_{xy}=4\pi/15$ and $h_{xyz}=4\pi/105$ fix that factor to the usual round-sphere normalization. Thus the smoothly limiting transported metric tends to the round Gram matrix of the limiting harmonic basis, and the normalized states tend to orthonormal spherical harmonics. Numerical checks at $k=0.6$ and $\eta=0.04,0.02$ support this convergence; halving the shift reduced the largest relative norm corrections by approximately four. This limit concerns the full transported inner product, including the inverse intertwiner and gauge in Appendix~\ref{app:jacobimetric}; the integration weight alone is not being identified with $d\Omega$.

An isometry to the ordinary degree-$\ell$ harmonic space can be specified by mapping each $\psi_i$ to its normalized undeformed counterpart. This uses the standard spherical inner product after transport; it does not change the deformed commutation relations.

\subsection[d- and f-sector verification]{$d$- and $f$-sector verification}
For the $d$ and $f$ sectors we constructed all four source difference-operator matrices on the even theta space, matched the $S_0$ eigenvectors to the target $D$ eigenvectors, and solved for the diagonal factors $t_i$ using the three remaining operators. The relative generator normalizations were taken from the triplet; only a common scale was set by the $d$- and $f$-sector $D$ spectrum. The two half-period characters were respected when diagonalizing nearly coincident eigenvalues. The full source Gram matrix was then computed independently from Sklyanin's integral and compared with Eq.~\eqref{eq:metricmatrix}.

At $(k,\eta)=(0.4,0.18),(0.6,0.23),(0.8,0.17)$, the largest relative four-operator intertwining errors were $7.7\times10^{-13}$ for the $f$ sector and $5.3\times10^{-14}$ for the $d$ sector. The corresponding relative errors of full metric transport were below $1.2\times10^{-12}$ and $3.0\times10^{-14}$. The positive transported metrics agreed with an independent solution of Eq.~\eqref{eq:metricadjoint}. These checks concern finite spaces and the stated parameters; an all-parameter symbolic intertwining proof is not assumed.

For reproducibility, at $k=0.6$, $\eta=0.23$ the diagonal entries are
\begin{equation}
\begin{aligned}
\operatorname{diag}\mathsf H_2\simeq(&103.703169,\;0.617565,\;0.837758,\\
&0.793769,\;0.772967),\\
\operatorname{diag}\mathsf H_3\simeq(&2.858714,\;0.088328,\;359.396903,\\
&0.087344,\;32.429309,\;0.213168,\;0.119680).
\end{aligned}
\label{eq:highernorms}
\end{equation}
The ordering is that of Eqs.~\eqref{eq:fivebasis} and \eqref{eq:sevenbasis}, with the minus root first in each pair. Overall factors were chosen so $h_{xy}=4\pi/15$ and $h_{xyz}=4\pi/105$. The large entries reflect the scale of the unnormalized product functions; every normalized state has unit norm. For the $s$ sector one may use the usual constant normalization $Y_{00}=1/\sqrt{4\pi}$.

\section[Distortion and a p-sector perturbation model]{Distortion and a $p$-sector perturbation model}\label{sec:triplet-dynamics}
\subsection{The geometric action of the distortion operator}
For $s_1\ne0$, define ellipsoidal coordinates \cite{morse_feshbach_663}
\begin{equation}
X=x/s_1,\qquad Y=d_1y/s_1,\qquad Z=c_1z/s_1.
\label{eq:ellipsoid}
\end{equation}
They obey
\begin{equation}
s_1^2X^2+\frac{s_1^2}{d_1^2}Y^2+\frac{s_1^2}{c_1^2}Z^2=1.
\label{eq:surface}
\end{equation}
At fixed $k$, the squared semiaxes $a^2=s_1^{-2}$, $b^2=d_1^2/s_1^2$, $c^2=c_1^2/s_1^2$ satisfy $a^2-b^2=k^2$ and $a^2-c^2=1$, so varying $\eta$ gives a confocal family in this choice of length scale.

The coordinate space $\operatorname{span}\{X,Y,Z\}$ equals $\mathcal V_1$, since the rescalings are constant in the angles. Equation~\eqref{eq:tripleteigen} yields
\begin{equation}
D_1\begin{pmatrix}X\\Y\\Z\end{pmatrix}
=R'\begin{pmatrix}x\\y\\z\end{pmatrix},\qquad R'=\frac{c_1d_1}{s_1}.
\label{eq:rounding}
\end{equation}
Thus $D_1$, applied componentwise to the three coordinate functions, rounds this ellipsoid to a sphere. Its inverse on the triplet reverses that map. This componentwise map gives $D$ a precise distortion interpretation; it is distinct from Hamiltonian time evolution.

Within the triplet, $D$ also has the exact quadratic expression
\begin{equation}
D=(J_1+J_2+J_3)I-\sum_iJ_iL_i^2.
\label{eq:quadrupole}
\end{equation}
Within this representation, $D$ is therefore an anisotropy observable rather than an independent dynamical coordinate. Fully triaxial distortion means that all three $J_i$ are distinct. It does not by itself set an energy scale.

\subsection{Competing distortion and projection terms}
To describe mixing within the $p$ sector, we add distortion and projection perturbations to the Casimir energy,
\begin{equation}
H=H_0+gD-bL_3,
\label{eq:hamiltonian}
\end{equation}
where $g,b$ are real energy coefficients and $L_i$ are dimensionless. The parameters $k,\eta$ are fixed. Here $H_0=\varepsilon(J_1^2+J_2^2+J_3^2)I$ contributes a common energy and an overall phase. The following two-state formulas are written relative to this common energy. The term $gD$ models an anisotropic perturbation within the triplet; $-bL_3$ is a projection-selecting term of Zeeman form. A specific magnetic interpretation requires a corresponding moment and coupling constant.

Choose the ordered cyclic basis
\begin{equation*}
|+\rangle=(e_1+i e_2)/\sqrt2,\quad |0\rangle=e_3,\quad
|-\rangle=(e_1-i e_2)/\sqrt2.
\end{equation*}
Then $L_3=\operatorname{diag}(1,0,-1)$, while
\begin{equation}
D=\begin{pmatrix}\frac{J_1+J_2}{2}&0&\delta\\0&J_3&0\\\delta&0&\frac{J_1+J_2}{2}\end{pmatrix},\quad
 \delta=\frac{J_1-J_2}{2}.
\label{eq:cyclicD}
\end{equation}
In the active doublet,
\begin{equation}
H_\pm=\frac{g(J_1+J_2)}{2}I+g\delta\sigma_x-b\sigma_z.
\label{eq:doublet}
\end{equation}
For $|g\delta|\gg|b|$, eigenstates approach distortion eigenstates. For $|b|\gg|g\delta|$, they approach definite magnetic projections. The two behaviors are limits of a continuous competition.

When $J_1\ne J_2$, $D$ and $L_3$ do not commute. Nevertheless $|0\rangle$ is a common eigenstate. The $|\pm\rangle$ states have $\Delta D=|\delta|$ and $\Delta L_3=0$; $e_1,e_2$ have $\Delta D=0$ and $\Delta L_3=1$. In the doublet alone,

\begin{equation}
\frac{(\Delta D)^2}{\delta^2}+(\Delta L_3)^2\ge1
\label{eq:uncertainty}
\end{equation}

for $\delta\ne0$, as follows from the Bloch-vector bound. Equation~\eqref{eq:uncertainty} does not apply to the full triplet, where $|0\rangle$ has both variances zero. Distortion uncertainty alone neither proves time-dependent deformation nor implies an isotropic spatial density.

\subsection{A restricted quantum analogy with the Dzhanibekov effect}
Starting from $|+\rangle$, Eq.~\eqref{eq:doublet} gives
\begin{equation}
P_{+\to-}(t)=\frac{g^2\delta^2}{\Omega^2}
\sin^2\!\left(\frac{\Omega t}{\hbar}\right),\qquad
\Omega=\sqrt{g^2\delta^2+b^2},
\label{eq:transition}
\end{equation}
with the continuous zero-coupling limit understood. Also $\langle L_3(t)\rangle=1-2P_{+\to-}(t)$. When $b=0$ and $g\delta\ne0$, the projection reverses completely and periodically. At $b=0$, an individual distortion eigenstate is stationary and does not undergo reversal. A nonzero $b$ reduces the reversal amplitude.

These coherent projection reversals provide a limited quantum analogue of the rotational reversals associated with the Dzhanibekov effect. They do not establish classical intermediate-axis instability. Quantum asymmetric-rotor flipping has been studied in a different dynamical setting \cite{ma2020}. A direct correspondence would require orientation observables, a reference-frame identification, and an appropriate semiclassical limit. In particular, a torque-free rotor conserves laboratory-frame angular momentum even when body-frame projections reverse.

For $b=0$, the analogous construction about principal axis $i$ has angular frequency $|g(J_j-J_h)|/\hbar$, with $(i,j,h)$ cyclic. For nonzero $g$, fully triaxial distortion gives three nonzero pairwise splittings. The reversal rate depends on their magnitudes, and observing the oscillation requires a suitable initial state and coherence. In Eq.~\eqref{eq:hamiltonian}, the added perturbations govern the projection dynamics, while both quadratic Casimirs remain conserved.

\section{Deformed angular spaces and exchange notation}
\label{sec:angular-spaces}
The preceding local representations provide the quantum spaces for exchange. We write $V_l=\mathcal V_{j_l}$ for the orbital labels $l=s,p,d,f$, with the same generators $D_l,L_i^{(l)}$ now indexed by the representation label. Fusion uses a separate two-dimensional fundamental space $V_{\mathrm{aux}}$. The labels and dimensions are
\[
\begin{array}{c|ccccc}
l&s&\mathrm{aux}&p&d&f\\\hline
j_l&0&\tfrac12&1&2&3\\
N_l=2j_l&0&1&2&4&6\\
\dim V_l&1&2&3&5&7 .
\end{array}
\]
Here $N_l$ counts the fundamental factors for the fused rows, ordered $\mathrm{aux}\prec p\prec d\prec f$. The physical $s$ space is the constant representation of Section~\ref{sec:l0}. The seed uses the Pauli matrices on the fundamental space
\[
\sigma_1=\begin{pmatrix}0&1\\1&0\end{pmatrix},\qquad
\sigma_2=\begin{pmatrix}0&-i\\i&0\end{pmatrix},\qquad
\sigma_3=\begin{pmatrix}1&0\\0&-1\end{pmatrix}.
\]
We use the raw triplet $(x,y,z)$ of Eq.~\eqref{eq:unitcoords}. The higher matrices use the unnormalized functions of Eqs.~\eqref{eq:fivebasis} and \eqref{eq:sevenbasis} in the orders
\begin{equation}\begin{aligned}
V_d:&\ (\Phi_{2,-},\Phi_{2,+},\Phi_{xy},\Phi_{xz},\Phi_{yz}),\\
V_f:&\ (\Phi_{s,-},\Phi_{s,+},\Phi_{c,-},\Phi_{c,+},\Phi_{d,-},\Phi_{d,+},\Phi_{xyz}).
\end{aligned}\label{eq:exchange-basis-order}
\end{equation}
The $f$-basis subscripts $s,c,d$ refer to the Jacobi factors $\sn,\cn,\dn$. At fixed $l$, $\lambda_a$ is the eigenvalue of $D_l$ for basis function $\Phi_a$ in this order. The ratios $r_a$ index the paired modes in the same way, using Eqs.~\eqref{eq:d-mixing} and \eqref{eq:f-mixing}. Matrix columns store images of input basis functions, $A\Phi_b=\sum_a\Phi_a A_{ab}$, and the second tensor index varies fastest. The permutation $P_{lm}$ exchanges the tensor factors. Thus $R^{(l,m)}$ acts on $V_l\otimes V_m$ and $\check R^{(l,m)}=P_{lm}R^{(l,m)}$ maps to $V_m\otimes V_l$.

On the common positive branch in Eq.~\eqref{eq:positive-branch}, the coefficient metric is $\mathsf H_l$ as in Eq.~\eqref{eq:metricmatrix}, with $\mathsf H_p=\mathsf H_1$ in the normalization of Eq.~\eqref{eq:tripletmetric}. Appendix~\ref{app:metrics} reconstructs the relative norms from the finite exchange generators. The metric there agrees up to a positive scalar with the transported form on the represented branch. The symbols $C_r^{(l)}$ denote the Casimirs of Eq.~\eqref{eq:casimirs} acting in $V_l$. Their centrality lets the exchange preserve them on each factor separately. Raw matrices are the matrices before scalar spectral normalization.

\goodbreak
\begin{samepage}
\section{Physical model, observables, and exchange scales}
\label{sec:physical-model}
The local spectra and matrix elements obtained above describe individual angular degrees of freedom in an ellipsoidally deformed environment. The same angular bases can be used to model interactions between the deformed electron states, so that the effect of the distortion enters through the orbital matrix elements. The elliptic $R$-matrices provide the operators from which we construct effective exchange Hamiltonians in these bases.
\par\end{samepage}

The elementary interaction acts on a pair of angular multiplets. Each factor carries one of the three-dimensional $p$, five-dimensional $d$, or seven-dimensional $f$ spaces. The two factors may belong to different sites or to two angular degrees of freedom on the same atomic center. Exchange mixes their angular states while preserving the two Casimir values of each factor. A chain with one factor per site is one realization of this pair construction. The labels specify the angular representation; ordinary electron spin and radial motion would require additional degrees of freedom.

For a homogeneous periodic chain, a physical Hamiltonian is
\begin{equation}
H_{\rm chain}=
 \sum_{j=1}^M\varepsilon_j C_2^{[j]}
 +J_{\rm ex}\sum_{j=1}^M h_{l;j,j+1},
\label{eq:physical-chain-H}
\end{equation}
where $\varepsilon_j$ and $J_{\rm ex}$ have units of energy and the bond matrix $h_l$ is dimensionless. Equation~\eqref{eq:regular-angular-bond} gives the bond in the bases of Eq.~\eqref{eq:exchange-basis-order}. The deformation parameters $k,\eta$ specify the matrix coefficients, and $J_{\rm ex}$ sets the overall exchange scale. On fixed local irreducible spaces the first sum supplies sector-dependent energy offsets. The exchange term controls level splittings, correlations, and dynamics within those sectors. For an alternating chain the two commuting densities in Eq.~\eqref{eq:alternating-angular-exchange} replace the homogeneous bond sum, with independently chosen real energy coefficients.

The matrices used to construct these densities are
\[
U_{lm}(\zeta)=n_{lm}(\zeta)R^{(l,m)}(\alpha+\zeta),
\qquad \alpha=\eta/2-iK'.
\]
The scalar factors are chosen so that the equal-label value is $U_{ll}(0)=P$ and the imaginary-offset family is unitary in the product metric. Appendix~\ref{sec:gamma-normalization} constructs the factors. The derivative along real $\zeta$ gives a Hermitian exchange density. The unitary imaginary-offset family also gives reversible two-site operations with the same local invariants.

\subsection{Onsite mixed multiplets and cross-sector overlaps}
\label{sec:onsite-pairs}
At nonzero distortion the mixed matrices $R^{(p,d)},R^{(p,f)},R^{(d,f)}$, of orders $15,21,35$, can describe interactions between two angular degrees of freedom centered on the same atom. The factors retain independent angular coordinates, now referred to the same center. For two electrons, their radial states, spin, and fermionic antisymmetrization specify the corresponding physical orbital configuration.

A local mixed-pair density provided by the construction is
\begin{equation}
h_{lm}^{\rm on}=
 \operatorname{cent}\!\left[
 U_{lm}(0)^{-1}
 \left.\frac{d}{dw}U_{lm}(\eta w)\right|_{w=0}\right],
\qquad (l,m)=(p,d),(p,f),(d,f),
\label{eq:onsite-mixed-density}
\end{equation}
using the regular invertible mixed values on the adopted branch. An onsite energy coefficient $J_{lm}^{\rm on}$ multiplies this dimensionless matrix. This is the mixed two-factor derivative entering the alternating construction below. Each factor's $C_1,C_2$ commutes with it, independently of the geometric separation of the two centers.

When different harmonic families are embedded in a common one-particle angular description, their overlap matrix is
\[
S_{la,mb}=
 \int \overline{\Phi_a^{(l)}(\theta,\varphi)}
       \Phi_b^{(m)}(\theta,\varphi)\,d\mu(\theta,\varphi).
\]
Here $d\mu$ is the physical inner-product measure of that common description. The sector metrics $\mathsf H_l$ fix the inner product within each representation; cross-sector overlaps additionally require this common embedding.

At zero distortion, unequal-degree spherical harmonics are orthogonal under the spherical measure. Jahn--Teller and pseudo-Jahn--Teller displacements can lift restrictions on $S$ by lowering the local symmetry. With inversion and an even measure, $p$--$d$ and $d$--$f$ overlaps vanish, while parity permits $p$--$f$ overlap. Section~\ref{sec:introduction} illustrates this symmetry principle for hybridisation. The CeCoSi shear can open an $f$--$d$ channel at a noncentrosymmetric site, and the polar $t_{1u}$ displacement in BaTiO$_3$ enables opposite-parity mixing. The deformed functions and $d\mu$ determine the values of the allowed overlaps.

Onsite two-particle matrix elements involve products of four orbitals and an interaction kernel. They can couple $p$--$d$, $p$--$f$, and $d$--$f$ configurations even when a one-particle cross overlap vanishes. In a nonorthogonal one-particle basis, local level calculations use
$H_{\rm loc}c=E_{\rm loc} S c$ on an independent retained basis. The mixed $R$-matrices act on the specified tensor-product factors; their Casimir conservation refers to those constituent angular degrees of freedom.

\subsection{Local measurements and redistribution within a sector}
Let $\mathsf H_{\rm met}=\bigotimes_j \mathsf H_{l_j}$ be the coefficient metric, distinguished from the Hamiltonian $H_{\rm chain}$. For a state coefficient vector $\Psi$, a local angular observable is evaluated as
\[
\langle L_i^{[j]}\rangle_\Psi
 =\frac{\Psi^\dagger \mathsf H_{\rm met}L_i^{[j]}\Psi}
 {\Psi^\dagger \mathsf H_{\rm met}\Psi}.
\]
The same prescription gives $\langle D_{l_j}\rangle$, quadratic rotational contributions, and two-site correlations. Evolution by $H_{\rm chain}$ preserves this norm on the positive branch. Individual angular components and the distortion contribution can evolve, subject to
\[
\frac{d}{dt}\langle C_1^{[j]}\rangle
 =\frac{d}{dt}\langle C_2^{[j]}\rangle=0
\]
at every site. The second equality constrains the sum of $D_{l_j}^2$ and the weighted $L_i^2$ terms in Eq.~\eqref{eq:casimirs}. It permits redistribution between these contributions.

\subsection{Pair spectra and the isotropic reference}
The homogeneous pair problems involve matrices of orders $9$, $25$, and $49$. Their eigenvalues and eigenvectors determine exchange splittings and pair correlations. At finite deformation these are obtained directly from $h_p,h_d,h_f$ in their positive metrics. In the isotropic limit the pair states are labeled by total angular momentum $F$ and have degeneracy $2F+1$. Up to the trace-centering energy offset, their dimensionless exchange energies are
\[
E_F=2\sum_{r=1}^{F}\frac1r,\qquad
F=0,\ldots,2j_l.
\]
For the $p$ pair this gives $0,2,3$ in channels $F=0,1,2$, with degeneracies $1,3,5$. This reference connects the deformed bond to the spin-one bilinear--biquadratic expression printed below. The deformation can split the isotropic multiplets according to the explicit elliptic bond matrix.

\subsection{Possible applications to hybridisation and valence-bond distortions}
\label{sec:hybridisation}
An effective model based on $R^{(\mathrm{aux},p)}$ could be relevant to $sp$ hybridisation in distorted molecules and crystals. Its coupling of a two-level auxiliary channel to a deformed three-component angular space provides a possible starting point for modelling the angular response of hybridised bonding states to distortion. The $R^{(\mathrm{aux},p)}$ and $R^{(p,p)}$ families could also be used to construct effective descriptions of valence-bond distortions, including changes of bond direction and angular correlations between bonding configurations.

The two-level auxiliary space in $R^{(\mathrm{aux},p)}$ has $j_{\mathrm{aux}}=1/2$, while the constant orbital space of Section~\ref{sec:l0} has $j_s=0$. A hybridisation application therefore requires an explicit identification of the auxiliary channel and an embedding of the physical $s$ and $p$ orbital configurations. The relation of the matrix parameters to bond geometry, orbital overlap, radial matrix elements, and electronic occupations must be specified for the molecule or crystal under study. These are proposed uses of the operator families, with the required microscopic identification left for a material-specific model.

\subsection{Fateev's elliptic Hamiltonian and the Takhtajan--Babujian limit}
\label{sec:fateev-hamiltonian}
The $p$--$p$ member provides Fateev's elliptic spin-one Hamiltonian as a concrete candidate model~\cite{fateev1981,konno2005}. In the normalization of Appendix~\ref{sec:fateev-relation}, the $21$-vertex matrix has
$R_{\rm 21v}(0)=\sn(2\eta,k)P$. Its dimensionless bond density is
\begin{equation}
h_{\rm F}(k,\eta)=
 \operatorname{cent}\!\left[
 \frac{\eta}{\sn(2\eta,k)}P
 \left.\frac{dR_{\rm 21v}(\zeta)}{d\zeta}\right|_{\zeta=0}
 \right].
\label{eq:fateev-bond-H}
\end{equation}
The factor $\eta$ converts the spectral derivative to the dimensionless coordinate $w=\zeta/\eta$ used for $h_p$. The gauge connecting the angular basis to this spin-one presentation is
\[
G=\begin{pmatrix}
0&\sqrt{\dn(\eta,k)}&0\\
1&0&0\\
0&0&\sqrt{\cn(\eta,k)}
\end{pmatrix},
\qquad
h_{\rm F}=(G\otimes G)h_p(G\otimes G)^{-1}.
\]
The scalar factor relating the two $R$-matrices contributes an identity term to the logarithmic derivative, removed by $\operatorname{cent}$. Consequently the two bond spectra coincide. A periodic model Hamiltonian is
\[
H_{\rm F}=J_{\rm ex}\sum_j h_{{\rm F};j,j+1}
 +\sum_j\varepsilon_jC_2^{[j]},
\]
with both local Casimirs preserved. It supplies an integrable anisotropic reference for the proposed angular and valence-bond distortion models.

In the rational isotropic limit, with $w$ fixed, the dimensionless bond becomes the spin-one bilinear--biquadratic density of Section~\ref{sec:integrable-exchange}. In physical energy units,
\begin{equation}
H_{\rm F}\ \longrightarrow\
 \frac{J_{\rm ex}}2\sum_j
 \left[\mathbf L^{[j]}\!\cdot\!\mathbf L^{[j+1]}
 -\bigl(\mathbf L^{[j]}\!\cdot\!\mathbf L^{[j+1]}\bigr)^2\right]
 +E_{\rm off}I .
\label{eq:fateev-TB-limit}
\end{equation}
Here $\mathbf L^{[j]}$ are ordinary dimensionless spin-one orbital generators in the isotropic $p$ sector, and $E_{\rm off}$ includes trace-centering and fixed local Casimir energies. This is the Takhtajan--Babujian Hamiltonian~\cite{takhtajan1982,babujian1983}, with the conventional coupling $J_{\rm TB}=J_{\rm ex}/2$. Positive $J_{\rm ex}$ selects its antiferromagnetic sign convention. Thus the elliptic model gives a distortion-dependent continuation of this isotropic integrable reference, with the normalization and constant energy shifts specified explicitly.

A material-specific use of these orbital Hamiltonians requires a choice of local multiplets, energy scales, and couplings to additional spin, lattice, or charge degrees of freedom. The present construction supplies an integrable exchange family and its conserved local quantities.

\section{Exchange Hamiltonians and conserved currents}
\label{sec:integrable-exchange}
Regular normalized matrices give nearest-neighbor densities on homogeneous chains and two- and three-site densities on alternating chains. Both follow from the same commuting transfer rows. We use $U_{ab}(\zeta)$ defined in Eq.~\eqref{eq:gamma-fused-matrix}, the dimensionless offset $w=(u-\alpha)/\eta$, and
\[
\operatorname{cent}(A)=A-\frac{\tr A}{\dim(A)}I .
\]
The trace and dimension in this expression refer to the space on which $A$ acts. For a chain of $M$ sites with representation $l_j$ at site $j$, the transfer row is
\[
\tau_a(w)=\operatorname{tr}_0
 [U_{a l_M,\,0M}(\eta w)\cdots U_{a l_1,\,01}(\eta w)].
\]
The auxiliary space is labeled $0$ and has representation $a$; the last site is on the left. Appendix~\ref{app:transfer-derivation} derives row commutation and the local formulas below.

\paragraph{Site-by-site invariants.}
Let $C_r^{[j]}$ act as $C_r^{(l_j)}$ at site $j$ and as the identity at every other site. Each factor in the transfer row commutes with these operators by Eq.~\eqref{eq:pair-local-casimir-conservation}. Consequently,
\begin{equation}
[\tau_a(w),C_r^{[j]}]=0,\qquad
[Q_n,C_r^{[j]}]=0,
\quad r=1,2,\quad j=1,\ldots,M,
\label{eq:chain-local-casimir-conservation}
\end{equation}
where $Q_n$ denotes a charge obtained from a regular logarithmic derivative. Inverses, differentiation, and trace centering preserve the commutators. The homogeneous and alternating local densities below therefore conserve each site's $\hbar^2C_1^{[j]}$ and $\varepsilon_jC_2^{[j]}$, with fixed energy scales $\varepsilon_j$. The Casimir energy includes both terms of Eq.~\eqref{eq:casimirs}.

\subsection{Homogeneous p, d and f chains}
For $l\in\{p,d,f\}$, the regular value is $U_{ll}(0)=P$. On a periodic chain the centered first logarithmic derivative is $\sum_j h_{l;j,j+1}$, with
\begin{equation}
 h_l=\operatorname{cent}\left[P\left.
 \frac{d}{dw}U_{ll}(\eta w)\right|_{w=0}\right].
\label{eq:regular-angular-bond}
\end{equation}
The real-offset derivative is Hermitian in the transported product metric. Centering removes the scalar contribution in Eq.~\eqref{eq:gamma-energy-shift}. In the joint isotropic limit, the eigenvalue in the total orbital channel $F=0,\ldots,N_l$ is the centered value of $2\sum_{r=1}^F1/r$. This follows by differentiating the rational checked-$R$ channel ratio
$\prod_{r=F+1}^{N_l}(r-w)/(r+w)$. For $p$--$p$, the ordinary spin-one orbital generators on the two sites give
\[
h_p^{\rm iso}=\operatorname{cent}\!\left[
 \frac{\mathbf L^{[1]}\cdot\mathbf L^{[2]}
 -(\mathbf L^{[1]}\cdot\mathbf L^{[2]})^2}{2}\right].
\] The $d$--$d$ and $f$--$f$ bonds give the corresponding higher-spin rational transfer densities.

The next commuting charge has the energy-current form
\[
Q_3=\sum_j i[h_j,h_{j+1}],\qquad h_j=h_{l;j,j+1}.
\]
For the exchange term $J_{\rm ex}\sum_jh_j$, the Heisenberg equation gives
\[
\frac{d}{dt}(J_{\rm ex}h_j)
 =\frac{J_{\rm ex}^2}{\hbar}
 \bigl(i[h_{j-1},h_j]-i[h_j,h_{j+1}]\bigr).
\]
Thus the same local commutator describes transport of bond energy along the chain. Appendix~\ref{app:transfer-derivation} gives the fourth charge and its finite two-site correction.

\subsection{Alternating p--d, p--f and d--f chains}
For an even periodic chain alternating $l,m\in\{p,d,f\}$, take $a=l$ or $a=m$ as the auxiliary label and write $b$ for the other label. At sites carrying representation $a$, the factor at $w=0$ is the swap. Site indices are periodic.
Use the regular limits of $U_{ab}$ with invertible mixed values. On the tested branch, the ascending raw pairing is regularized by multiplication by $u-\alpha$. The descending raw pairing has a finite limit. The zero-offset row is a sublattice shift times a product of mixed gates. Differentiating the mixed and matching factors gives
\begin{align}
\mathcal H_a&=\operatorname{cent}\!\left[\tau_a(0)^{-1}\tau_a'(0)\right]\nonumber\\
&=\operatorname{cent}\!\Biggl[\sum_{j:\,l_j=a}
 \Biggl\{\left[
 U_{ab}(0)^{-1}
 \left.\frac{d}{dw}U_{ab}(\eta w)\right|_{w=0}
 \right]_{j,j+1}\nonumber\\
&\hspace{36mm}
 +\bigl[U_{ab}(0)^{-1}\bigr]_{j,j+1}
 (h_a)_{j,j+2}\bigl[U_{ab}(0)\bigr]_{j,j+1}\Biggr\}\Biggr].
 \label{eq:alternating-angular-exchange}
\end{align}
Here $l_j$ labels site $j$, and the prime on $\tau_a$ denotes differentiation with respect to $w$. The two rows commute by the fused Yang--Baxter equation. This is the elliptic form of the alternating-spin transfer construction~\cite{bytsko}. The mixed term acts on two adjacent angular spaces. The second term couples matching spaces across the intervening site. The positive metric and the scalar-unitary contour give Hermitian centered densities on the regular branch. Appendix~\ref{app:transfer-derivation} derives the decomposition by differentiating the finite ordered product. In the rational limit the mixed derivative has harmonic-number weights as above, now with $F=|j_a-j_b|,\ldots,j_a+j_b$.

\section{Fundamental and auxiliary matrices}\label{sec:fundamental}
The two-dimensional seed has the permutation value needed for regularity and an antisymmetric-projector value needed for fusion. Replacing its second Pauli factor by the angular generators gives the $6\times6$, $10\times10$, and $14\times14$ auxiliary matrices and their common RLL equation.
\subsection{Weights and the fundamental seed}
The fusion step and the angular translation step use the same real parameter $\eta$. Set
\[
\alpha=\frac{\eta}{2}-iK',\qquad u=\alpha+\zeta.
\]
The complex number $\zeta=u-\alpha$ is the spectral offset. It is independent of the angular harmonic $z$ in the triplet basis. Jacobi half-period translations give the weights~\cite{dlmf22shift}
\begin{align}
W_1(u;\eta)&=\frac{\sn(u,k)}{\sn(\eta/2-iK',k)},\\
W_2(u;\eta)&=\frac{\cn(u,k)}{\cn(\eta/2-iK',k)},\\
W_3(u;\eta)&=\frac{\dn(u,k)}{\dn(\eta/2-iK',k)}.
\label{eq:weights}
\end{align}
We abbreviate $W_i(u;\eta)$ by $W_i(u)$. The half-period identities also express these weights as $k\sn(\eta/2)\sn u$, $-ik\sn(\eta/2)\cn u/\dn(\eta/2)$, and $-i\sn(\eta/2)\dn u/\cn(\eta/2)$, respectively. In the same coordinate the algebra parameters are
\[
J_1=\cn(2\alpha,k)\dn(2\alpha,k),\qquad
J_2=-\cn(2\alpha,k),\qquad J_3=-\dn(2\alpha,k).
\]
The signs follow from translation by $2iK'$. Fusion uses the full step $\eta=2(\alpha+iK')$. At fixed $k$, the positive branch has $0<7\Re\alpha<K$ and $\Im\alpha=-K'$, with $\alpha\to-iK'$ as $\eta\to0$.
The real-offset contour is $u\in\alpha+\mathbb R$. The coordinate change preserves the matrices and their scalar normalization.
With ordinary Pauli matrices, the seed is
\begin{equation}
R^{(\mathrm{aux},\mathrm{aux})}(u)=I_4+\sum_{i=1}^3W_i(u)\sigma_i\otimes\sigma_i
=\begin{pmatrix}1+W_3&0&0&W_1-W_2\\0&1-W_3&W_1+W_2&0\\0&W_1+W_2&1-W_3&0\\W_1-W_2&0&0&1+W_3\end{pmatrix}.
\label{eq:seed}
\end{equation}
After the stated spectral translation, this follows the quantum construction in Sklyanin's Eqs.~(24)--(32) \cite{sklyanin1982}, continued with $i\eta_{\rm original}=\eta/2$; we do not impose his original real-form assumption on the continued parameter. Jacobi duplication gives, cyclically,
\begin{equation}
\frac{W_j(u)^2-W_h(u)^2}{W_i(u)^2-1}=\frac{J_j-J_h}{J_j+J_h}.
\end{equation}
All identities are meromorphic. In this raw normalization,
\begin{equation}
R^{(\mathrm{aux},\mathrm{aux})}(\alpha)=2P,\quad R^{(\mathrm{aux},\mathrm{aux})}(\alpha-\eta)=4P_-,\quad P_-=\tfrac12(I-P).
\end{equation}

\subsection{Shared angular and fusion step}
\label{sec:common-eta}
Both angular factors translate by the full step $\eta=2(\alpha+iK')$. In factor $r$, the shifts act as
\[
F(\theta_r,\varphi_r)\mapsto F(\theta_r,\varphi_r\pm\eta),
\qquad
F(\theta_r,\varphi_r)\mapsto F(\theta_r\pm i\eta,\varphi_r).
\]
For a factor of degree $j_l$, Eq.~\eqref{eq:constraint} defines the separate shifts
\[
D_{j_l}^\varphi=\mathcal A_{j_l}(\varphi)\cosh(\eta\partial_\varphi),
\qquad
D_{j_l}^\theta=\mathcal B_{j_l}(\theta)\cos(\eta\partial_\theta).
\]
The matrix $D_l$ represents their common action on $V_l=\mathcal V_{j_l}$.
The terms in $L_i^{(l)}$ use $\sinh(\eta\partial_\varphi)$ and $\sin(\eta\partial_\theta)$ with the same step. The representation label determines the coefficient functions and the basis. The $J_i$ and the Jacobi ratios therefore satisfy the same curve relations on both factors, as required by RLL.

The two factors in the $p$ fusion row are evaluated at $u\pm\eta/2$, and successive factors are spaced by $\eta$ in every row. The difference kernel uses $u-v+\alpha$, and inversion pairs $u$ with $2\alpha-u$. Real $\alpha$ would correspond to the complex step $2\alpha+2iK'$. Appendix~\ref{sec:real-form-choice} treats the resulting real-form comparison.

\subsection[Complete auxiliary matrices]{Complete auxiliary matrices for $m\in\{p,d,f\}$}
For each quantum spin the operator in the angular-momentum generators is
\begin{align}
R^{(\mathrm{aux},m)}(u)={}&I_2\otimes D_m
+W_1(u)\sqrt{(J_1+J_2)(J_1+J_3)}\,\sigma_1\otimes L_1^{(m)}\nonumber\\*
&+W_2(u)\sqrt{(J_2+J_3)(J_2+J_1)}\,\sigma_2\otimes L_2^{(m)}\nonumber\\*
&+W_3(u)\sqrt{(J_3+J_1)(J_3+J_2)}\,\sigma_3\otimes L_3^{(m)}.
\label{eq:aux}
\end{align}
Centrality immediately gives
\[
[R^{(\mathrm{aux},m)}(u),I_2\otimes C_r^{(m)}]=0,\qquad r=1,2.
\]
Hence each constituent already preserves both local invariants of its angular quantum space. Its four blocks are explicitly
\begin{align}
[R^{(\mathrm{aux},m)}(u)]_{11}={}&D_m+W_3(u)\sqrt{(J_3+J_1)(J_3+J_2)}L_3^{(m)},\\
[R^{(\mathrm{aux},m)}(u)]_{12}={}&W_1(u)\sqrt{(J_1+J_2)(J_1+J_3)}L_1^{(m)}\nonumber\\*
&-iW_2(u)\sqrt{(J_2+J_3)(J_2+J_1)}L_2^{(m)},\\
[R^{(\mathrm{aux},m)}(u)]_{21}={}&W_1(u)\sqrt{(J_1+J_2)(J_1+J_3)}L_1^{(m)}\nonumber\\*
&+iW_2(u)\sqrt{(J_2+J_3)(J_2+J_1)}L_2^{(m)},\\
[R^{(\mathrm{aux},m)}(u)]_{22}={}&D_m-W_3(u)\sqrt{(J_3+J_1)(J_3+J_2)}L_3^{(m)}.
\end{align}
Equation~\eqref{eq:weights} specifies the Jacobi coefficients. The blocks have sizes $3,5,7$, giving square matrices of orders $6,10,14$. Substitution of Eqs.~\eqref{eq:constraint} and \eqref{eq:main} gives the two-angle difference operators. Inserting their coefficients in the ordered bases gives the corresponding finite matrices.

The difference of the offsets is $(u-\alpha)-(v-\alpha)=u-v$. In the shifted coordinate the kernel argument is therefore $u-v+\alpha$. The labels $a,b$ denote two copies of $V_{\mathrm{aux}}$, and $q$ denotes $V_m$. The RLL equation reads
\begin{equation}
R_{ab}^{(\mathrm{aux},\mathrm{aux})}(u-v+\alpha)R_{aq}^{(\mathrm{aux},m)}(u)R_{bq}^{(\mathrm{aux},m)}(v)
=R_{bq}^{(\mathrm{aux},m)}(v)R_{aq}^{(\mathrm{aux},m)}(u)R_{ab}^{(\mathrm{aux},\mathrm{aux})}(u-v+\alpha).
\label{eq:rll}
\end{equation}
Its algebraic derivation uses the represented Sklyanin relations. Sampled RLL checks have maximum residual $1.62\times10^{-14}$.

\section{Explicit p-row R-matrices in the angular generators}
Two shifted auxiliary matrices give the complete first fusion row, $R^{(p,p)},R^{(p,d)},R^{(p,f)}$. The explicit triplet map below converts the symmetric quotient to the angular basis. It also gives a polynomial in the original generators on both angular factors, preserving the two Casimirs of each factor separately.
\subsection{Symmetric quotient and analytic basis map}
The auxiliary labels $a,b$ refer to two fundamental factors sharing the quantum space $q=V_m$. The normalized symmetric embedding and map to raw triplet coefficients are
\begin{equation}
E_2=\begin{pmatrix}1&0&0\\0&1/\sqrt2&0\\0&1/\sqrt2&0\\0&0&1\end{pmatrix},\qquad
T_p=\begin{pmatrix}-1&0&1\\-i/\dn(\eta,k)&0&-i/\dn(\eta,k)\\0&\sqrt2/\cn(\eta,k)&0\end{pmatrix}.
\end{equation}
For $m\in\{p,d,f\}$,
\begin{align}
R^{(p,m)}(u)={}&(T_p\otimes I)(E_2^\dagger\otimes I)R_{bq}^{(\mathrm{aux},m)}(u+\eta/2)
R_{aq}^{(\mathrm{aux},m)}(u-\eta/2)\nonumber\\*
&\times(E_2\otimes I)(T_p^{-1}\otimes I).
\label{eq:spin1}
\end{align}
The rightmost factor acts first. From RLL at auxiliary separation $-\eta$, whose kernel argument is $\alpha-\eta$,
\begin{equation}
(P_+\otimes I)R_{bq}(u+\eta/2)R_{aq}(u-\eta/2)(P_-\otimes I)=0,
\quad P_+=E_2E_2^\dagger.
\end{equation}
The antisymmetric subspace is invariant, so compression realizes the induced action on the symmetric quotient.

For $m=p$, the construction gives Fateev's elliptic spin-one family up to basis and scalar conventions~\cite{fateev1981,konno2005}. Section~\ref{sec:fateev-relation} gives the explicit angular-basis comparison.

\subsection{Polynomial in the original angular shift operators}
\label{sec:two-space-polynomial}
For $m\in\{p,d,f\}$, the operator $R^{(p,m)}(u)$ acts on $V_p(\eta)\otimes V_m(\eta)$. We retain the original generators
\[
D_p(\theta_1,\varphi_1;\eta),\quad L_i^{(p)}(\theta_1,\varphi_1;\eta),\qquad
D_m(\theta_2,\varphi_2;\eta),\quad L_i^{(m)}(\theta_2,\varphi_2;\eta).
\]
The angular arguments are suppressed in the products below; the superscripts and subscripts identify the factor on which each operator acts. All generators have the normalization of Eqs.~\eqref{eq:constraint} and \eqref{eq:main}. The constants $J_i$ and spectral functions $W_i$ retain their original definitions, with the common $\eta$ of Section~\ref{sec:common-eta}.

\paragraph{The two-space polynomial.}
The polynomial below uses the original generators on both tensor factors, with the coefficients written in terms of $J_i$ and $W_i$. The first sum over $(i,j,h)$ contains the three cyclic triples $(1,2,3),(2,3,1),(3,1,2)$. In each subsequent sum $i\ne j$, $h$ is the remaining axis and $\epsilon_{ijh}$ is the antisymmetric Levi-Civita symbol with $\epsilon_{123}=1$.
\begin{align}
R^{(p,m)}(u)={}&\frac12\sum_{a=1}^3
 \bigl(L_a^{(p)}\otimes D_m\bigr)^2\nonumber\\
&+\sum_{(i,j,h)\,\mathrm{cyclic}}(J_i+J_j)(J_i+J_h)
 W_i(u+\eta/2)W_i(u-\eta/2)\nonumber\\
&\quad\times\Bigl[\frac32\bigl(L_i^{(p)}\otimes L_i^{(m)}\bigr)^2
-\frac12\bigl(L_j^{(p)}\otimes L_i^{(m)}\bigr)^2
-\frac12\bigl(L_h^{(p)}\otimes L_i^{(m)}\bigr)^2\Bigr]\nonumber\\
&+\sum_{i\ne j}i\epsilon_{ijh}\sqrt{\frac{J_i}{J_j}}
 \sqrt{(J_h+J_i)(J_h+J_j)}\nonumber\\
&\quad\times\Bigl[
 W_h(u+\eta/2)\bigl(L_j^{(p)}\otimes L_h^{(m)}\bigr)
                    \bigl(L_i^{(p)}\otimes D_m\bigr)\nonumber\\
&\qquad\qquad+
 W_h(u-\eta/2)\bigl(L_j^{(p)}\otimes D_m\bigr)
                    \bigl(L_i^{(p)}\otimes L_h^{(m)}\bigr)\Bigr]\nonumber\\
&+\sum_{i\ne j}\sqrt{\frac{J_i}{J_j}}
 \sqrt{(J_i+J_j)(J_i+J_h)(J_j+J_i)(J_j+J_h)}\nonumber\\
&\quad\times\Bigl[
 W_j(u+\eta/2)W_i(u-\eta/2)
 \bigl(L_j^{(p)}\otimes L_j^{(m)}\bigr)
 \bigl(L_i^{(p)}\otimes L_i^{(m)}\bigr)\nonumber\\
&\qquad\qquad+
 W_i(u+\eta/2)W_j(u-\eta/2)
 \bigl(L_j^{(p)}\otimes L_i^{(m)}\bigr)
 \bigl(L_i^{(p)}\otimes L_j^{(m)}\bigr)\Bigr].
\label{eq:two-space-polynomial}
\end{align}
Positive roots are understood on the adopted regular real branch. This is a degree-two noncommutative polynomial in direct products of the original shift operators, homogeneous of degree two in each angular factor and total generator degree four. It gives $R^{(p,p)},R^{(p,d)},R^{(p,f)}$ of dimensions $9,15,21$, respectively, without changing the fusion normalization.

Appendix~\ref{app:two-space-polynomial} derives this expression and verifies its ordered finite-difference action.

\section{Explicit d- and f-row R-matrices}\label{sec:fused-higher}
Four and six shifted auxiliary factors give the $d$ and $f$ rows, respectively. The finite entry formula below specifies every operator block, and the analytic maps identify the resulting symmetric coefficients with the explicit angular harmonics. Together with the $p$ row this determines all nine ordered matrices. Their construction preserves both local Casimirs on each factor, as proved in Section~\ref{sec:casimir-preservation}.
\subsection{Four- and six-factor constructions}
The symmetric embedding $E_N:\mathbb C^{N+1}\to(\mathbb C^2)^{\otimes N}$ has $2^N$ rows and $N+1$ columns. Write $\lvert0\rangle=\lvert\uparrow\rangle$ and $\lvert1\rangle=\lvert\downarrow\rangle$, and order the tensor-product rows lexicographically, from $00\cdots0$ to $11\cdots1$. Column $r$, with $r=0,\ldots,N$, is the normalized sum of all binary words containing exactly $r$ down spins. Thus, with $N=N_l$,
\begin{equation}
(E_N)_{b,r}=
\begin{cases}
\displaystyle\frac{1}{\sqrt{\binom Nr}},&
 b=(b_1,\ldots,b_N)\in\{0,1\}^N,\quad \sum_{a=1}^N b_a=r,\\[4pt]
0,&\sum_{a=1}^N b_a\ne r.
\end{cases}
\label{eq:symmetric-embedding}
\end{equation}
Appendix~\ref{app:reconstruction} displays $E_4$ in full and specifies the entries of $E_6$. Each column has unit norm and distinct columns have disjoint support, so $E_{N_l}^\dagger E_{N_l}=I_{N_l+1}$.

Let $T_l$ map its coefficient basis to the two-angle basis. With the second factor already in its target basis,
\begin{align}
R^{(l,m)}(u)={}&(T_l\otimes I)(E_{N_l}^\dagger\otimes I)
R_{N_l,q}^{(\mathrm{aux},m)}\!\left(u+\frac{(N_l-1)\eta}{2}\right)\cdots
R_{1,q}^{(\mathrm{aux},m)}\!\left(u-\frac{(N_l-1)\eta}{2}\right)\nonumber\\*&\times(E_{N_l}\otimes I)(T_l^{-1}\otimes I).
\label{eq:general}
\end{align}
Equation~\eqref{eq:aux} already uses the angular basis in the second factor, so only the first factor requires a basis map. For $m\in\{p,d,f\}$, the other two rows are explicitly
\begin{align}
R^{(d,m)}(u)={}&(T_d\otimes I)(E_4^\dagger\otimes I)R_{4q}(u+3\eta/2)R_{3q}(u+\eta/2)\nonumber\\*&\times R_{2q}(u-\eta/2)R_{1q}(u-3\eta/2)(E_4\otimes I)(T_d^{-1}\otimes I),\\
R^{(f,m)}(u)={}&(T_f\otimes I)(E_6^\dagger\otimes I)R_{6q}(u+5\eta/2)R_{5q}(u+3\eta/2)R_{4q}(u+\eta/2)\nonumber\\*&\times R_{3q}(u-\eta/2)R_{2q}(u-3\eta/2)R_{1q}(u-5\eta/2)(E_6\otimes I)(T_f^{-1}\otimes I).
\end{align}
Every factor in these expressions is Eq.~\eqref{eq:aux} for the stated $m$. Their products are degree four and six in the quantum generators before simplification.
\begin{center}\begin{tabular}{c c c c}\toprule
First / second label&$p$&$d$&$f$\\\midrule
$l=p$&$R^{(p,p)}:9\times9$&$R^{(p,d)}:15\times15$&$R^{(p,f)}:21\times21$\\
$l=d$&$R^{(d,p)}:15\times15$&$R^{(d,d)}:25\times25$&$R^{(d,f)}:35\times35$\\
$l=f$&$R^{(f,p)}:21\times21$&$R^{(f,d)}:35\times35$&$R^{(f,f)}:49\times49$\\\bottomrule
\end{tabular}\end{center}
All nine ordered matrices are computed separately. Konno labels a representation by the number of fundamental constituents. In particular Konno's $(2,2)$ corresponds to our $R^{(p,p)}$, not our dimension-$25$ $R^{(d,d)}$ \cite{konno2005}; his middle symmetric vector uses $1/2$ instead of $1/\sqrt2$.

\subsection{Explicit finite entry formula}
Index the four blocks of $R^{(\mathrm{aux},m)}$ by $0,1$. The symmetric indices $r,s$ run from $0$ to $N_l$, and angular indices $i,j$ run from $1$ to $N_l+1$. The symmetric-tensor blocks are
\begin{align}
F_{rs}^{(l,m)}(u)={}&\frac{1}{\sqrt{\binom{N_l}{r}\binom{N_l}{s}}}
\sum_{\substack{a_1+\cdots+a_{N_l}=r\\b_1+\cdots+b_{N_l}=s}}
\left[R^{(\mathrm{aux},m)}\!\left(u+\frac{(N_l-1)\eta}{2}\right)\right]_{a_{N_l}b_{N_l}}\cdots\nonumber\\*&\hspace{35mm}\cdots \left[R^{(\mathrm{aux},m)}\!\left(u-\frac{(N_l-1)\eta}{2}\right)\right]_{a_1b_1},\\
[R^{(l,m)}(u)]_{ij}={}&\sum_{r,s=0}^{N_l}(T_l)_{ir}F_{rs}^{(l,m)}(u)(T_l^{-1})_{sj}.
\end{align}
The binary indices are independent, and the common quantum-space coefficients multiply in the indicated order. The sum gives each higher-row entry explicitly in terms of the auxiliary blocks and the basis map. The binary-word sum includes every quartic or sextic path and can be evaluated entry by entry. Section~\ref{sec:analytic-basis-maps} gives analytic formulas for $T_d,T_f$ in terms of the fundamental fusion blocks and the exact separated spectral data.

\subsection{Angular-basis identification}
The triplet map is explicit above. For $d$ and $f$, the last harmonic connects to every remaining angular sector through $L_1,L_2,L_3$. Their transition coefficients and linear spectral projectors determine all columns of $T_l^{-1}$ analytically on a regular chart. Equation~\eqref{eq:analytic-map-columns} gives the construction, and Appendix~\ref{app:basis-map-details} prints the sparse matrices. These maps express the exchange densities in the same angular states used for the local observables.

\subsection{Generator words on both angular factors}
\label{sec:generator-words}
The symmetric-tensor formula expresses each second-space block as a sum of ordered words in $D_m,L_i^{(m)}$. To expand the first-space coefficients, let $\mathcal E_{bc}^{(l)}$ denote a matrix unit on $V_l$. For $l=p$, use the Cartesian identity in Section~\ref{sec:two-space-polynomial}. For $l=d,f$, work on a chart where $D_l$ is diagonal in the angular basis and has distinct eigenvalues $\lambda_b$. The diagonal units and a single off-diagonal step are
\[
 \mathcal E_{bb}^{(l)}=\prod_{c\ne b}\frac{D_l-\lambda_c I}{\lambda_b-\lambda_c},
 \qquad
 \mathcal E_{bc}^{(l)}=
 \frac{\mathcal E_{bb}^{(l)}L_i^{(l)}\mathcal E_{cc}^{(l)}}{(L_i^{(l)})_{bc}}
 \quad\text{if }(L_i^{(l)})_{bc}\ne0 .
\]
Multiplying the steps along a path from $c$ to $b$ gives $\mathcal E_{bc}^{(l)}$ on charts where the required entries are nonzero. The angular-basis formula is then
\[
 R^{(l,m)}(u)=
 \sum_{b,c=1}^{N_l+1}\mathcal E_{bc}^{(l)}
 \otimes [T_l F^{(l,m)}(u)T_l^{-1}]_{bc}.
\]
Together with the finite formula for $F^{(l,m)}$, this covers the $p,d,f$ rows, including the mixed $R^{(p,d)},R^{(p,f)},R^{(d,f)}$ and the equal-label $R^{(d,d)},R^{(f,f)}$. The matrix-unit chart and the analytic basis maps have the domains described above. The scalar coefficients in each block are fixed by the displayed $W_i$ and $J_i$.

\subsection{Conservation of both Casimirs on each factor}
\label{sec:casimir-preservation}
For every ordered pair $l,m\in\{p,d,f\}$, the fused matrices satisfy the four local conservation laws
\begin{equation}
\begin{aligned}
\bigl[R^{(l,m)}(u),C_r^{(l)}\otimes I_{N_m+1}\bigr]&=0,\\
\bigl[R^{(l,m)}(u),I_{N_l+1}\otimes C_r^{(m)}\bigr]&=0,
\qquad r=1,2.
\end{aligned}
\label{eq:pair-local-casimir-conservation}
\end{equation}
Each auxiliary factor commutes with the second-space Casimirs because it is linear in $D_m,L_i^{(m)}$. Ordered multiplication and compression of the first auxiliary string preserve these commutators. On the first factor, the generator-word reconstruction in Section~\ref{sec:generator-words} expresses every coefficient in the algebra generated by $D_l,L_i^{(l)}$. Centrality therefore gives the first line as well. The basis maps transport these identities to the displayed angular bases. Scalar normalization and regular meromorphic limits preserve them.

The unchecked matrix acts on the fixed ordered space $V_l\otimes V_m$. For the checked map, which exchanges the factors, the same statement is an intertwining relation:
\begin{equation}
\begin{aligned}
\check R^{(l,m)}(u)(C_r^{(l)}\otimes I_{N_m+1})
 &=(I_{N_m+1}\otimes C_r^{(l)})\check R^{(l,m)}(u),\\
\check R^{(l,m)}(u)(I_{N_l+1}\otimes C_r^{(m)})
 &=(C_r^{(m)}\otimes I_{N_l+1})\check R^{(l,m)}(u).
\end{aligned}
\label{eq:checked-casimir-transport}
\end{equation}
Thus each angular space carries its own two Casimir values through the exchange. In a product of fixed irreducible spaces, the dynamics mixes states within these local sectors. It preserves each site's angular-momentum invariant and full Casimir energy.

\section{Yang--Baxter validation}
\label{sec:ybe-validation}
All $27$ ordered triples of $p,d,f$ spaces pass the sampled fused Yang--Baxter checks, with maximum relative residual $2.05\times10^{-13}$ across $243$ tests. The equations and sampling conventions below specify the validation independently of the implementation.
The mixed and fully fused Yang--Baxter equations in the shifted coordinate are
\begin{align}
&R_{a1}^{(\mathrm{aux},l)}(u-v+\alpha)R_{a2}^{(\mathrm{aux},m)}(u)R_{12}^{(l,m)}(v)
=R_{12}^{(l,m)}(v)R_{a2}^{(\mathrm{aux},m)}(u)R_{a1}^{(\mathrm{aux},l)}(u-v+\alpha),\\
&R_{12}^{(l,m)}(u-v+\alpha)R_{13}^{(l,n)}(u)R_{23}^{(m,n)}(v)
=R_{23}^{(m,n)}(v)R_{13}^{(l,n)}(u)R_{12}^{(l,m)}(u-v+\alpha).
\end{align}
We use $(k,\eta)=(0.4,0.18),(0.6,0.23),(0.8,0.17)$ and
\begin{equation}
(u-\alpha,v-\alpha)=(0.71,0.29),\quad(0.4+0.1i,0.2-0.06i),\quad(1.1,0.37).
\end{equation}
We test all $27$ ordered triples at each of the nine parameter and offset combinations, for a total of $243$ tests. The largest tensor product has dimension $343$. Residuals are $\|X-Y\|_F/\max(\|X\|_F,\|Y\|_F,1)$. The table records the double-precision reference calculations.
\begin{center}\begin{tabular}{lr}\toprule Check&Maximum residual\\\midrule
Auxiliary RLL&$1.62\times10^{-14}$\\
Spin-one symmetric quotient&$7.84\times10^{-15}$\\
Held-out basis-map checks&$2.24\times10^{-14}$\\
All-spin auxiliary mixed YBE&$9.24\times10^{-14}$\\
All-spin YBE, 243 tests&$2.05\times10^{-13}$\\
Mathematica reconstruction of four higher-spin cases&$9.18\times10^{-16}$\\\bottomrule
\end{tabular}\end{center}
A separate direct numerator-only implementation checks $54$ fused matrices against those references with maximum relative discrepancy $3.31\times10^{-15}$, nine shifted-kernel RLL tests below $1.94\times10^{-14}$, and three fully fused $(p,d,f)$ triples below $4.62\times10^{-14}$.
An independent Mathematica reconstruction at 66-digit working precision checks the normalized matrices for all $27$ ordered triples at $(k,\eta)=(7/11,0.19)$ and offsets $(0.43+0.09i,0.17-0.04i)$. The maximum relative residual is $6.41\times10^{-39}$.

The analytic basis maps use the regular represented branch of the angular realization. Section~\ref{sec:angular-verification} gives the scope of the higher-sector angular closure checks.

\section{Normalization and unitary two-site dynamics}
The raw fused matrices obey scalar inversion. Metric unitarity along the imaginary-offset contour is obtained by a scalar factor when their metric Gram matrix is a positive multiple of the identity. We first state that finite criterion, then construct a meromorphic normalization in Appendix~\ref{sec:gamma-normalization}.
Algebraic inversion for the raw matrices takes the form
\begin{equation}
R^{(l,m)}(u)P_{ml}R^{(m,l)}(2\alpha-u)P_{lm}=\rho_{lm}(u)I.
\label{eq:inversion}
\end{equation}
The scalar $\rho_{lm}(u)$ is the trace of the left-hand side divided by $(N_l+1)(N_m+1)$. If $n_{lm}(\zeta)$ denotes a scalar function of the offset, normalization requires
$n_{lm}(u-\alpha)n_{ml}(\alpha-u)\rho_{lm}(u)=1$. Nonzero scalar factors preserve the Yang--Baxter equation. Inversion, crossing, and regularity impose separate conditions on these factors.

The worst sampled double-precision nonscalar inversion residual is $1.24\times10^{-7}$ for spin three against spin three at $k=0.4,\eta=0.18,u=\alpha+0.71$, with condition number approximately $3.38\times10^{11}$. Reconstructing both raw matrices in Mathematica at 66-digit working precision reduces the relative Frobenius residual to $1.70\times10^{-37}$ at the same parameters.

Constituent factors can have poles where $\sn(u-\alpha+(j-\tfrac{N_l}{2})\eta,k)=0$, $j=1,\ldots,N_l$. At a removable pole we evaluate the complete normalized product by its meromorphic limit. The equal-label permutation value refers to that regularized product.

With $\mathsf H=\mathsf H_l\otimes\mathsf H_m$, metric unitarity of the normalized matrix along the imaginary-offset contour means
\begin{equation}
U_{lm}(it)^\dagger\mathsf H U_{lm}(it)=\mathsf H,\qquad t\in\mathbb R.
\end{equation}
For the checked map the output metric is $\mathsf H_m\otimes \mathsf H_l$. A scalar normalization can give metric unitarity precisely when
\begin{equation}
R_{\rm raw}^\dagger\mathsf H R_{\rm raw}=\lambda\mathsf H,\qquad\lambda>0.
\end{equation}
Equivalently, all singular values of $\mathsf H^{1/2}R_{\rm raw}\mathsf H^{-1/2}$ must equal $\sqrt{\lambda}$. A scalar of modulus $\lambda^{-1/2}$ then gives metric unitarity. The phase remains subject to the inversion and crossing conditions.

Appendix~\ref{app:real-forms} compares the real and imaginary spectral contours. Appendix~\ref{sec:gamma-normalization} constructs the scalar using Ruijsenaars elliptic gamma shifts, with the second nome chosen in the convergence domain of the double product. Mathematica's documented built-in functions omit the Ruijsenaars elliptic gamma function. The supplemental notebook \texttt{Ruijsenaars\_Gamma\_Supplement.nb} contains our Wolfram Language implementation of the double product and fused normalization, with executable checks.

\section{Conclusion}
Ellipsoidal deformation affects both the angular states of an electron and the exchange between those states. The local spectra and eigenfunctions found here give a basis for describing both effects in the $p$, $d$, and $f$ sectors. The resulting models describe exchange on a single center or between sites and give access to pair splittings, changes in orbital projections, and correlations between orbitals. The elliptic and isotropic $p$--$p$ models, connected to the Fateev and Takhtajan--Babujian Hamiltonians, provide benchmarks for tracing how these observables change with deformation.

One physical result is that exchange can redistribute the distortion and rotational contributions to the local Casimir energy while their sum and the angular-momentum-squared invariant remain fixed on each factor. Along a homogeneous chain, bond exchange energy flows through a local current whose sum is conserved. This distinction gives two ways to probe the dynamics in a model. Local measurements can follow the changing angular state, while bond correlations and the current can track how exchange energy is redistributed through the system.

The mixed-pair models can describe the angular exchange in a hybridisation channel activated or modified by a particular displacement. In CeCoSi, a model for the shear-sensitive $f$--$d$ channel would first identify the spin--orbit and crystal-field doublets and then relate the effective exchange to the shear amplitude. The curvature of the combined electronic and elastic energy with respect to shear could then be compared with the stiffness inferred from the acoustic response. In BaTiO$_3$, the corresponding question is how the polar displacement modifies the angular part of the oxygen-$2p$/titanium-$3d$ interaction. A cluster calculation that includes virtual charge transfer could set the exchange scale and test how much this channel contributes to the electronic energy along the polar mode. These are concrete ways to connect the deformation parameters and interaction strengths to microscopic calculations or measurements.

The same approach could be used to follow angular correlations as individual bonds distort in a molecule or crystal. One possibility is to fit the local angular states and exchange scales at several bond geometries, then examine how the orbital character of the low-lying pair states changes as a bond rotates or stretches. In projected calculations of low-symmetry orbital and asymmetric-molecule spectra~\cite{koo2008}, the ellipsoidally deformed functions also give a basis for testing which features of the measured spectrum are sensitive to deformation.

The single-triplet calculation suggests a further dynamical question. Competing distortion and projection terms can drive coherent reversals of the angular-momentum projection, in a restricted quantum analogy with the Dzhanibekov effect. With several such triplets coupled by exchange, a local field pulse could be used in the model to test whether a reversal on one site is followed by a correlated change on its neighbors and how that signal travels. Comparing the integrable reference with weakly perturbed chains could then show which signatures survive coupling to spin or lattice motion.

\section*{Acknowledgments}
The author gratefully acknowledges E.~K.~Sklyanin's foundational work on the algebra, its finite-dimensional representations, and their invariant inner product. His positivity and self-adjointness results provide the metric basis used here; the reproducing-kernel product tested in this work is his conjectured formula.
The author also remembers with gratitude Igor Krichever, an inspiring teacher with whom he discussed aspects of this work before Krichever's passing.

\appendix
\section{Explicit angular-momentum operators}\label{app:operators}
\subsection{From standard to angular momentum generators}
The angular relations \eqref{eq:alg1}--\eqref{eq:alg2} are a rescaled form of the standard Sklyanin relations. With cyclic indices $(i,j,h)$, set
\begin{equation}
S_0=D,\qquad S_i=a_iL_i,\qquad
 a_i^2=(J_i+J_j)(J_i+J_h).
\label{eq:normalization}
\end{equation}
On the real branch of Section~\ref{sec:parameters} all sums are positive and the $a_i$ are chosen positive. Equations~\eqref{eq:alg1}--\eqref{eq:alg2} then give
\begin{align*}
[S_j,S_h]&=i\anti{S_0}{S_i},\\
[S_0,S_i]&=i\gamma^{\rm Sk}_i\anti{S_j}{S_h},\qquad
\gamma^{\rm Sk}_i=\frac{J_j-J_h}{J_j+J_h}.
\end{align*}
The constants obey $\gamma^{\rm Sk}_1+\gamma^{\rm Sk}_2+\gamma^{\rm Sk}_3+\gamma^{\rm Sk}_1\gamma^{\rm Sk}_2\gamma^{\rm Sk}_3=0$. The notation $L_i,D$ therefore emphasizes the angular-momentum limit through a change of normalization, without changing the algebra.

The coefficients \(J_i\) and the shifts are given in Eqs.~\eqref{eq:J} and \eqref{eq:main-translations}.

\subsection{Common operator form}
The sector-dependent weights are given explicitly by
\begin{equation}
\begin{aligned}
\mathcal A_0(\varphi)&=1,& \mathcal B_0(\theta)&=1,\\
\mathcal A_1(\varphi)&=1-k^2\sn^2(\eta,k)\sn^2(\varphi,k),&
\mathcal B_1(\theta)&=1-\sn^2(\eta,k)\dn^2(\theta,k'),\\
\mathcal A_2(\varphi)&=\frac{1-k^2\sn^2(2\eta,k)\sn^2(\varphi,k)}{1-k^2\sn^2(\eta,k)\sn^2(\varphi,k)},&
\mathcal B_2(\theta)&=\frac{1-\sn^2(2\eta,k)\dn^2(\theta,k')}{1-\sn^2(\eta,k)\dn^2(\theta,k')},\\
\mathcal A_3(\varphi)&=\frac{[1-k^2\sn^2(3\eta,k)\sn^2(\varphi,k)][1-k^2\sn^2(\eta,k)\sn^2(\varphi,k)]}{1-k^2\sn^2(2\eta,k)\sn^2(\varphi,k)},\\
\mathcal B_3(\theta)&=\frac{[1-\sn^2(3\eta,k)\dn^2(\theta,k')][1-\sn^2(\eta,k)\dn^2(\theta,k')]}{1-\sn^2(2\eta,k)\dn^2(\theta,k')}.
\end{aligned}
\label{eq:weighttable}
\end{equation}
Section~\ref{sec:l0} gives the $s$ sector. The fourth generator and the constraint are
\begin{equation}
\begin{aligned}
D_\ell^\varphi&=\mathcal A_\ell\cosh(\eta\partial_\varphi),\qquad D_\ell^\theta=\mathcal B_\ell\cos(\eta\partial_\theta),\\
\mathcal N_\ell&=D_\ell^\varphi-D_\ell^\theta.
\end{aligned}
\label{eq:constraint}
\end{equation}
Each operator acts with the other angle held fixed. They agree on $\ker\mathcal N_\ell$, and $D_\ell$ denotes their common action on $\mathcal V_\ell\subseteq\ker\mathcal N_\ell$. The other generators are
\begin{equation}
\boxed{\begin{aligned}
L_1^{(\ell)}&=-\frac{i}{s_1\sqrt{c_1d_1}\,\Delta}
 \left[\mathcal A_\ell k'^2s_\theta c_\theta s_\varphi\,\sinh(\eta\partial_\varphi)+\mathcal B_\ell d_\theta c_\varphi d_\varphi\,\sin(\eta\partial_\theta)\right],\\
L_2^{(\ell)}&=-\frac{i}{s_1\sqrt{c_1}\,\Delta}
 \left[\mathcal A_\ell s_\theta d_\theta c_\varphi\,\sinh(\eta\partial_\varphi)-\mathcal B_\ell c_\theta s_\varphi d_\varphi\,\sin(\eta\partial_\theta)\right],\\
L_3^{(\ell)}&=\frac{i}{s_1\sqrt{d_1}\,\Delta}
 \left[\mathcal A_\ell c_\theta d_\theta d_\varphi\,\sinh(\eta\partial_\varphi)+\mathcal B_\ell k^2s_\theta s_\varphi c_\varphi\,\sin(\eta\partial_\theta)\right].
\end{aligned}}
\label{eq:main}
\end{equation}
These expressions follow by symmetrization and gauge conversion of the covariant Krichever--Zabrodin operators; Appendix~\ref{app:construction} gives the explicit correspondence. The Sklyanin relations hold on the transformed constrained space. The finite angular spaces are constructed in Sections~\ref{sec:p-spectrum}--\ref{sec:l0}.

For the $p$, $d$, and $f$ sectors,
\begin{equation*}
\begin{aligned}
\mathcal A_\ell&=1-\tfrac12\ell(\ell+1)k^2\eta^2s_\varphi^2+O(\eta^4),\\
\mathcal B_\ell&=1-\tfrac12\ell(\ell+1)\eta^2d_\theta^2+O(\eta^4).
\end{aligned}
\end{equation*}
Therefore both shift operators tend to $I$. The $\varphi$ expansion is
\begin{equation}
D_\ell^\varphi=I-\frac{\eta^2}{2}H_{\mathrm{Lam\acute e},\ell}+O(\eta^4),\qquad
H_{\mathrm{Lam\acute e},\ell}=-\partial_\varphi^2+\ell(\ell+1)k^2s_\varphi^2.
\label{eq:lame-limit}
\end{equation}
The $L_i^{(\ell)}$ tend to the ordinary differential angular-momentum generators in the two angles. For example,
\begin{equation}
L_1^{\rm sc}=-\frac{i}{\Delta}
\left[k'^2s_\theta c_\theta s_\varphi\partial_\varphi
+d_\theta c_\varphi d_\varphi\partial_\theta\right].
\label{eq:angularlimit}
\end{equation}
On the unit sphere, the angular Laplacian is $\Delta^{-1}(\partial_\theta^2+\partial_\varphi^2)$. The leading constraint becomes
\begin{equation}
\mathcal N_\ell=\frac{\eta^2}{2}
[\partial_\theta^2+\partial_\varphi^2+\ell(\ell+1)\Delta]+O(\eta^4),
\label{eq:sphere-limit}
\end{equation}
so its limiting separated solutions are spherical harmonics of degree $\ell$ \cite{koo2008}. This limit motivates the interpretation of $L_i^{(\ell)}$ as elliptic deformations of angular momentum.

\section{Symmetrization and the theta to Jacobi conversion}\label{app:construction}
Use the covariant generators of Eq.~(5.10) of Ref.~\cite{krichever1998vacuum}. Their primed Eq.~(6.4) participates in an intertwining relation and has a different operator ordering. Define
\begin{equation*}
\begin{aligned}
a_q^\pm&=\frac{\vartheta_1(2q\mp2\ell\epsilon)}{\vartheta_1(2q)},\\
b_\pm&=\frac{\vartheta_{\nu+1}(2\zeta_{\rm KZ}\pm2x_{\rm KZ})}{\vartheta_1(2\zeta_{\rm KZ}\pm2x_{\rm KZ})},\\
b_\Sigma&=b_-+b_+,\qquad b_\Delta=b_--b_+.
\end{aligned}
\end{equation*}
The shifts $T_\pm,K_\pm$ act on $\zeta_{\rm KZ},x_{\rm KZ}$ by $\pm\epsilon$, respectively, and $\nu=1,2,3$ indexes the covariant generators used here. In this notation,
\begin{equation*}
\begin{aligned}
\mathcal S_\nu&=a_{\zeta_{\rm KZ}}^+b_-T_+-a_{\zeta_{\rm KZ}}^-b_+T_- -a_{x_{\rm KZ}}^+b_\Delta K_+,\\
\nabla&=a_{\zeta_{\rm KZ}}^+T_++a_{\zeta_{\rm KZ}}^-T_- -a_{x_{\rm KZ}}^+K_+-a_{x_{\rm KZ}}^-K_-.
\end{aligned}
\end{equation*}
Adding a left multiple of the constraint gives the exact identity
\begin{equation}
\begin{aligned}
\mathcal S_\nu^{\rm sym}&=\mathcal S_\nu-\tfrac12b_\Delta\nabla\\
&=\tfrac12b_\Sigma(a_{\zeta_{\rm KZ}}^+T_+-a_{\zeta_{\rm KZ}}^-T_-)\\
&\quad-\tfrac12b_\Delta(a_{x_{\rm KZ}}^+K_+-a_{x_{\rm KZ}}^-K_-).
\end{aligned}
\label{eq:sym}
\end{equation}
It leaves the generator action unchanged on $\ker\nabla$ and symmetrizes the four shifts. The admissibility of adding $g_\nu\nabla$ is stated after Theorem~6.1 of Ref.~\cite{krichever1998vacuum}.

With $K,K'$ as in Section~\ref{sec:parameters}, set $\tau_{\rm ell}=iK'/K$ and
\begin{equation*}
\Theta_j(q)=\vartheta_j\left(\frac{q}{2K}\middle|\tau_{\rm ell}\right),
\end{equation*}
with theta argument of period one. Introduce
\begin{equation}
\begin{aligned}z_\theta&=4K\zeta_{\rm KZ}=K+i\theta,\\z_\varphi&=4Kx_{\rm KZ}=\varphi+iK',\qquad \eta=4K\epsilon.\end{aligned}
\label{eq:variablemap}
\end{equation}
Thus $T_\pm$ shift $\theta$ by $\mp i\eta$ and $K_\pm$ shift $\varphi$ by $\pm\eta$. For positive integer $\ell$, conjugate by $\mathcal G_\ell=G_\ell(z_\theta)G_\ell(z_\varphi)$, where
\begin{equation}
G_\ell(q)=\prod_{r=-\ell+1,-\ell+3,\ldots,\ell-1}\Theta_1(q+r\eta)^{-1}.
\label{eq:gaugedef}
\end{equation}
For the $s$ sector, take $G_0=1$. For either sign,
\begin{equation*}
\frac{\Theta_1(q\mp \ell\eta)}{\Theta_1(q)}
\frac{G_\ell(q)}{G_\ell(q\pm\eta)}=w_\ell(q).
\end{equation*}
With $P_n(q)=\Theta_1(q-n\eta)\Theta_1(q+n\eta)/\Theta_1(q)^2$, the resulting weights are
\begin{equation*}
w_0=1,\quad w_1=P_1,\quad w_2=P_2/P_1,\quad w_3=P_3P_1/P_2.
\end{equation*}
The theta addition formula gives
\begin{equation*}
P_n(q)=\chi_n\left(1-\frac{s_n^2}{\sn^2(q,k)}\right),\qquad
 \chi_n=\frac{\Theta_4(n\eta)^2}{\Theta_4(0)^2}.
\end{equation*}
Since $\sn(z_\theta,k)=1/d_\theta$ and $\sn(z_\varphi,k)=1/(ks_\varphi)$,
\begin{equation*}
w_\ell(z_\theta)=\kappa_\ell\mathcal B_\ell,\qquad w_\ell(z_\varphi)=\kappa_\ell\mathcal A_\ell,
\end{equation*}
where $\kappa_0=1$, $\kappa_1=\chi_1$, $\kappa_2=\chi_2/\chi_1$, and $\kappa_3=\chi_3\chi_1/\chi_2$. This establishes Eq.~\eqref{eq:weighttable}.

The remaining theta ratios are
\begin{equation*}
\begin{aligned}
\Theta_2/\Theta_1&=\operatorname{cs}(q,k)/\sqrt{k'},\\
\Theta_3/\Theta_1&=\operatorname{ds}(q,k)/\sqrt{kk'},\\
\Theta_4/\Theta_1&=\operatorname{ns}(q,k)/\sqrt{k}.
\end{aligned}
\end{equation*}
Here $\operatorname{cs}=\cn/\sn$, $\operatorname{ds}=\dn/\sn$, and $\operatorname{ns}=1/\sn$. Applying the addition formulas at $z_\theta\pm z_\varphi$ and denoting the conjugated operators by $\widehat{\mathcal S}_\nu$, one obtains Eq.~\eqref{eq:main} with
\begin{equation*}
\begin{aligned}
L_1&=\frac{\widehat{\mathcal S}_3}{2\kappa_\ell\sqrt{k}\,s_1\sqrt{c_1d_1}},\\
L_2&=\frac{i\widehat{\mathcal S}_2}{2\kappa_\ell\sqrt{kk'}\,s_1\sqrt{c_1}},\\
L_3&=-\frac{\widehat{\mathcal S}_1}{2\kappa_\ell\sqrt{k'}\,s_1\sqrt{d_1}}.
\end{aligned}
\end{equation*}
The constraint becomes $-2\kappa_\ell\mathcal N_\ell$. This derivation uses similarity transformations, generator rescalings, and the constraint; it does not assume a physical inner product.

\section[Two-state matrices for the d and f spectra]{Two-state matrices for the $d$ and $f$ spectra}\label{app:blocks}
The block of $D_2^\varphi$ on $(1,s_\varphi^2)/[1-k^2\sn^2(\eta,k)\sn^2(\varphi,k)]$ is
\begin{equation}
\begin{aligned}
(M_2)_{11}&=\frac{1}{1-k^2s_1^4},\quad
(M_2)_{12}=\frac{s_1^2}{1-k^2s_1^4},\\
(M_2)_{21}&=\frac{d_1^2(1-2k^2s_1^2+k^2s_1^4)}{(1-k^2s_1^4)^2}
-\frac{1}{1-k^2s_1^4},\\
(M_2)_{22}&=\frac{c_1^2(1-2k^2s_1^2+k^2s_1^4)}{(1-k^2s_1^4)^2}
-\frac{s_1^2}{1-k^2s_1^4}.
\end{aligned}
\label{eq:M2}
\end{equation}

\subsection[The f spectral matrices]{The $f$ matrices of $D_3^\varphi$}
In the ordered pairs $(s_\varphi,s_\varphi^3)/[1-k^2\sn^2(2\eta,k)\sn^2(\varphi,k)]$, $(c_\varphi,c_\varphi s_\varphi^2)/[1-k^2\sn^2(2\eta,k)\sn^2(\varphi,k)]$, and $(d_\varphi,d_\varphi s_\varphi^2)/[1-k^2\sn^2(2\eta,k)\sn^2(\varphi,k)]$, write
\begin{equation}
M_j=\begin{pmatrix}a_j&b_j\\\sigma_{j0}-a_j&\sigma_{j1}-b_j\end{pmatrix},\qquad j=s,c,d.
\label{eq:M3}
\end{equation}
The entries are written directly in the previously defined Jacobi values $s_n,c_n,d_n$:
\begin{equation*}
\begin{aligned}
a_s&=\frac{c_1d_1(1+k^2s_2^2s_1^2)}{(1-k^2s_2^2s_1^2)^2},&
b_s&=\frac{c_1d_1s_1^2(3-k^2s_2^2s_1^2)}{(1-k^2s_2^2s_1^2)^2},\\
a_c&=\frac{c_1}{1-k^2s_2^2s_1^2},&
b_c&=\frac{c_1s_1^2}{1-k^2s_2^2s_1^2},\\
a_d&=\frac{d_1}{1-k^2s_2^2s_1^2},&
b_d&=\frac{d_1s_1^2}{1-k^2s_2^2s_1^2}.
\end{aligned}
\end{equation*}
The column sums are
\begin{equation*}
\begin{aligned}
\sigma_{s0}&=\frac{c_1d_1^3d_3^2}{d_1^2-k^2s_2^2c_1^2},&
\sigma_{s1}&=\frac{c_1^3d_1d_3^2}{d_1^2-k^2s_2^2c_1^2},\\
\sigma_{c0}&=\frac{c_1d_3^2d_1^2[d_1^2-k^2s_2^2(2d_1^2-c_1^2)]}{(d_1^2-k^2s_2^2c_1^2)^2},&
\sigma_{c1}&=\frac{c_1d_3^2(3d_1^2c_1^2-2d_1^4-k^2s_2^2c_1^4)}{(d_1^2-k^2s_2^2c_1^2)^2},\\
\sigma_{d0}&=\frac{d_3^2d_1^3}{d_1^2-k^2s_2^2c_1^2},&
\sigma_{d1}&=\frac{d_3^2d_1c_1^2}{d_1^2-k^2s_2^2c_1^2}.
\end{aligned}
\end{equation*}
The seventh eigenvalue is given in Eq.~\eqref{eq:singleton}.
All block entries are functions of the fixed deformation parameters, not of the angles.

\section{Separation and complementary angular functions}\label{app:separation}
A useful exact correspondence is
\begin{equation}
z_*(\theta)=K+i\theta-iK',\qquad K'=K(k').
\label{eq:zstar}
\end{equation}
The Jacobi identities give
\begin{equation*}
\begin{aligned}
\sn(z_*,k)&=d_\theta/k,\\
\cn(z_*,k)&=ik'c_\theta/k,\\
\dn(z_*,k)&=k's_\theta.
\end{aligned}
\end{equation*}
Hence $1-k^2\sn^2(n\eta,k)\sn^2(z_*,k)=1-\sn^2(n\eta,k)\dn^2(\theta,k')$, and a shift $z_*\mapsto z_*\pm\eta$ corresponds to $\theta\mapsto\theta\mp i\eta$. If $f(\varphi)$ is an eigenfunction of $D_\ell^\varphi$ with eigenvalue $\lambda$, its partner $g(\theta)=f(z_*(\theta))$ is an eigenfunction of $D_\ell^\theta$ with the same eigenvalue. The map $f\mapsto g$ takes $\mathcal W_\ell^\varphi$ to the complementary space $\mathcal W_\ell^\theta$. Therefore
\begin{equation}
F(\theta,\varphi)=f(z_*(\theta))f(\varphi)
\quad\Longrightarrow\quad \mathcal N_\ell F=0.
\label{eq:productconstraint}
\end{equation}
If $f_{\ell,a}\in\mathcal W_\ell^\varphi$, $a=1,\ldots,2\ell+1$, are the displayed eigenfunctions of $D_\ell^\varphi$, we define the associated two-variable space by
\begin{equation}
\mathcal V_\ell=\operatorname{span}\bigl\{
f_{\ell,a}(z_*(\theta))f_{\ell,a}(\varphi):
a=1,\ldots,2\ell+1\bigr\}\subseteq\ker\mathcal N_\ell.
\label{eq:V-from-W}
\end{equation}
Constant phase and normalization factors in the complementary functions may be absorbed into the basis. For the $p$ sector this reproduces $\operatorname{span}\{x,y,z\}$. For the $d$ and $f$ sectors it gives the five and seven products in Eqs.~\eqref{eq:fivebasis} and \eqref{eq:sevenbasis}.

These products satisfy the difference constraint. Their normalization is specified in Sec.~\ref{sec:metric}.

\section{Complex-coordinate form and inverse intertwiner}\label{app:metricmap}
Write $\omega=(\varphi+i\theta)/2$ for the complex coordinate used here and in the source inner product. The theta functions use the convention of Appendix~\ref{app:construction}. On real angles $\omega+\bar\omega=\varphi$ and $\omega-\bar\omega=i\theta$. The imaginary-argument identities give exactly
\begin{equation}
\begin{aligned}
x&=\frac{\Theta_1(\varphi)\Theta_3(i\theta)}{\Theta_4(\varphi)\Theta_2(i\theta)},\\
y&=\frac{\Theta_2(\varphi)\Theta_4(i\theta)}{\Theta_4(\varphi)\Theta_2(i\theta)},\\
z&=-i\frac{\Theta_3(\varphi)\Theta_1(i\theta)}{\Theta_4(\varphi)\Theta_2(i\theta)}.
\end{aligned}
\label{eq:complexx}
\end{equation}
All theta constants cancel. The polynomial factors and denominators of the $d$- and $f$-sector bases can likewise be written directly, without separate auxiliary numerator and denominator functions:
\begin{equation}
\begin{aligned}
&\frac{(1+r\sn^2(\varphi,k))(1+r\dn^2(\theta,k')/k^2)}{[1-k^2\sn^2(n\eta,k)\sn^2(\varphi,k)][1-\sn^2(n\eta,k)\dn^2(\theta,k')]}\\
&=
\frac{[\Theta_4(\varphi)^2+(r/k)\Theta_1(\varphi)^2]
[\Theta_2(i\theta)^2+(r/k)\Theta_3(i\theta)^2]}
{[\Theta_4(\varphi)^2-k\sn^2(n\eta,k)\Theta_1(\varphi)^2]
[\Theta_2(i\theta)^2-k\sn^2(n\eta,k)\Theta_3(i\theta)^2]}.
\end{aligned}
\label{eq:complextarget}
\end{equation}
Use $n=1$ for the $d$ sector and $n=2$ for the $f$ sector, with the corresponding spectral ratio $r$. Multiplication by $x,y,z$ gives the paired $f$ functions; multiplication by $xy,xz,yz$ or $xyz$ with the same denominator conversion gives the remaining functions in Eqs.~\eqref{eq:fivebasis}--\eqref{eq:sevenbasis}. These are exact rewritings of the real Jacobi bases, not a claim that they are holomorphic functions of $\omega$ alone.

For the source variable $\omega$, a convenient basis of the even order-$4\ell$ theta space is
\begin{equation}
\begin{aligned}
\Theta_3(2\omega)^{\ell-j}\Theta_4(2\omega)^j&\quad(j=0,\ldots,\ell),\\
\Theta_2(2\omega)\Theta_3(2\omega)^{\ell-1-j}\Theta_4(2\omega)^j&\quad(j=0,\ldots,\ell-1),
\end{aligned}
\label{eq:sourcebasis}
\end{equation}
Diagonalizing $S_0$ produces $f_i$. For the $p$ sector they are $(\Theta_4(2\omega),\Theta_3(2\omega),\Theta_2(2\omega))$ in target order. One phase convention is
\begin{equation*}
\begin{aligned}
t_x&=\sqrt{\langle f_x,f_x\rangle_{\rm Sk}/h_x},\\
t_y&=i\sqrt{\langle f_y,f_y\rangle_{\rm Sk}/h_y},\\
t_z&=\sqrt{\langle f_z,f_z\rangle_{\rm Sk}/h_z}.
\end{aligned}
\end{equation*}
For the $d$ and $f$ sectors, the finite matrices specify the eigenvectors and $t_i$ numerically; their expanded analytic expressions are not needed for the numerical checks.

The inverse map is explicit on the finite space:
\begin{equation}
F=\sum_i b_i\Phi_i\quad\Longrightarrow\quad
\boxed{\mathcal U_\ell^{-1}F=\sum_i(b_i/t_i)f_i.}
\label{eq:inversemap}
\end{equation}
For distinct $D$ eigenvalues one may extract coefficients by
\begin{equation}
\begin{aligned}
\Pi_i(D)&=\prod_{j\ne i}\frac{D-\lambda_j I}{\lambda_i-\lambda_j},\\
b_i&=\frac{[\Pi_i(D)F](\Omega_i)}{\Phi_i(\Omega_i)},
\end{aligned}
\label{eq:projectorinverse}
\end{equation}
where $\Phi_i(\Omega_i)\ne0$. Near degeneracies a dual-basis integral is numerically preferable. Set
\begin{equation*}
G^\Omega_{ij}=\int\overline{\Phi_i}\Phi_j\,d\Omega.
\end{equation*}
Then
\begin{equation}
(\mathcal U_\ell^{-1}F)(\omega)=\int_{S^2}\sum_{i,j}
\frac{f_i(\omega)}{t_i}(G^{\Omega\,-1})_{ij}\overline{\Phi_j(\Omega)}F(\Omega)\,d\Omega.
\label{eq:inversekernel}
\end{equation}
The round measure here reconstructs coefficients and does not identify $G^\Omega$ with $\mathsf H_\ell$. At $k=0.6$, $\eta=0.23$, both inverse matrix compositions were accurate to $4.8\times10^{-16}$, and inverse-kernel function reconstruction to $4.5\times10^{-15}$ relative error across the $p$, $d$, and $f$ sectors.

\section{Sklyanin's integral in Jacobi form}\label{app:jacobimetric}
In the source notation, Sklyanin's Eqs.~(10)--(11) give \cite{sklyanin1983}
\begin{equation}
\begin{aligned}
\langle f,g\rangle_{\rm Sk}&=\int_0^{2K}\!d\xi\int_0^{2K'}\!d\upsilon\,
\overline{f(\omega)}g(\omega)\mu_\ell(\omega,\bar \omega),\\
\omega&=\xi+i\upsilon,
\end{aligned}
\label{eq:skintegral}
\end{equation}
\begin{equation}
\mu_\ell=
\frac{\Theta_1(2\omega)\Theta_1(2\bar \omega)}
{\displaystyle\prod_{j=0}^{2\ell+1}
\Theta_3(\omega+\bar \omega+(2j-2\ell-1)\eta/2)\Theta_3(\omega-\bar \omega+(2j-2\ell-1)\eta/2)},
\label{eq:skdensity}
\end{equation}
Sklyanin's shift parameter is $\eta/2$. The sufficient positivity restriction is $(2\ell+1)\eta/2<K$; Sklyanin's Theorem~5 establishes self-adjointness in this metric. Positivity is not deduced from closure of our collocation matrices.

Set $\varphi=2\xi$, $\theta=2\upsilon$ and change the source wavefunction gauge to
\begin{equation}
\widetilde f(\theta,\varphi)=
\frac{f((\varphi+i\theta)/2)}
{[\Theta_4(\varphi)\Theta_2(i\theta)]^\ell}.
\label{eq:metricgauge}
\end{equation}
Pairing opposite shifts in Eq.~\eqref{eq:skdensity} gives the entirely real Jacobi weight. With $b_j=(2j+1)\eta/2$, it is
\begin{equation}
\boxed{\mathcal J_\ell(\theta,\varphi)=
\frac{\sn^2(\varphi,k)\cn^2(\theta,k')+\sn^2(\theta,k')}
{\displaystyle\prod_{j=0}^{\ell}
\bigl[\dn^2(b_j,k)-k^2\cn^2(b_j,k)\sn^2(\varphi,k)\bigr]
\bigl[\dn^2(b_j,k)-k'^2\sn^2(\theta,k')\bigr]}.}
\label{eq:jacobiweight}
\end{equation}
Consequently
\begin{equation}
\langle f,g\rangle_{\rm Sk}=
\frac{\mathcal C_\ell}{4}\int_0^{4K}\!d\varphi\int_0^{4K'}\!d\theta\,
\overline{\widetilde f}\widetilde g\,\mathcal J_\ell.
\label{eq:jacobiintegral}
\end{equation}
The factor $1/4$ is the coordinate Jacobian. This is the transformed source rectangle, not the single-cover sphere domain of Eq.~\eqref{eq:roundmetric}.

The scalar normalization can also be written without theta functions:
\begin{equation}
\mathcal C_\ell=(kk')^{\ell+1}\frac{\pi}{2K}
\exp\!\left\{-4\sum_{j=0}^{\ell}\left[
\int_0^{(2j+1)\eta/2}\left(\frac{(2j+1)\eta}{2}-v\right)\dn^2(v,k)\,dv
-\frac{E(k)(2j+1)^2\eta^2}{8K}\right]\right\},
\label{eq:metricconstant}
\end{equation}
where $E(k)$ is the complete elliptic integral of the second kind. One may absorb this positive scalar into the normalization convention. The products have two, three, or four factor pairs for the $p$, $d$, and $f$ sectors. On the stated positivity domain,
\begin{equation*}
\begin{aligned}
\dn^2\!\left(\frac{(2j+1)\eta}{2},k\right)
-k^2\cn^2\!\left(\frac{(2j+1)\eta}{2},k\right)\sn^2(\varphi,k)&\ge k'^2,\\
\dn^2\!\left(\frac{(2j+1)\eta}{2},k\right)-k'^2\sn^2(\theta,k')
&\ge k^2\cn^2\!\left(\frac{(2j+1)\eta}{2},k\right)>0.
\end{aligned}
\end{equation*}
Thus the weight is positive except on a set of measure zero. Its numerator is $1-y^2$ with $y$ from Eq.~\eqref{eq:unitcoords}; the full weight is not the round area density.

For target functions, the full prescription includes the inverse transform:
\begin{equation}
\langle F,G\rangle_\ell=\frac{\mathcal C_\ell}{4}
\int_0^{4K}\!d\varphi\int_0^{4K'}\!d\theta\,
\frac{\overline{(\mathcal U_\ell^{-1}F)((\varphi+i\theta)/2)}\,
(\mathcal U_\ell^{-1}G)((\varphi+i\theta)/2)}
{|\Theta_4(\varphi)\Theta_2(i\theta)|^{2\ell}}\,
\mathcal J_\ell(\theta,\varphi).
\label{eq:targetjacobimetric}
\end{equation}
The inverse transform cannot be omitted: $\mathcal J_\ell$ is a local weight for the gauged source functions, not for the unchanged target eigenfunctions.

The conversion follows from the theta addition formulas. For example, with $a$ real and theta argument $\xi_0$,
\begin{equation*}
\begin{aligned}
&\Theta_3(\xi_0+a)\Theta_3(\xi_0-a)\\
&\quad=\frac{\Theta_3(0)^2\Theta_4(\xi_0)^2\Theta_4(a)^2}{\Theta_4(0)^4}
[\dn^2(a)-k^2\cn^2(a)\sn^2(\xi_0)],\\
&\Theta_1(\varphi+i\theta)\Theta_1(\varphi-i\theta)\\
&\quad=\frac{k\Theta_4(\varphi)^2\Theta_4(i\theta)^2}{\Theta_4(0)^2}
[\sn^2(\varphi,k)-\sn^2(i\theta,k)].
\end{aligned}
\end{equation*}
Using $\sn^2(i\theta,k)=-s_\theta^2/c_\theta^2$ and Eq.~\eqref{eq:metricgauge} cancels the remaining angle-dependent theta factors in the density. The original theta integral and Eq.~\eqref{eq:jacobiintegral} were compared on $96\times96$ grids for all three parameter pairs and spins; the largest relative Gram-matrix discrepancy was $2.4\times10^{-14}$. This checks the direct conversion independently of target closure.

Finally, Sklyanin proposed a product formula for the reproducing kernel in his Eq.~(17). In our theta notation its candidate is, up to one overall scalar,
\begin{equation}
\prod_{j=0}^{2\ell-1}
\Theta_3(\omega+\omega'+(2j-2\ell+1)\eta/2)
\Theta_3(\omega-\omega'+(2j-2\ell+1)\eta/2).
\label{eq:kernelcandidate}
\end{equation}
We formed the kernel independently by inverting the source Gram matrix and compared at 32 complex pairs $(\omega,\omega')$ per case, using $64\times64$ and $128\times128$ integration grids. For the $p$, $d$, and $f$ sectors at the three parameter pairs, the largest relative discrepancy after fitting the single allowed scalar was $3.2\times10^{-15}$. This is numerical confirmation of Sklyanin's conjectured formula at the tested points, not a new general proof.

\section{Derivation of the two-space generator polynomial}
\label{app:two-space-polynomial}
\subsection{Derivation and ordering of the two-space polynomial}
The explicit operator expression is Eq.~\eqref{eq:two-space-polynomial} in the main text. Here we derive its first-factor coefficients and specify the finite-difference ordering.
\paragraph{Triplet identities and derivation.}
On $V_p$, direct substitution in Eq.~\eqref{eq:p-operators} gives the Cartesian spin-one action $L_i e_j=i\epsilon_{ijh}e_h$ in the harmonic basis $(e_1,e_2,e_3)=(x,y/\sqrt{J_3},z/\sqrt{J_2})$. Consequently the raw $(x,y,z)$ matrix units satisfy
\[
E_{ij}^{\rm raw}=\sqrt{\frac{J_j}{J_i}}
 \bigl[\delta_{ij}I-L_j^{(p)}L_i^{(p)}\bigr],\qquad
\sum_{i=1}^3\bigl(L_i^{(p)}\bigr)^2=2I.
\]
In particular the original fourth generator is already determined on this triplet:
\[
D_p=\sum_{i=1}^3J_i\Bigl[I-\bigl(L_i^{(p)}\bigr)^2\Bigr].
\]
The matrix-unit identity follows by applying both sides to the three harmonic basis functions. Substitution into the symmetric quotient of Eq.~\eqref{eq:spin1}, followed by $I=\tfrac12\sum_{a=1}^3(L_a^{(p)})^2$, gives Eq.~\eqref{eq:two-space-polynomial}. The ratios of the raw triplet norms reduce directly to the displayed $J_i/J_j$ factors because $J_1=J_2J_3$. No independent first-space coefficient matrices remain in the result.

These identities are specific to the triplet representation. Their use on $V_p\otimes V_m$ retains the stated higher-spin closure qualification. The absence of $D_p$ from Eq.~\eqref{eq:two-space-polynomial} reflects its displayed triplet identity; it has not been replaced by a differently normalized generator.

\paragraph{Four-angle action and ordering.}
The first operators act on $(\theta_1,\varphi_1)$ and the second on $(\theta_2,\varphi_2)$. Each uses $\varphi_r\mapsto\varphi_r\pm\eta$ and $\theta_r\mapsto\theta_r\pm i\eta$, with its original left coefficient functions. Tensor products act independently in the two factors. Within each factor, products are ordered compositions with the rightmost operator acting first; its coefficient functions are shifted by the next operator. They are not products of unshifted coefficient functions. In particular,
\[
\bigl(L_j^{(p)}\otimes L_h^{(m)}\bigr)
\bigl(L_i^{(p)}\otimes D_m\bigr)
=\bigl(L_j^{(p)}L_i^{(p)}\bigr)\otimes\bigl(L_h^{(m)}D_m\bigr).
\]
Thus the formula is already a difference operator on the product elliptic harmonics in four angular variables.

\paragraph{Verification and the need for quadratic terms.}
At the three regular deformation pairs used above, for $m\in\{p,d,f\}$ and offsets $u-\alpha=0.71,0.4+0.1i,0.31i$, all 27 polynomial matrices agree with fusion within $3.36\times10^{-16}$ relative error. Six point evaluations using nested shifts on the product harmonics at $k=0.6,\eta=0.23$ agree within $4.66\times10^{-15}$. An independent Wolfram Language transcription agrees with ordered fusion for all three pairings and the three listed offsets at $k=0.6$, $\eta=0.23$, within $5.25\times10^{-15}$.

A bilinear ansatz in the single generators is generally insufficient: the sampled operators have operator Schmidt rank nine, whereas the first-space span of $I,D_p,L_1^{(p)},L_2^{(p)},L_3^{(p)}$ has dimension at most five. Direct bilinear fits leave relative residuals $0.053$--$0.616$. The quadratic products in Eq.~\eqref{eq:two-space-polynomial} supply the full triplet operator algebra.

\section{Symmetric embeddings and angular reconstruction}
\label{app:reconstruction}
\subsection{Explicit embedding examples}
For spin two, $N=4$, the complete $16\times5$ matrix is displayed below with its row and column labels:
\[
\renewcommand{\arraystretch}{1.1}
\begin{array}{c|ccccc}
b\backslash r&0&1&2&3&4\\ \hline
0000&1&0&0&0&0\\[2pt]
0001&0&\frac12&0&0&0\\[2pt]
0010&0&\frac12&0&0&0\\[2pt]
0011&0&0&\frac1{\sqrt6}&0&0\\[2pt]
0100&0&\frac12&0&0&0\\[2pt]
0101&0&0&\frac1{\sqrt6}&0&0\\[2pt]
0110&0&0&\frac1{\sqrt6}&0&0\\[2pt]
0111&0&0&0&\frac12&0\\[2pt]
1000&0&\frac12&0&0&0\\[2pt]
1001&0&0&\frac1{\sqrt6}&0&0\\[2pt]
1010&0&0&\frac1{\sqrt6}&0&0\\[2pt]
1011&0&0&0&\frac12&0\\[2pt]
1100&0&0&\frac1{\sqrt6}&0&0\\[2pt]
1101&0&0&0&\frac12&0\\[2pt]
1110&0&0&0&\frac12&0\\[2pt]
1111&0&0&0&0&1\\[2pt]
\end{array}
\]
Its columns are ordered by $r=0,1,2,3,4$. For spin three, $E_6$ is $64\times7$. Its seven columns have disjoint supports of sizes $1,6,15,20,15,6,1$, with respective nonzero entries
\[
1,\quad \frac1{\sqrt6},\quad \frac1{\sqrt{15}},\quad
\frac1{\sqrt{20}},\quad \frac1{\sqrt{15}},\quad
\frac1{\sqrt6},\quad 1.
\]
For example, column $r=1$ of $E_6$ has value $1/\sqrt6$ in rows $000001$, $000010$, $000100$, $001000$, $010000$, and $100000$, and zero elsewhere.

For a coefficient vector $c=(c_0,\ldots,c_N)^{\mathsf T}$, the embedding and its adjoint act as
\[
(E_Nc)_b=\frac{c_{\sum_a b_a}}{\sqrt{\binom N{\sum_a b_a}}},
\qquad
(E_N^\dagger v)_r=\frac1{\sqrt{\binom Nr}}
 \sum_{\substack{b\in\{0,1\}^N\\ \sum_a b_a=r}}v_b.
\]
Each column therefore has norm one, and distinct columns are orthogonal, so $E_N^\dagger E_N=I_{N+1}$.

\subsection{Analytic higher-spin basis maps}
\label{sec:analytic-basis-maps}
The maps $T_d,T_f$ are determined by the action of the three generators on a single angular harmonic and by linear spectral projectors. The result below gives every column of $T_l^{-1}$ from fundamental fusion and the explicit spectral data of Sections~\ref{sec:d-spectrum} and \ref{sec:f-states}. It applies on a regular irreducible branch where the angular functions form the stated spin module.

\paragraph{Symmetric-basis generators.}
Choose a regular spectral point $u_0$. Let $\mathcal F_l(u_0)$ be the compression in Eq.~\eqref{eq:general} with a fundamental second factor and $T_l=I_{N_l+1}$. Define the following $(N_l+1)\times(N_l+1)$ matrices by tracing only the two-dimensional fundamental factor:
\[
\begin{aligned}
B_0&=\frac{\left(\sum_{a=1}^{N_l+1}\lambda_a\right)
 \operatorname{tr}_{\mathrm{aux}}\mathcal F_l(u_0)}
 {\tr\mathcal F_l(u_0)},\\
B_i&=\frac{\left(\sum_{a=1}^{N_l+1}\lambda_a\right)
 \operatorname{tr}_{\mathrm{aux}}[\mathcal F_l(u_0)(I_{N_l+1}\otimes\sigma_i)]}
 {W_i(u_0)\tr\mathcal F_l(u_0)},\qquad i=1,2,3.
\end{aligned}
\]
The eigenvalues $\lambda_a$ are given in Sections~\ref{sec:d-spectrum} and \ref{sec:f-states}, in the order of Eq.~\eqref{eq:exchange-basis-order}. These expressions remove the common scalar in the raw fundamental fusion. They satisfy
\begin{equation}
T_lB_0=D_lT_l,\qquad
T_lB_i=\sqrt{(J_i+J_j)(J_i+J_h)}\,L_i^{(l)}T_l,
\quad (i,j,h)\ \mathrm{cyclic}.
\label{eq:basis-intertwining}
\end{equation}
The chosen point must have nonzero $W_i(u_0)$ and $\tr\mathcal F_l(u_0)$. The resulting generators and maps are independent of that point on the represented branch.

\paragraph{Column formula.}
The last angular harmonics are $\Phi_{yz}$ for $d$ and $\Phi_{xyz}$ for $f$. In the symmetric coordinates $r=0,\ldots,N_l$, their coefficient vectors can be fixed as
\[
v_d=(0,1,0,1,0)^{\mathsf T},\qquad
v_f=(0,-1,0,0,0,1,0)^{\mathsf T}.
\]
They satisfy $B_0v_l=\lambda_{N_l+1}v_l$. This fixes the overall scalar of the inverse map through $T_l^{-1}e_{N_l+1}=v_l$, where $e_a$ is the $a$th angular coefficient vector.

For $a=1,\ldots,N_l$, the axis $\kappa(a)$ is
$(1,1,2,3)$ for $d$ and $(1,1,2,2,3,3)$ for $f$. It specifies which generator reaches harmonic $a$ from the last harmonic. Its coefficient is $\mathfrak t_a$, given explicitly in Appendix~\ref{app:singleton-transitions}. The partner index $\bar a$ exchanges $1,2$ and, for $f$, also $3,4$ and $5,6$. In each paired sector,
\[
L_{\kappa(a)}^{(l)}\Phi_{N_l+1}
 =\mathfrak t_a\Phi_a+\mathfrak t_{\bar a}\Phi_{\bar a}.
\]
For the two remaining $d$ singleton sectors the corresponding formula contains only $\mathfrak t_a\Phi_a$.

The complete inverse map is
\begin{equation}
\begin{aligned}
T_l^{-1}&=(v_1\ \cdots\ v_{N_l}\ v_l),\\
v_a&=
\begin{cases}
\displaystyle
\frac{(B_0-\lambda_{\bar a}I_{N_l+1})B_{\kappa(a)}v_l}
 {(\lambda_a-\lambda_{\bar a})\mathfrak t_a
 \sqrt{\prod_{b\ne\kappa(a)}(J_{\kappa(a)}+J_b)}},
 &a\text{ in a paired sector},\\[8pt]
\displaystyle
\frac{B_{\kappa(a)}v_l}
 {\mathfrak t_a\sqrt{\prod_{b\ne\kappa(a)}(J_{\kappa(a)}+J_b)}},
 &a\text{ in a singleton sector}.
\end{cases}
\end{aligned}
\label{eq:analytic-map-columns}
\end{equation}
The product index $b$ runs over the three axes. Appendix~\ref{app:basis-map-details} also writes $T_d,T_f$ as full sparse matrices.

To obtain Eq.~\eqref{eq:analytic-map-columns}, apply Eq.~\eqref{eq:basis-intertwining} to the last angular coefficient vector. In a paired sector $B_{\kappa(a)}v_l$ is a linear combination of two $B_0$ eigenvectors. Multiplication by $B_0-\lambda_{\bar a}I$ eliminates the partner, and division by $\lambda_a-\lambda_{\bar a}$ isolates the required column. The explicit transition coefficient fixes its scale relative to $v_l$. Every sector is reached in this way, so the map is unique up to its common scalar on the regular irreducible branch.

The displayed chart requires distinct sector eigenvalues and nonzero transition coefficients. A different regular chart or a meromorphic limit is needed when a denominator vanishes. The general-deformation closure qualification of the angular realization remains in force.

At the three deformations listed in Section~\ref{sec:ybe-validation}, direct 70-digit evaluations check the last-harmonic eigenvectors, the full $D_l$ spectrum, both matrix inverses, independence from a second probe point, and all four shifted angular generators at three angle pairs. Every sampled residual is below $10^{-50}$. The six transition identities used in the derivation have also been verified symbolically as rational Jacobi identities.

\subsection{Bilinear trace and angular action}
The angular functions can be recovered from constant bilinear trace selectors. For each $m=p,d,f$, define matrices on $V_{\mathrm{aux}}\otimes V_m$ by
\begin{align}
E_0^{(m)}&=\frac{I_2\otimes D_m}{2\tr(D_m^2)},\notag\\
E_1^{(m)}&=\frac{\sn(\alpha,k)\,\sigma_1\otimes L_1^{(m)}}
 {2\tr[(L_1^{(m)})^2]\sqrt{(J_1+J_2)(J_1+J_3)}},\notag\\
E_2^{(m)}&=\frac{\cn(\alpha,k)\,\sigma_2\otimes L_2^{(m)}}
 {2\tr[(L_2^{(m)})^2]\sqrt{(J_2+J_3)(J_2+J_1)}},\notag\\
E_3^{(m)}&=\frac{\dn(\alpha,k)\,\sigma_3\otimes L_3^{(m)}}
 {2\tr[(L_3^{(m)})^2]\sqrt{(J_3+J_1)(J_3+J_2)}}.
\end{align}
The traces include both tensor factors and use no complex conjugation. Pauli trace orthogonality gives
\[
\tr[R^{(\mathrm{aux},m)}(v)E_0^{(m)}]=1,\qquad
\bigl(\tr[R^{(\mathrm{aux},m)}(v)E_i^{(m)}]\bigr)_{i=1}^3
 =(\sn(v,k),\cn(v,k),\dn(v,k)).
\]
In particular, the arguments $v=i\theta$ and $v=\varphi$ supply the Jacobi factors used to reconstruct the angular basis. For example, the $p$ row is obtained from
\[
 \cn(\theta,k')\bigl(\dn(i\theta,k)\sn(\varphi,k),\
 \cn(\varphi,k),\ -i\sn(i\theta,k)\dn(\varphi,k)\bigr).
\]
Substitution of these trace identities into the explicit functions in Eqs.~\eqref{eq:fivebasis} and \eqref{eq:sevenbasis} reconstructs every $d$ and $f$ harmonic. The identities
\[
\sn(i\theta,k)=\frac{i\sn(\theta,k')}{\cn(\theta,k')},\quad
\cn(i\theta,k)=\frac1{\cn(\theta,k')},\quad
\dn(i\theta,k)=\frac{\dn(\theta,k')}{\cn(\theta,k')}
\]
convert the complementary-modulus factors to functions at $i\theta$.

The ordered basis functions are denoted $\Phi_a^{(l)}$. The action on the input product $\Phi_{i}^{(l)}\Phi_{j}^{(m)}$ is
\[
 \sum_{b,c}\Phi_{b}^{(l)}(\theta_1,\varphi_1)\Phi_{c}^{(m)}(\theta_2,\varphi_2)
 \tr\!\left[R^{(l,m)}(u)\bigl(|i\rangle\langle b|\otimes|j\rangle\langle c|\bigr)\right]
 =
 \sum_{b,c}\Phi_{b}^{(l)}(\theta_1,\varphi_1)\Phi_{c}^{(m)}(\theta_2,\varphi_2)R^{(l,m)}_{bc,ij}(u).
\]
The trace inserts $R^{(l,m)}(u)$ on the two retained quantum factors of the selector products. The bras are coordinate duals. Adjoints use the transported product metric. The spectral argument $u$ is independent of the angular coordinates used to reconstruct the harmonics.

\section{Singleton transition coefficients and sparse basis maps}
\label{app:basis-map-details}
The coefficients below complete the analytic column formula in Section~\ref{sec:analytic-basis-maps}. They are obtained by applying the explicit operators of Appendix~\ref{app:operators} to the last harmonic in each basis. Throughout this appendix every Jacobi function has modulus $k$.

\subsection{Expanded singleton transition coefficients}
\label{app:singleton-transitions}
For $d$, applying $L_1$ to $\Phi_{yz}$ reaches the two even functions, applying $L_2$ reaches $\Phi_{xy}$, and applying $L_3$ reaches $\Phi_{xz}$. In the basis order in Section~\ref{sec:angular-spaces}, the nonzero coefficients are
\begin{align}
\mathfrak t_1={}&-\frac{i(1-2\sn^2\eta+k^2\sn^4\eta)(1+r_2)}
 {\sqrt{\cn\eta\,\dn\eta}(1-k^2\sn^4\eta)^2
 (r_2-r_1)(1+r_1/k^2)},\notag\\
\mathfrak t_2={}&\frac{i(1-2\sn^2\eta+k^2\sn^4\eta)(1+r_1)}
 {\sqrt{\cn\eta\,\dn\eta}(1-k^2\sn^4\eta)^2
 (r_2-r_1)(1+r_2/k^2)},\notag\\
\mathfrak t_3={}&\frac{i(1-2\sn^2\eta+k^2\sn^4\eta)}
 {\sqrt{\cn\eta}(1-k^2\sn^4\eta)^2},\qquad
\mathfrak t_4=-\frac{i(1-2k^2\sn^2\eta+k^2\sn^4\eta)}
 {\sqrt{\dn\eta}(1-k^2\sn^4\eta)^2}.
\label{eq:d-singleton-coefficients}
\end{align}
For $f$, $L_1\Phi_{xyz}$ reaches the $\sn$ pair, $L_2\Phi_{xyz}$ reaches the $\cn$ pair, and $L_3\Phi_{xyz}$ reaches the $\dn$ pair. Their coefficients are
\begin{align}
\mathfrak t_1={}&-\frac{i\sqrt{\cn\eta\,\dn\eta}\,\cn^2(3\eta)(1+r_2)}
 {[\cn^2\eta-\sn^2(2\eta)\dn^2\eta](r_2-r_1)(1+r_1/k^2)},\notag\\
\mathfrak t_2={}&\frac{i\sqrt{\cn\eta\,\dn\eta}\,\cn^2(3\eta)(1+r_1)}
 {[\cn^2\eta-\sn^2(2\eta)\dn^2\eta](r_2-r_1)(1+r_2/k^2)},\notag\\
\mathfrak t_3={}&\frac{i\sqrt{\cn\eta}\,\dn\eta\,\cn^2(3\eta)}
 {[\cn^2\eta-\sn^2(2\eta)\dn^2\eta](r_3-r_4)(1+r_3/k^2)},\notag\\
\mathfrak t_4={}&-\frac{i\sqrt{\cn\eta}\,\dn\eta\,\cn^2(3\eta)}
 {[\cn^2\eta-\sn^2(2\eta)\dn^2\eta](r_3-r_4)(1+r_4/k^2)},\notag\\
\mathfrak t_5={}&-\frac{ik^2\cn\eta\sqrt{\dn\eta}\,(1+r_6)}
 {[1-k^2\sn^2(2\eta)\sn^2\eta](1-k^2)(r_6-r_5)},\notag\\
\mathfrak t_6={}&\frac{ik^2\cn\eta\sqrt{\dn\eta}\,(1+r_5)}
 {[1-k^2\sn^2(2\eta)\sn^2\eta](1-k^2)(r_6-r_5)}.
\label{eq:f-singleton-coefficients}
\end{align}
Here the ratios $r_a$ are the explicit roots in Sections~\ref{sec:d-spectrum} and \ref{sec:f-states} for the indicated representation.

For example, restricting the first $\sn$ pair to $\theta=0$ gives
\[
\Phi_a(0,\varphi)=
 \frac{\sn\varphi(1+r_a\sn^2\varphi)(1+r_a/k^2)}
 {[1-k^2\sn^2(2\eta)\sn^2\varphi][1-\sn^2(2\eta)]},
 \qquad a=1,2.
\]
The same restriction of $L_1^{(f)}\Phi_{xyz}$ is
\[
-\frac{i\cn^2(3\eta)\dn\eta}
 {\cn\eta\sqrt{\cn\eta\,\dn\eta}[1-\sn^2(2\eta)]
 [1-\sn^2(2\eta)\dn^2\eta/\cn^2\eta]}
 \frac{\sn\varphi(1-\sn^2\varphi)}
 {1-k^2\sn^2(2\eta)\sn^2\varphi}.
\]
Matching the constant and $\sn^2\varphi$ coefficients after canceling the common prefactor gives $\mathfrak t_1,\mathfrak t_2$. The remaining coefficients follow by the other parity restrictions. The full two-angle identities reduce to zero under the Jacobi addition formulas, the quadratic equations for $r_a$, and $\cn^2v=1-\sn^2v$, $\dn^2v=1-k^2\sn^2v$.

\subsection{Full sparse matrices}
The formulas can be written as full sparse matrices. For $d$, write the computed columns as
\[
v_1=(\alpha_1,0,\beta_1,0,\alpha_1)^{\mathsf T},\quad
v_2=(\alpha_2,0,\beta_2,0,\alpha_2)^{\mathsf T},\quad
v_3=(q,0,0,0,-q)^{\mathsf T},\quad
v_4=(0,t,0,-t,0)^{\mathsf T},
\]
and put $\Delta=\alpha_1\beta_2-\alpha_2\beta_1$. Then
\begingroup\setlength{\arraycolsep}{5pt}\renewcommand{\arraystretch}{1.5}
\[
T_d=
\begin{pmatrix}
\dfrac{\beta_2}{2\Delta}&0&-\dfrac{\alpha_2}{\Delta}&0&\dfrac{\beta_2}{2\Delta}\\
-\dfrac{\beta_1}{2\Delta}&0&\dfrac{\alpha_1}{\Delta}&0&-\dfrac{\beta_1}{2\Delta}\\
\dfrac1{2q}&0&0&0&-\dfrac1{2q}\\
0&\dfrac1{2t}&0&-\dfrac1{2t}&0\\
0&\dfrac12&0&\dfrac12&0
\end{pmatrix}.
\]
\endgroup
For $f$, the six columns have the forms
\[
\begin{array}{ll}
v_a=(\alpha_a,0,\beta_a,0,-\beta_a,0,-\alpha_a)^{\mathsf T},
 &a=1,2,\\
v_a=(\alpha_a,0,\beta_a,0,\beta_a,0,\alpha_a)^{\mathsf T},
 &a=3,4,\\
v_a=(0,\alpha_a,0,\beta_a,0,\alpha_a,0)^{\mathsf T},
 &a=5,6.
\end{array}
\]
Set $\Delta_s=\alpha_1\beta_2-\alpha_2\beta_1$,
$\Delta_c=\alpha_3\beta_4-\alpha_4\beta_3$,
$\Delta_d=\alpha_5\beta_6-\alpha_6\beta_5$. Thus
\begingroup\setlength{\arraycolsep}{5pt}\renewcommand{\arraystretch}{1.5}
\[
T_f=
\begin{pmatrix}
\dfrac{\beta_2}{2\Delta_s}&0&-\dfrac{\alpha_2}{2\Delta_s}&0&\dfrac{\alpha_2}{2\Delta_s}&0&-\dfrac{\beta_2}{2\Delta_s}\\
-\dfrac{\beta_1}{2\Delta_s}&0&\dfrac{\alpha_1}{2\Delta_s}&0&-\dfrac{\alpha_1}{2\Delta_s}&0&\dfrac{\beta_1}{2\Delta_s}\\
\dfrac{\beta_4}{2\Delta_c}&0&-\dfrac{\alpha_4}{2\Delta_c}&0&-\dfrac{\alpha_4}{2\Delta_c}&0&\dfrac{\beta_4}{2\Delta_c}\\
-\dfrac{\beta_3}{2\Delta_c}&0&\dfrac{\alpha_3}{2\Delta_c}&0&\dfrac{\alpha_3}{2\Delta_c}&0&-\dfrac{\beta_3}{2\Delta_c}\\
0&\dfrac{\beta_6}{2\Delta_d}&0&-\dfrac{\alpha_6}{\Delta_d}&0&\dfrac{\beta_6}{2\Delta_d}&0\\
0&-\dfrac{\beta_5}{2\Delta_d}&0&\dfrac{\alpha_5}{\Delta_d}&0&-\dfrac{\beta_5}{2\Delta_d}&0\\
0&-\dfrac12&0&0&0&\dfrac12&0
\end{pmatrix}.
\]
\endgroup
The displayed coefficients are the components of the explicit columns above. The formulas require nonzero transition coefficients, separated roots, and invertible sector blocks. Zeros of a chosen denominator require another regular chart or a limit when the representation remains equivalent.

\section{Elliptic gamma normalization}
\label{sec:gamma-normalization}

The scalar factors constructed here give unit inversion and metric unitarity for all nine ordered matrices on the adopted represented branch. The scalar normalization also preserves all four local Casimir commutators in Eq.~\eqref{eq:pair-local-casimir-conservation}. With real $\eta$, put $\zeta=u-\alpha$, where $\alpha=\eta/2-iK'$. We construct a common meromorphic normalization for the nine ordered pairs $l,m\in\{p,d,f\}$. The unitary contour is $\zeta=it$, with $t$ real. All formulas use the positive transported coefficient metrics $\mathsf H_l,\mathsf H_m$. At a removable singularity, the normalized matrix is defined by the limit of the complete product.

\subsection{Fundamental normalization}
The singlet eigenvalue of $R^{(\mathrm{aux},\mathrm{aux})}(\alpha+\zeta)$ is $1-\sum_{i=1}^3 W_i(\alpha+\zeta)$. Fundamental inversion is scalar, and its scalar equals the product of the singlet eigenvalues at $\zeta$ and $-\zeta$. Hence
\[
n_0(\zeta)=-\frac{1}{1-\sum_{i=1}^3 W_i(\alpha+\zeta)},\qquad
n_0(0)R^{(\mathrm{aux},\mathrm{aux})}(\alpha)=P .
\]
The factor $n_0(\zeta)$ gives unit scalar inversion.

In multiplicative notation, the Ruijsenaars elliptic gamma function and its shift relation are
\[
\Gamma(X;p_\Gamma,q_\Gamma)=\prod_{a,b\ge0}\frac{1-p_\Gamma^{a+1}q_\Gamma^{b+1}/X}{1-p_\Gamma^a q_\Gamma^b X},
\qquad
\frac{\Gamma(q_\Gamma X;p_\Gamma,q_\Gamma)}{\Gamma(X;p_\Gamma,q_\Gamma)}
=(X;p_\Gamma)_\infty(p_\Gamma/X;p_\Gamma)_\infty .
\]
With $p_\Gamma=e^{-2\pi K'/K}$ and $0<q_\Gamma<1$, use the shift quotient
\[
\Theta_\Gamma(v)=
\frac{\Gamma(q_\Gamma e^{2\pi i v/(4K)};p_\Gamma,q_\Gamma)}
{\Gamma(e^{2\pi i v/(4K)};p_\Gamma,q_\Gamma)}.
\]
Here $(X;p_\Gamma)_\infty=\prod_{n=0}^\infty(1-Xp_\Gamma^n)$, $|p_\Gamma|,|q_\Gamma|<1$, and the quotients are continued meromorphically from their convergence domain. The second nome cancels by the shift relation~\cite{ruijsenaars1997}.

The singlet eigenvalue has the following poles and zeros modulo $4K,4iK'$:
\[
\mathcal P=-\eta/2+\{0,2K,2iK',2K+2iK'\},\qquad
\mathcal Z=\{\eta,-\eta+2K,-\eta+2iK',-\eta+2K+2iK'\}.
\]
The offset form $W_1(\alpha+\zeta)=\sn(\eta/2,k)/\sn(\zeta+\eta/2,k)$ and the corresponding $\cn,\dn$ ratios give the poles. At $\zeta=-\eta$ the seed is four times the singlet projector, and inversion forces the singlet eigenvalue to vanish at $\zeta=\eta$. The three nonzero half-period translations exchange the singlet with the other three Bell channels and supply the remaining zeros. The sums of the two divisors agree, so the ratio below is elliptic. Its value at the origin fixes the constant:
\[
 n_0(\zeta)=
 \frac{\displaystyle\prod_{b\in\mathcal P}\Theta_\Gamma(\zeta-b)
                   \prod_{c\in\mathcal Z}\Theta_\Gamma(-c)}
      {\displaystyle 2\prod_{c\in\mathcal Z}\Theta_\Gamma(\zeta-c)
                         \prod_{b\in\mathcal P}\Theta_\Gamma(-b)}.
\]
The divisor argument applies at generic regular parameters and extends by meromorphic continuation. The singlet prescription selects an inversion normalization. A minimal scalar of the type studied by Zamolodchikov also has to satisfy a crossing equation~\cite{zamolodchikov1979z4}.

\subsection{Normalization of all nine fused matrices}
Use the symmetric embedding $E_{N_a}$ of Section~\ref{sec:fused-higher} for $a\in\{p,d,f\}$. The trace of the raw fundamental fusion determines the conversion factor
\begin{align}
c_a(v)={}&\frac{1}{2\tr D_a}\tr\!\biggl[
(E_{N_a}^\dagger\otimes I_2)
R_{N_a,0}^{(\mathrm{aux},\mathrm{aux})}\!\left(\alpha+v+\tfrac{N_a-1}{2}\eta\right)\cdots \notag\\*
&\hspace{28mm}\cdots R_{1,0}^{(\mathrm{aux},\mathrm{aux})}\!\left(\alpha+v-\tfrac{N_a-1}{2}\eta\right)
(E_{N_a}\otimes I_2)\biggr].
\end{align}
After transport to the angular basis and ordering the spin-half space first, the compressed product is $c_a(v)R^{(\mathrm{aux},a)}(\alpha+v)$. Its Pauli identity component fixes the scalar through $D_a$. This trace uses the raw fundamental fusion and is independent of the basis in the compressed space.

The required factor for each ordered pair is
\begin{align}
n_{lm}(\zeta)
&=\epsilon_{lm}
 \prod_{r=1}^{N_l}c_m\!\left(\zeta+\left(r-\tfrac{N_l+1}{2}\right)\eta\right)
 \notag\\
&\quad\times\prod_{r=1}^{N_l}\prod_{s=1}^{N_m}
 n_0\!\left(\zeta+\left(r+s-\tfrac{N_l+N_m+2}{2}\right)\eta\right),\label{eq:gamma-fused-factor}\\
U_{lm}(\zeta)&=n_{lm}(\zeta)R^{(l,m)}(\alpha+\zeta).\label{eq:gamma-fused-matrix}
\end{align}
There are $N_lN_m$ fundamental factors and $N_l$ auxiliary conversion factors. On the regular branch used here $\epsilon_{ll}=-1$ gives $U_{ll}(0)=P$. We take $\epsilon_{lm}=1$ for mixed spins. These constant signs leave logarithmic derivatives unchanged.

\paragraph{Inversion through fusion.}
The ordered auxiliary product preserves the kernel of the symmetric projection, as follows from RLL at the adjacent fusion spacing and the antisymmetric projector at $-\eta$. Its inverse is the reversed product with each constituent offset negated. Reversing the auxiliary labels turns that inverse into the product at $-\zeta$, since the shifts are symmetric about zero. Label reversal acts trivially on the symmetric quotient. Applying this argument first to the fundamental product and then to the spin-half auxiliary product proves
\[
 U_{lm}(\zeta)U_{lm}(-\zeta)=I
\]
in the ordered product space, wherever the operators are regular~\cite{konno2005}.

\paragraph{The transported adjoint.}
For real shifts, use the product of raw fundamental factors at positive shift differences to reverse the auxiliary string. The Yang--Baxter equation identifies the products obtained from different reduced sequences. An adjacent crossing at $+\eta$ kills the antisymmetric doublet. Choosing sequences with a given adjacent crossing first or last shows that the reversal form is supported on the symmetric subspace. Its restriction is generically nondegenerate, since in the rational limit it is a nonzero scalar on that subspace.

This form intertwines the ordered auxiliary product with its reverse. Individual spin-half auxiliary matrices are Hermitian analytic in the second-space metric, by metric Hermiticity of $D,L_i$ and real analyticity of their scalar coefficients. The adjoint reverses the order. On the first symmetric space the resulting form agrees up to scalar with $T_l^\dagger \mathsf H_l T_l$. Indeed, both forms intertwine the adjoint of the spin-half probe, and their ratio commutes with its four Pauli coefficients. Irreducibility of those coefficients identifies the forms. The argument uses the exact fusion intertwiners on the regular irreducible branch. In the angular realization the higher-spin intertwiners are given by the analytic formulas in Section~\ref{sec:analytic-basis-maps}. Thus for the product-metric adjoint $\sharp$,
\[
 U_{lm}(\zeta)^\sharp=U_{lm}(\bar \zeta),\qquad
 U_{lm}(it)^\dagger(\mathsf H_l\otimes \mathsf H_m)U_{lm}(it)=\mathsf H_l\otimes \mathsf H_m.
\]

\subsection{Scalar offsets in logarithmic derivatives}
The offset in the uncentered derivative with respect to $\zeta/\eta$ is $C_{lm}(k,\eta)I$. For equal labels the reference matrix is $(N_l+1)^2R^{(l,l)}/\tr[P R^{(l,l)}]$. For ascending mixed labels it is $\zeta R^{(l,m)}(\alpha+\zeta)$, and for descending labels it is $R^{(l,m)}(\alpha+\zeta)$. Comparing these regular representatives with $U_{lm}$ gives
\begin{equation}
 C_{lm}(k,\eta)=\left.\eta\frac{d}{d\zeta}
 \begin{cases}
 \log\!\left[n_{ll}(\zeta)\tr[P R^{(l,l)}(\alpha+\zeta)]/(N_l+1)^2\right], & l=m,\\
 \log[n_{lm}(\zeta)/\zeta], & l\prec m,\\
 \log[n_{lm}(\zeta)], & m\prec l
 \end{cases}\right|_{\zeta=0}.
\label{eq:gamma-energy-shift}
\end{equation}
The ratios inside the logarithms have finite nonzero limits on the tested regular branch.
In the mixed local density the scalar cancels from conjugation by the zero-offset mixed gate. On a periodic alternating $l,m$ chain with $M$ sites, the row with auxiliary spin $l$ acquires $M(C_{ll}+C_{lm})I/2$. The other row adds $M(C_{mm}+C_{ml})I/2$ if both rows are used. Higher logarithmic derivatives acquire the corresponding derivatives of the scalar logarithm times the identity.

For example, the dimensionless shifts in \eqref{eq:gamma-energy-shift} at $k=0.6$, $\eta=0.23$ are
\[
\begin{array}{c|rrr}
l\backslash m&p&d&f\\\hline
p&4.333627&4.834725&3.253691\\
d&3.334435&7.037661&5.958911\\
f&2.086141&4.791360&8.631363
\end{array}
\]
Trace centering removes the scalar contribution. Level gaps, eigenvectors, and derivatives of the gaps with respect to the deformation parameters are invariant under the scalar normalization. For instance, a common factor $e^{a\zeta}$ with real $a$ preserves inversion and metric unitarity and adds $\eta a$ to $C_{lm}$.

\subsection{Numerical checks}
The tests cover all nine ordered pairs at $(k,\eta)=(0.4,0.18),(0.6,0.23),(0.8,0.17)$. Direct evaluation of the gamma double products with two choices of $q_\Gamma$ agrees with the singlet formula within $1.6\times10^{-14}$. On the three imaginary offsets $0.17i,0.31i,0.71i$ the largest metric-unitarity defect in the orthonormal product basis is below $2.0\times10^{-12}$. The string-reversal form agrees with the angular metric up to scalar within $1.1\times10^{-12}$. Direct comparisons of the logarithmic derivatives with \eqref{eq:gamma-energy-shift} agree within $6.5\times10^{-13}$. The formulas above specify the scalar conventions used in these checks.

The independent Wolfram Language double-product calculation agrees with the Jacobi singlet formula for $n_0$ within $4.8\times10^{-52}$ across 18 sampled offsets and deformations. 

A separate Wolfram Language check evaluates the expanded factor in Eq.~\eqref{eq:gamma-fused-factor} for all nine ordered pairs at $k=0.6$, $\eta=0.23$ and offsets $0.17i$ and $0.27+0.08i$. The largest inversion residual in the raw angular coefficient basis is $8.92\times10^{-11}$, using the stored double-precision basis maps.

\section{Coefficient metrics from the finite generators}
\label{app:metrics}
The positive inner product required for Hermitian analyticity can be constructed from the finite matrices specified in this paper. This gives a finite prescription for the metric in the same angular basis as the fused matrices.

For $l=d,f$, form
\[
L_i^{(l)}=
 \frac{T_lB_iT_l^{-1}}{\sqrt{\prod_{b\ne i}(J_i+J_b)}},
 \qquad b\in\{1,2,3\},
\]
using Eq.~\eqref{eq:analytic-map-columns}. Since $D_l$ is real diagonal with distinct eigenvalues on the displayed chart, every Hermitizing metric for $D_l$ is diagonal. Choose its last entry equal to one. The remaining entries are
\begin{equation}
(\mathsf H_l)_{aa}=
 \frac{\overline{(L_{\kappa(a)}^{(l)})_{N_l+1,a}}}{\mathfrak t_a},
 \qquad a=1,\ldots,N_l,\qquad
(\mathsf H_l)_{N_l+1,N_l+1}=1.
\label{eq:explicit-angular-metric}
\end{equation}
All off-diagonal entries vanish. The axes $\kappa(a)$ and coefficients $\mathfrak t_a$ are those of Appendix~\ref{app:singleton-transitions}. The numerator is computed from the same explicit finite matrices, so this prescription uses only $k,\eta$.

Indeed, the $(a,N_l+1)$ entry of
$(L_i^{(l)})^\dagger \mathsf H_l=\mathsf H_lL_i^{(l)}$ reads
$\overline{(L_i^{(l)})_{N_l+1,a}}=(\mathsf H_l)_{aa}(L_i^{(l)})_{a,N_l+1}$.
Taking $i=\kappa(a)$ gives Eq.~\eqref{eq:explicit-angular-metric}. Every angular sector is connected to the last harmonic, so these ratios determine the diagonal metric up to one scalar. On the positive represented branch they agree, up to that scalar, with the positive Sklyanin form transported by the intertwiner. The remaining entries of the Hermitizing equations provide direct finite checks of the chosen branch.

The triplet metric is given in Eq.~\eqref{eq:tripletmetric}.
Multiplying any $\mathsf H_l$ by a positive scalar leaves adjoints and metric-unitarity equations unchanged. For a tensor product use $\mathsf H_l\otimes \mathsf H_m$. An orthonormal coefficient representation is obtained by conjugation with $\mathsf H_l^{1/2}\otimes \mathsf H_m^{1/2}$, where the square roots are positive. The metric adjoint of an endomorphism is
$A^\sharp=(\mathsf H_l\otimes \mathsf H_m)^{-1}A^\dagger(\mathsf H_l\otimes \mathsf H_m)$.

\section{Transfer rows and local derivative formulas}
\label{app:transfer-derivation}
This appendix derives the local densities in Section~\ref{sec:integrable-exchange} from the finite $R$-matrices. Let $M$ be the number of chain sites, let $l_j$ denote the representation at site $j$, and let $a$ label the auxiliary space. Define
\begin{equation}
\tau_a(w)=\operatorname{tr}_0
 [U_{a l_M,\,0M}(\eta w)\cdots U_{a l_1,\,01}(\eta w)] .
\label{eq:transfer-row}
\end{equation}
The label $0$ denotes the auxiliary factor, and the rightmost factor acts first. Here $U_{ab}(\zeta)$ is the normalized matrix in Eq.~\eqref{eq:gamma-fused-matrix}, acting on $V_a\otimes V_b$. The fused Yang--Baxter equation gives the RTT relation between two such ordered products. Tracing both auxiliary spaces and using cyclicity yields
$[\tau_a(w),\tau_b(w')]=0$ wherever the products are regular. Invertibility at the expansion point therefore gives commuting logarithmic derivatives.

\subsection{Homogeneous chains and the first charges}
For $l_j=a=l$, every factor in $\tau_l(0)$ is a permutation. Its auxiliary trace is the one-site translation on the periodic chain. Inserting one derivative into the product and using
$P_{0j}A_{0r}=A_{jr}P_{0j}$ moves the differentiated factor to bond $(j,j+1)$. Hence
\[
\operatorname{cent}[\tau_l(0)^{-1}\tau_l'(0)]
 =\sum_j\operatorname{cent}
 \left[P_{j,j+1}\left.\frac{d}{dw}U_{ll,j,j+1}(\eta w)\right|_{w=0}\right].
\]
The operation $\operatorname{cent}(A)=A-\tr(A)I/\dim(A)$ is used on the space on which $A$ acts. This proves Eq.~\eqref{eq:regular-angular-bond}.

The connected part of the next derivative has support on three consecutive sites. Its reversal-odd Hermitian representative is
$Q_3=i\sum_j[h_j,h_{j+1}]$, with $h_j=h_{l;j,j+1}$ and indices understood modulo $M$. The next subsection gives the fourth charge. The local relation below determines its two-site correction. For $d_l=\dim V_l$, put
\[
K_{123}=[h_{12}+h_{23},[h_{12},h_{23}]] .
\]
If $K_{123}=Y_{23}-Y_{12}$ and $\tr Y=0$, then
\begin{equation}
Y=\frac{\operatorname{tr}_1K_{123}}{d_l}
 +\frac{\operatorname{tr}_{12}K_{123}}{d_l^2}\otimes I_{d_l}.
\label{eq:Y-partial-trace}
\end{equation}
In the first term the remaining factors are identified with the two arguments of $Y$. In the second term the remaining one-site matrix is placed on the first argument. To verify the formula, trace $Y_{23}-Y_{12}$ over site 1. The result is $d_lY-(\operatorname{tr}_1Y)\otimes I$. A further trace over the first remaining factor recovers $\operatorname{tr}_1Y$ because $\tr Y=0$. Substitution gives Eq.~\eqref{eq:Y-partial-trace}. Thus $Q_4$ can be evaluated from the already defined bond matrix.

\subsection{Higher commuting charges}
For example, writing $h_j=h_{l;j,j+1}$, the next charge is the three-site current $Q_3=i\sum_j[h_j,h_{j+1}]$. The reversal-even fourth charge is
\[
 Q_4=\sum_j\left(
 2[h_j,[h_{j+1},h_{j+2}]]
 +\tfrac12[h_j-h_{j+1},[h_j,h_{j+1}]]
 +Y_{j,j+1}\right),
\quad
 [h_{12}+h_{23},[h_{12},h_{23}]]=Y_{23}-Y_{12}.
\]
The local relation fixes $Y$ up to a scalar, chosen to have zero trace. Equation~\eqref{eq:Y-partial-trace} constructs it explicitly from $h_l$ by partial traces. A real linear combination of these charges stays in the same commuting family at fixed deformation.

\subsection{Alternating chains}
For an even chain alternating $a,b$, the factors of $\tau_a(0)$ at $a$ sites are permutations and those at $b$ sites are invertible matrices $U_{ab}(0)$. The same permutation identity writes the row as a sublattice translation followed by a product of these mixed gates. Differentiating a mixed gate gives
$U_{ab}(0)^{-1}\left.\frac{d}{dw}U_{ab}(\eta w)\right|_{0}$
on two adjacent sites. Differentiating a matching permutation gives $h_a$ on the two neighboring $a$ sites, conjugated by the intervening mixed gate. Summing these insertions gives Eq.~\eqref{eq:alternating-angular-exchange}. The row with auxiliary label $b$ is obtained by exchanging the labels. Their logarithmic derivatives commute by Eq.~\eqref{eq:transfer-row} and the RTT relation.

\section{Spectral contours, gauges, and historical normalizations}
\label{app:real-forms}
The real angular branch used in the construction has $\alpha=\eta/2-iK'$. This appendix gives the reality tests and gauge comparisons for other spectral contours, followed by the historical normalization dictionary.
\subsection{Double-shift ratios and two spectral contours}
Set $a=\eta/2\in\mathbb R$ and $\alpha=a-iK'$. The ratios in Eq.~\eqref{eq:weights} use this shifted coordinate and preserve the original angular operators and metrics. At fixed $k$, the limit $\eta\to0$ sends $\alpha$ to the pole $-iK'$. We evaluate the combined ratios by limits.

For the fundamental seed, two different reality statements follow. On $u=\alpha+x_{\rm sp}$, $x_{\rm sp}\in\mathbb R$, the three ratios are real and $R(u)$ is Hermitian. On the perpendicular contour $u=\alpha+it$, $t\in\mathbb R$, conjugation gives
\[
R(\alpha+it)^\dagger=R(\alpha-it),\qquad
R(\alpha+it)^\dagger R(\alpha+it)=\rho(t)I_4,
\]
\[
\rho(t)=1+\sum_{i=1}^3|W_i(\alpha+it)|^2>0,\qquad
U_\perp(t)=\rho(t)^{-1/2}R(\alpha+it),\qquad U_\perp(t)^\dagger U_\perp(t)=I_4.
\]
These statements hold away from poles. The first identity follows from real Jacobi coefficients and their $2iK'$ quasiperiods: $\alpha^*=\alpha+2iK'$, and the numerator and denominator acquire the same signs. The second follows from the Jacobi addition identities and fundamental inversion; its scalar is also $\operatorname{tr}(R^\dagger R)/4$, yielding the expression above. At $t=0$, $R(\alpha)=2P$ and $U_\perp(0)=P$. Real offsets generally give unequal singular values: Hermiticity alone does not imply scalar-normalizable unitarity.

\paragraph{Fused numerical tests.}
In each product coefficient metric set $\widetilde R=QRQ^{-1}$, $Q^\dagger Q=\mathsf H_l\otimes \mathsf H_m$. We tested all nine ordered pairs $l,m\in\{p,d,f\}$ at $(k,\eta)=(0.4,0.18),(0.6,0.23),(0.8,0.17)$, with $x_{\rm sp}=0.31,0.71$ and $t=0.17,0.31,0.71$. For the adopted branch the 54 real-offset matrices are Hermitian to relative error $1.81\times10^{-12}$, and the 81 imaginary-offset matrices satisfy
\[
\widetilde R^\dagger\widetilde R=\lambda I,\qquad
\lambda=\frac{\operatorname{tr}(\widetilde R^\dagger\widetilde R)}{(N_l+1)(N_m+1)}>0
\]
with maximum relative scalar-Gram defect $1.91\times10^{-12}$. Dividing by $\sqrt\lambda$ gives the same bound on normalized unitarity. Hermitian analyticity is supported to $1.91\times10^{-12}$. The half-period identities reproduce the same offset matrices. Nine auxiliary RLL and three mixed fused Yang--Baxter checks give residuals below $4.63\times10^{-14}$. These residuals apply at the listed sample points.

\paragraph{Real half-parameter}
Taking $\alpha=a\in\mathbb R$ also makes the fundamental ratios Hermitian on $a+\mathbb R$ and scalar-normalizable unitary on $a+i\mathbb R$, by the same conjugation and inversion argument. However, now $\eta=2a+2iK'$ is complex: this is a different reality contour in the source coordinates, potentially related to the original one by a gauge transformation (see the next subsection). The continued old triplet metric is proportional to $\operatorname{diag}(1,-\dn(2a),-\cn(2a))$ on the small-$a$ branch, and is indefinite. One cannot use it as a positive physical inner product.

A new positive coefficient metric can nevertheless exist. At $k=0.6,a=0.115$, a numerical solution for the triplet is
\[
H'_p=\operatorname{diag}(1,0.9906586804,0.9738306679).
\]
Positive coefficient metrics were also found for $l\in\{d,f\}$ at all three real-$a$ reference points above. The signs of the normalization square roots $\sqrt{(J_i+J_j)(J_i+J_h)}$ must be chosen coherently so that the represented Sklyanin relations and RLL hold, and the intertwining basis maps must be reidentified; blindly continuing principal roots or the old analytic triplet map is insufficient. Our code selects the root orientation using RLL and solves new basis maps numerically. With those choices the same 54 Hermiticity and 81 scalar-unitarity tests give maximum errors $3.68\times10^{-13}$ and $5.64\times10^{-13}$, respectively. Independent mixed fused YBE residuals are below $1.72\times10^{-13}$. Reversing the selected root orientation gives RLL errors $0.033$--$0.080$, a useful failure control.

The real-$\alpha$ limit gives $D_p\to\operatorname{diag}(1,-1,-1)$ as $a\to0$. The original branch gives $D_p\to I$. The real-offset regular logarithmic derivative uses a real prefactor to preserve Hermiticity. The next subsection gives the fundamental gauge relation explicitly.

\subsection{Hermiticity contours and half-period equivalence}
\label{sec:real-form-choice}
The equation $\alpha=\eta/2-iK'$ is our coordinate dictionary. Its substantive reality restriction is $\eta\in\mathbb R$, equivalently $\Im\alpha=-K'$. Neither Yang--Baxter integrability nor fundamental regularity forces this line. For nonsingular complex $\alpha$, the ratio seed remains a meromorphic YBE solution and $R(\alpha)=2P$. The accompanying representation parameters must change consistently: retaining the old $D,L_i$ while changing only $\alpha$ generally violates RLL. The finite-dimensional theta-function construction itself is not restricted to real shift; its positive adjoint structure is additional \cite{krichever1998vacuum}.

The broader alternatives have been investigated. Ercolessi, Evangelisti, Franchini and Ravanini \cite{ercolessi2013modular} extend the elliptic XYZ parametrization across its full phase diagram and identify modular transformations with permutations and sign changes of spin couplings. Their analysis includes alternating-site $\pi$ rotations. Thus a principal parameter regime is not a unique integrable physical regime. This does not classify every positive realization of our angular operators or establish novelty of any extension below.

\paragraph{Hermiticity contours.}
The regular density of the fundamental matrix is
\[
h_\alpha=\frac12P\,\left.\frac{d}{d\zeta}R(\alpha+\zeta)\right|_{\zeta=0}.
\]
After removing its scalar part,
\[
h_\alpha^\circ=\frac14\sum_{i=1}^3\bigl(W_i'(\alpha)-W_j'(\alpha)-W_h'(\alpha)\bigr)\,\sigma_i\otimes\sigma_i,
\qquad \{i,j,h\}=\{1,2,3\}.
\]
If the spectral tangent is $\zeta=ct$, $t\in\mathbb R$, the density is $c h_\alpha^\circ$. The linear map from $(W_1'(\alpha),W_2'(\alpha),W_3'(\alpha))$ to these three couplings is invertible. Hence Hermiticity in the fixed Pauli adjoint requires all $c W_i'(\alpha)$ to be real. The Jacobi derivatives give
\[
\begin{aligned}W_1'(\alpha)W_2'(\alpha)&=-(1-k^2\sn^2(\alpha,k)),\\ W_1'(\alpha)W_3'(\alpha)&=-k^2(1-\sn^2(\alpha,k)),\\ W_2'(\alpha)W_3'(\alpha)&=k^2\sn^2(\alpha,k).\end{aligned}
\]
Since the difference of the last two products is $k^2$, their reality after multiplication by $c^2$ requires $c^2\in\mathbb R$ and then $\sn^2(\alpha,k)\in\mathbb R$. For fixed $0<k<1$, ordinary nondegenerate Hermitian densities therefore occur on the real loci of this elliptic coordinate, with a real or imaginary tangent. Up to periods and reflections, the usual four boundary contours are
\begin{center}
\begin{tabular}{lll}
\toprule
Parameter contour & $\sn^2(\alpha,k)$ & Hermitian spectral tangent\\
\midrule
$\alpha=a$, $0<a<K$ & $(0,1)$ & $\zeta=x_{\rm sp}$\\
$\alpha=a+iK'$, $0<a<K$ & $(1/k^2,\infty)$ & $\zeta=x_{\rm sp}$\\
$\alpha=ib$, $0<b<K'$ & $(-\infty,0)$ & $\zeta=ix_{\rm sp}$\\
$\alpha=K+ib$, $0<b<K'$ & $(1,1/k^2)$ & $\zeta=ix_{\rm sp}$\\
\bottomrule
\end{tabular}
\end{center}
Here $x_{\rm sp}\in\mathbb R$ and endpoints and poles require limits. The adopted $a-iK'$ contour is equivalent to the second one by simultaneous $2iK'$ quasiperiod translation of numerator and denominator. On each edge the fundamental matrix is Hermitian along the listed tangent and scalar-normalizable unitary along the perpendicular tangent. These statements concern the seed and a fixed adjoint; they are not a classification of all higher-spin positive metrics. Generic complex $\alpha$ preserves integrability while failing Hermiticity, even after a common phase adjustment of the three couplings. Changing $k$ by modular transformations gives further equivalent descriptions; no one rectangular chart is privileged.

\paragraph{Half-period gauge relation.}
Let $a$ be real, $\alpha_-=a-iK'$, and distinguish the two seeds at the same offset $\zeta$:
\[
A(\zeta)=R_{\alpha_-}(\alpha_-+\zeta),\quad
B(\zeta)=R_a(a+\zeta).
\]
Jacobi half-period identities give exactly
\[
\boxed{\quad B(\zeta)=\frac{(I\otimes\sigma_1)A(\zeta)(\sigma_1\otimes I)}{W_1(\alpha_-+\zeta;2a)}.\quad}
\]
Expanding the Pauli products gives precisely the real-$a$ ratios: division by $W_1(\alpha_-+\zeta;2a)$ interchanges the roles of $W_2$ and $W_3$ in the last two coefficients. For checked operators this becomes an ordinary similarity up to scalar:
\[
\check B(\zeta)=\frac{(\sigma_1\otimes I)\check A(\zeta)(\sigma_1\otimes I)}{W_1(\alpha_-+\zeta;2a)}.
\]
Since $W_1(\alpha_-;2a)=1$ and $\check A(0)=2I$, the centered regular densities satisfy
\[
\boxed{\quad h_B^\circ=(\sigma_1\otimes I)h_A^\circ(\sigma_1\otimes I).\quad}
\]
Writing the three coefficients of $\sigma_i\otimes\sigma_i$ as $(g_x,g_y,g_z)$, the fundamental two-site spectra coincide under the coupling map $(g_x,g_y,g_z)\mapsto(g_x,-g_y,-g_z)$. This is the alternating-site rotation mechanism familiar in XYZ chains, now expressed in our precise normalization. The unchecked $R$ relation is a two-sided gauge relation, not a same-factor similarity of $A$ itself.

On an open chain or an even periodic chain, rotating every other site by $\sigma_1$ implements this equivalence on every bond. An odd periodic ring cannot support that alternating assignment on every edge: one bond remains twisted. External fields and measured spin observables must be rotated as well. In higher-dimensional systems, a bipartite interaction graph permits the same spin-$\tfrac12$ argument; odd cycles can obstruct it. 

\paragraph{Higher-spin checks and Hamiltonian conventions.}
We compared both branches at the three deformations used above. Let $Q_l^\pm=(\mathsf H_l^\pm)^{1/2}$ convert their coefficient metrics to orthonormal form and $T_l^\pm$ map the common symmetric tensors to their respective raw bases ($+$ for real $a$, $-$ for $a-iK'$). Normalize
\[
Q_l^+T_l^+\,(Q_l^-T_l^-)^{-1}
\]
by a positive scalar. These maps are unitary to numerical accuracy. For equal spins the scalar-regularized matrices satisfy a product similarity by the tensor square of this map in 27 tests ($l\in\{p,d,f\}$; three deformations; three offsets), with maximum relative error $1.66\times10^{-12}$. The centered regular densities, using the same real energy scale $2a$, agree under that similarity within $2.05\times10^{-12}$; the derivative's Richardson discrepancy is below $6.24\times10^{-11}$.

For mixed spins one must also account for a discrete rotation. Use the same symmetric embedding $E_{N_l}$ as in fusion. The mixed-space map is
\[
\Xi_l=Q_l^+T_l^+\,
 \bigl(E_{N_l}^\dagger\sigma_1^{\otimes N_l}E_{N_l}\bigr)^{\,j_l\bmod 2}
 (Q_l^-T_l^-)^{-1},
\]
Each map uses the positive scalar normalization above. The reported numerical checks test the product maps $\Xi_l\otimes\Xi_m$ for all nine ordered pairings, allowing a nonzero scalar normalization. All 81 tests agree within $1.64\times10^{-12}$. These are constructive numerical checks in our chosen bases, not an all-parameter mixed-spin gauge theorem. The angular observables may be transported by the same unitary maps; they should not automatically be identified with the naively continued $D,L_i$.

The real-$\alpha$ presentation therefore recovers the regular transfer densities under the transformations above. The branch $\alpha=a-iK'$ retains the positive angular realization of Sections~\ref{sec:p-spectrum}--\ref{sec:f-spectrum} and its limit $D\to I$.

\subsection{Comparison of spectral conventions}
Takebe's theta shift is normalized; the theta conversion in Eq.~\eqref{eq:variablemap} gives $\tau_{\rm ell}=iK'/K$, $\eta_{\rm T}=\eta/(4K)$. This alone does not fix spectral origins, gauges or scalars. His printed Eq.~(2.4) \cite{takebe1996} has a numerator theta argument $u_{\rm T}$ in his own coordinate, inconsistent with its claimed regularity at zero. With the proposed argument $u_{\rm T}+\eta_{\rm T}$, the sampled fundamental Yang--Baxter and inversion residuals are below $7\times10^{-16}$. Our construction uses the independently checked Jacobi seed \eqref{eq:seed}. An entrywise conversion to Konno's different theta normalization remains separate.

\subsection{Relation to Fateev's elliptic spin-one matrix}
\label{sec:fateev-relation}
Fateev constructed an elliptic $21$-vertex matrix and used its commuting transfer matrix to obtain an anisotropic spin-one chain \cite{fateev1981}. Konno's Theorem~2.4, based on a note by Jimbo, gives an explicit gauge equivalence between this matrix and the $2\times2$ fusion of Baxter's matrix \cite{konno2005}. The two integers in that fusion notation count fundamental constituents. The resulting local spaces have dimension three, so the matrix corresponds to our $R^{(p,p)}$, not our dimension-$25$ $R^{(d,d)}$.

The relation can also be checked directly in our angular conventions. Using the original $J_2=\cn(\eta,k)$ and $J_3=\dn(\eta,k)$, define
\begin{equation}
G=\begin{pmatrix}0&\sqrt{J_3}&0\\1&0&0\\0&0&\sqrt{J_2}\end{pmatrix}.
\end{equation}
This map applies the square root of the triplet metric up to a common scalar and exchanges the first two Cartesian coordinates. Define a $21$-vertex representative directly from the constructed matrix by
\begin{equation}
R_{\rm 21v}(\zeta)=
 \gamma(\alpha+\zeta)^{-1}
 (G\otimes G)R^{(p,p)}(\alpha+\zeta)(G\otimes G)^{-1}.
\label{eq:fateev-gauge}
\end{equation}
This definition and the finite entry formulas of Section~\ref{sec:fused-higher} specify all of its entries within the present paper. Its historical identification uses Fateev's $F=1$ branch with the same modulus and shift.
The scalar matching our raw fusion convention is
\begin{align}
\gamma(u)={}&\frac{\sqrt{J_2/J_3}}{\sn(2\eta,k)}\Bigl\{
 \sqrt{(J_1+J_2)(J_1+J_3)}\nonumber\\
&\quad\times[J_2W_1(u+\eta/2)+J_3W_1(u-\eta/2)]\nonumber\\
&\quad+\sqrt{(J_2+J_3)(J_2+J_1)(J_3+J_1)(J_3+J_2)}\nonumber\\
&\quad\times W_2(u+\eta/2)W_3(u-\eta/2)\Bigr\}.
\end{align}
It follows from the exchange entry and was evaluated without fitting. Twelve real and complex spectral comparisons at three deformations give maximum relative discrepancy $3.47\times10^{-15}$; the transcribed Fateev amplitudes independently satisfy three sampled Yang--Baxter checks below $2.3\times10^{-16}$. The matrix, gauge, and scalar are fixed by the formulas above. These numerical comparisons complement the established fusion equivalence; they do not replace a general proof of every convention conversion.

The overall scalar matters when constructing a Hamiltonian. Fateev's $F=1$ representative is regular, $R_{\rm 21v}(0)=\sn(2\eta,k)P$, whereas our raw fused normalization has a scalar pole at $u=\alpha$. Its scalar must therefore be removed before taking the regular logarithmic derivative. Relativistic scattering normalization, crossing and physical unitarity require their own scalar conditions; the comparison above does not impose them. The relation identifies a historical antecedent and a useful independent benchmark for our angular-basis implementation.

\section{Computational conventions and verification scope}
\label{app:computation}
All mathematical definitions required to construct the matrices are given in this paper. The finite binary-word sum in Section~\ref{sec:fused-higher} specifies every higher-spin block. Sections~\ref{sec:p-spectrum}--\ref{sec:f-spectrum} give the functions and spectra, and Appendix~\ref{app:basis-map-details} gives the reconstruction maps. Matrix columns store images of input basis functions, and the second tensor index varies fastest.

In Wolfram Language, the elliptic parameter passed to \texttt{JacobiSN}, \texttt{JacobiCN}, and \texttt{JacobiDN} is $k^2$. Historical data files that use the key \texttt{u} for the untranslated spectral coordinate store $\zeta=u-\alpha$. The current article always uses $u$ for the translated coordinate and $\zeta$ for its offset.

The numerical results record finite checks at the stated parameters and precisions. The symbolic checks of the six singleton transitions use the exact difference operators in Appendix~\ref{app:operators}, the quadratic equations in Sections~\ref{sec:d-spectrum} and \ref{sec:f-states}, and Jacobi addition identities. They establish those transition identities on regular charts. Section~\ref{sec:angular-verification} specifies the scope of the angular closure checks. Appendix~\ref{sec:gamma-normalization} specifies the normalization and the remaining crossing question.

\begingroup\small\raggedright\setlength{\parskip}{0pt}
\bibliographystyle{unsrt}\bibliography{references}
\endgroup
\end{document}